\documentclass[twocolumn,,amssymb, nobibnotes, aps, prd, nofootinbib, longbibliography]{revtex4-2}

\usepackage{amsmath}
\usepackage{amsfonts}
\usepackage{amssymb}
\usepackage{booktabs}
\usepackage[shortlabels]{enumitem}
\usepackage[T1]{fontenc}
\usepackage{graphicx}
\usepackage[colorlinks=true, linkcolor=blue, urlcolor=blue, citecolor=blue]{hyperref}
\usepackage{siunitx}
\usepackage{slashed}
\usepackage{tikz}
\usepackage{xcolor}
\usepackage{multirow}
\usepackage{xspace}

\usetikzlibrary{shapes.geometric, arrows}
\tikzstyle{line}=[draw] 

\definecolor{theblue}{HTML}{95D0FC}
\definecolor{theorange}{HTML}{FBDD7E}
\definecolor{thepale}{HTML}{FEFFCA}

\definecolor{hotpink}{rgb}{1.0, 0.41, 0.71}

\newcommand{\A}{\mathcal{A}}

\newcommand{\bmax}{batchmax}

\newcommand{\BtSG}{B_{\textrm{tS}/\textrm{G}}}
\newcommand{\BSGL}{B_{\textrm{S}/\textrm{GL}}}
\newcommand{\BSGLtL}{B_{\textrm{S}/\textrm{GLtL}}}
\newcommand{\dl}{\Delta \lambda}
\newcommand{\dmax}{\texttt{distromax}}

\newcommand{\F}{\mathcal{F}}
\newcommand{\gauss}{\textrm{Gauss}}
\newcommand{\gumbel}{\textrm{Gumbel}}

\newcommand{\hyp}[1]{\mathcal{H}_{\textrm{#1}}}

\newcommand{\Pro}{\mathrm{P}}
\newcommand{\pdf}{p}

\newcommand{\mprop}{\textrm{MaxProp}}

\newcommand{\M}{\mathcal{M}}
\newcommand{\N}{\mathcal{N}}

\newcommand{\slambda}{\slashed{\lambda}}

\newcommand{\starf}{f^{*}}
\newcommand{\starx}{x^{*}}
\newcommand{\T}{\mathrm{T}}
\newcommand{\trans}{\mathcal{T}}

\newcommand{\xit}{\xi_{\textrm{T}}}
\newcommand{\Ft}{2\F_{\max}}
\newcommand{\Nlambda}{N_{\lambda}}

\begin{document}

\title{Empirically estimating the distribution of the loudest candidate\\from a gravitational-wave search}

\author{Rodrigo Tenorio} 
\email{rodrigo.tenorio@ligo.org}
\affiliation{Departament de F\'isica, Institut d'Aplicacions Computacionals i de Codi Comunitari (IAC3), Universitat de les Illes Balears, 
and Institut d'Estudis Espacials de Catalunya (IEEC), Carretera de Valldemossa km 7.5, E-07122 Palma, Spain}
\author{Luana M. Modafferi} 
\affiliation{Departament de F\'isica, Institut d'Aplicacions Computacionals i de Codi Comunitari (IAC3), Universitat de les Illes Balears, 
and Institut d'Estudis Espacials de Catalunya (IEEC), Carretera de Valldemossa km 7.5, E-07122 Palma, Spain}
\author{David Keitel}
\affiliation{Departament de F\'isica, Institut d'Aplicacions Computacionals i de Codi Comunitari (IAC3), Universitat de les Illes Balears, 
and Institut d'Estudis Espacials de Catalunya (IEEC), Carretera de Valldemossa km 7.5, E-07122 Palma, Spain}
\author{Alicia M. Sintes}
\affiliation{Departament de F\'isica, Institut d'Aplicacions Computacionals i de Codi Comunitari (IAC3), Universitat de les Illes Balears, 
and Institut d'Estudis Espacials de Catalunya (IEEC), Carretera de Valldemossa km 7.5, E-07122 Palma, Spain}

\date{\today}

\begin{abstract}
    Searches for gravitational-wave signals are often based on maximizing a detection statistic 
    over a bank of waveform templates,
    covering a given parameter space with a variable level of correlation.
    Results are often evaluated using a noise-hypothesis test,  
    where the background is characterized by the sampling distribution of the loudest template.
    In the context of continuous gravitational-wave searches,
    properly describing said distribution is an open problem:
    current approaches focus on a particular detection statistic and neglect template-bank correlations.
    We introduce a new approach using extreme value theory to describe the distribution of the loudest
    template's detection statistic
    in an arbitrary template bank.
    Our new proposal automatically generalizes to a wider class of detection statistics,
    including (but not limited to) line-robust statistics and transient continuous-wave signal hypotheses,
    and improves the estimation of the expected maximum detection statistic at a negligible computing cost.
    The performance of our proposal is demonstrated on simulated data
    as well as by applying it to different kinds of (transient) 
    continuous-wave searches using O2 Advanced LIGO data. 
    We release an accompanying Python software package, \dmax{}, implementing our new developments.
\end{abstract}

\maketitle

\section{Introduction}
\label{sec:introduction}
The search for gravitational-wave (GW) signals can be formulated as a multi-hypothesis test between a
background-noise hypothesis
and a set of signal hypotheses, each asserting the presence of a signal 
with a specific set of parameters~\cite{jaynes_2003}. 
Actual search implementations, however, usually split this process into three stages:
a detection stage, which simply assesses the presence of a feature in the datastream 
unlikely to be caused by noise (null-hypothesis test);  
a validation stage, 
in which candidates are sieved through a set of vetoes to discard any instrumental causes;
and a parameter-estimation stage, in which a proper Bayesian hypothesis test is carried out to
infer the actual parameters of any detected signal.
This division is motivated by the increasing computing cost of each stage~\cite{Owen:1995tm, Brady:1998nj},
as a simple null-hypothesis test (usually assuming Gaussian noise) is orders of magnitude more affordable
than a single parameter-estimation stage.

The standard detection stage consists in performing a finite number of detection statistic evaluations
over the parameter-space region of interest, 
usually using matched-filtering against a bank of waveform \emph{templates}~\cite{Sathyaprakash:1991mt, Owen:1995tm, 
Owen:1998dk, Allen:2005fk, Prix:2007ks, Harry:2009ea, Messenger:2008ta, Allen:2021yuy, Wagner:2021hgv}. 
Loud templates, i.e. those scoring a high detection statistic, are deemed ``signal candidates'' 
and selected for the validation stage.
The detection statistic can be usually interpreted as a Bayes factor, 
assessing the preference of the data for a particular signal hypothesis (represented by the template at hand) 
versus the background-noise hypothesis. 
Thus, the detection stage is a multi-hypothesis test in disguise in which parameter-space
\emph{marginalization} has been approximated to zeroth-order by \emph{maximization}~\cite{Prix:2009tq}.

Loudest candidates from a template bank fall generally into one of two categories:
The strongest excursions away from the background, 
such as an instrumental feature~\cite{2018PhRvD..97h2002C, Davis:2020nyf} 
or a very clear GW signal (such as GW150914~\cite{LIGOScientific:2016aoc}),
are usually comparatively simple to deal with,
as strong candidates tend to show characteristic signatures according to their cause.
But weaker outliers that are in principle compatible with both a weak signal
or an extreme event of the general noise background
require a more careful analysis.

While much of the statistical framework used in this work is generally applicable,
we mainly focus on the search for continuous gravitational-wave signals (CWs)~\cite{universe},
produced by long-standing quadrupolar deformations,
such as in the case of non-axisymmetric spinning neutron stars (NS)~\cite{Sieniawska_2019}.
From the point of view of the current generation of advanced detectors
(Advanced LIGO \cite{AdvancedLIGO}, Advanced Virgo~\cite{AdvancedVirgo}, and KAGRA~\cite{KAGRA:2018plz}),
they belong in the weak-signal regime, meaning they are expected to blend into the background distribution.
Characterizing the expected distribution of extreme background candidates is, thus, a simple approach to
identify interesting outliers in a search and quantify their significance.

Pioneering work on describing the distribution of the loudest candidate from a CW search
using the $\F$-statistic~\cite{JKS1998,PhysRevD.72.063006} was presented in~\cite{Wette:2009uea} 
and later extended in~\cite{Papa:2020vfz, Wette:2021tbv}.
Despite its wide applicability in the CW literature (see e.g.~\cite{Abadie:2010hv, LIGOScientific:2013wcb, 
PhysRevD.91.064007, LIGOScientific:2016ahk, Zhu:2016ghk, Abbott:2017pqa, 2019PhRvD.100f4058K}), 
basic assumptions of the method make it insufficient for realistic
template banks with a certain degree of correlation between neighbouring templates~\cite{Dreissigacker:2018afk}.
Latest developments on the subject used extreme value theory (EVT) to propose a suitable ansatz to circumvent 
the problems posed by template-bank correlations~\cite{Tenorio:2021njf};
but the concrete method requires re-evaluating full template banks many times
(similar in  spirit to that in~\cite{Papa:2020vfz})
and is thus computationally unsuitable for wide parameter-space searches.

This work proposes \dmax{},
a new method to describe the distribution of the loudest candidate stemming from a generic GW search. 
The method generalizes with respect to previous approaches presented 
in~\cite{Wette:2009uea, Wette:2021tbv} in two main aspects. First, the method is robust to
typical degrees of template-bank correlations arising either due to the overlap of nearby templates or mild 
non-Gaussianities in the data. Second, the method is applicable to a wider class of detection statistics, 
including other $\F$-statistic-based detection statistics such as 
line-robust statistics~\cite{Keitel:2013wga,Keitel:2015ova} or transient CW search statistics~\cite{Prix:2011qv},
as well as detection statistics from other search approaches.
An implementation of \dmax{} is publicly available as a homonymous Python package \cite{distromax}.

The paper is structured as follows: 
Section~\ref{sec:cw_searches} introduces basic data-analysis tools for CW searches
and discusses the origin of parameter-space correlations.
Section~\ref{sec:practical_applications} describes the quantitative effect of parameter-space correlations 
on the distribution of the loudest candidate, 
comparing standard approaches in the field to extreme value theory results. 
Section~\ref{sec:loudest_outlier} introduces \dmax{} to estimate 
the distribution of the loudest outlier of a search
and discusses its basic phenomenology on synthetic data.
In Section~\ref{sec:real_data}, we apply \dmax{} to the results of a search on O2 Advanced LIGO data
for (transient) CW signals. 
Appendix~\ref{sec:extreme_value_theory} collects basic results in extreme value theory 
and provides further references for the interested reader.
Appendix~\ref{sec:loud_disturbances} proposes a simple method to deal with narrow-band noise disturbances,
common in realistic CW searches. 
The robustness of \dmax{} to the presence of weak CW signals is discussed in Appendix~\ref{sec:injections}.
 
\section{Continuous wave searches}
\label{sec:cw_searches}
In this section, we revisit the basics of CW searches to frame our discussion of \dmax{}. 
Section~\ref{subsec:Fnoise} reintroduces the $\F$-statistic and explicitly constructs
its distribution under the noise hypothesis; 
Sec.~\ref{subsec:correlations} uses the explicit construction to discuss the two possible
origins of parameter-space correlations affecting a template bank; 
Sec.~\ref{subsec:Fsignal} completes the analysis deriving the standard result for the distribution of 
the $\F$-statistic under the signal hypothesis.

The response of a ground-based GW detector to a passing CW or long-duration CW-like transient (tCW) 
is given by the linear combination of four linear filters \cite{JKS1998,Prix:2011qv}
\begin{equation}
    s(t;\A, \lambda, \trans{}) = w(t;\trans{}) \sum_{\mu=0}^{3} \A_{\mu} \; h_{\mu}(t; \lambda) \;,
    \label{eq:h_of_t_A}
\end{equation}
where $\A$ represents the source's amplitude parameters, namely
GW amplitude $h_0$, inclination angle $\iota$, 
polarization angle $\psi$, and initial phase $\phi_0$,
which can be combined into the 
so-called JKS decomposition \mbox{$\{\A_{\mu}, \mu = 0, 1, 2, 3\}$}; 
and $\lambda$ describes the phase-evolution parameters,
namely the GW frequency and spindown $\{f_0, f_1, f_2, \dots\}$, 
the sky position $\vec{n}$, and possibly binary 
orbital parameters if the source orbits a companion.
The time-dependent quadratures $h_{\mu}(t; \lambda)$ 
encompass the detector's antenna pattern effects on the 
signal. 
The window function $w(t;\trans{})$ is a time-dependent amplitude modulation 
parametrized by the 
transient parameters $\trans{}$ to account for tCW signals \cite{Prix:2011qv}. 
The standard CW signal model is recovered for $w(t;\trans{}) = 1 \; \forall t$.

Given a datastream $x$, 
the detection problem consists in deciding between the background noise hypothesis $\hyp{N}$, 
under which the data stream contains only Gaussian noise $x = n$, 
and the signal hypothesis $\hyp{S}$, 
according to which there is a (t)CW signal with a defined set of parameters 
$x = n + s(\lambda, \A, \trans{})$. 
Further hypotheses accounting for different non-Gaussian populations,
such as narrow instrumental artifacts in the data~\cite{2018PhRvD..97h2002C},
can be also included in the analysis~\cite{Keitel:2013wga, Keitel:2015ova},
although the usual approach is to apply post-processing veto
strategies targeting specific types of disturbances~\cite{SanchodelaJordana:2008dc, Leaci:2015iuc, 
PhysRevD.91.064007, Zhu:2017ujz}.

\subsection{$\F$-statistic under the noise hypothesis}
\label{subsec:Fnoise}

A basic tool to conduct CW searches is the $\F$-statistic, first introduced in 
\cite{JKS1998,PhysRevD.72.063006} as a maximum-likelihood estimator with respect to amplitude parameters $\A$, 
and later re-introduced in a Bayesian context~\cite{Prix:2009tq, Whelan:2013xka, 
Dhurandhar:2017rlr, Bero:2018xyq}. 
The basic idea is to exploit the linear dependency of Eq.~\eqref{eq:h_of_t_A} on $\A$ to analytically marginalize 
the matched-filtering likelihood using a suitable set of priors. 
The result can be readily expressed as a quadratic form \cite{CFSv2}
\begin{equation}
    2\F(\lambda) = \sum_{\mu, \nu = 0}^{3}{ x_{\mu}(\lambda) \; \M^{-1}_{\mu \nu}(\lambda) \; x_{\nu}(\lambda) }\;,
    \label{eq:F_M}
\end{equation}
where $x_{\mu}$ are the projections of the data stream $x$ onto the four quadrature functions
\begin{equation}
    x_{\mu}(\lambda) = \langle h_{\mu}(\lambda), x\rangle 
    \label{eq:x_mu}
\end{equation}
and $\M^{-1}(\lambda)$ is the inverse Gram matrix associated to the four quadrature functions
\begin{equation}
    \M_{\mu \nu}(\lambda) = \langle h_{\mu}(\lambda), h_{\nu}(\lambda) \rangle \;.
\end{equation}
The functional scalar product~\cite{Finn:1992wt}
\begin{equation}
    \langle x, y \rangle = 4 \; \mathfrak{R}\int_{0}^{\infty} \mathrm{d}f\;
    \frac{x(f) \; y^{*}(f)}{S_{\textrm{n}}(f)}
    \label{eq:scalar_product}
\end{equation}
accounts for the presence of correlated noise in the data stream through the single-sided power spectral
density (PSD) $S_{\textrm{n}}$. Current implementations of Eq.~\eqref{eq:scalar_product} make use of the
so called $\F$-statistic \emph{atoms} \cite{CFSv2},
evaluated over individual Short Fourier Transforms (SFTs) of the data.
These could be simply described as a set of 
complex-valued spectrograms (from now on \emph{atomic spectrograms}) containing both phase and 
amplitude information, whose proper combination results in an efficient computation of Eq.~\eqref{eq:F_M}.

Under the noise hypothesis $\hyp{N}$, the data stream is composed of zero-mean Gaussian noise 
and Eq.~\eqref{eq:x_mu} implies the four projections \mbox{$\{n_{\mu}(\lambda)\}$} are drawn from a 
\mbox{4-dimensional} Gaussian distribution with covariance matrix $\M(\lambda)$. Hence, 
\begin{equation}
    \{n_{\mu}(\lambda)\} \sim \gauss(0, \M(\lambda)) \;,
    \label{eq:n_Gauss}
\end{equation}
and $n_{\mu}(\lambda)$ values can be constructed as a linear combination of four zero-mean unit-variance 
Gaussian random variables
\begin{equation}
    n_{\mu}(\lambda) = \sum_{\nu=0}^{3} L_{\mu \nu}(\lambda) g_{\nu}[\lambda]\;,
    \label{eq:Gauss_Cholesky}
\end{equation}
where \mbox{$g_{\nu}[\lambda]\sim \gauss(0, 1)$} and $L$ is a $4\times4$ matrix such that
\mbox{$L L^{\T} = \M$} (e.g.~Cholesky decomposition). 
Here the square brackets indicate that Gaussian numbers are to be
drawn independently for each template $\lambda$,
but their distribution does not depend on $\lambda$;
as opposed to round brackets, which represent deterministic relations.

Introducing these results into Eq.~\eqref{eq:F_M},
\begin{equation}
    2\F(\lambda)= \sum_{\mu=0}^{3} g_{\nu}[\lambda]^2 \;,
    \label{eq:obvious_chi2}
\end{equation}
we obtain $2\F(\lambda)$ as the Euclidean norm of a \mbox{4-dimensional} Gaussian vector.
Consequently, the probability distribution associated to $2\F$ under the noise hypothesis $\hyp{N}$
for a fixed template $\lambda$ is given by a $\chi^2$ distribution with four degrees of freedom
\begin{equation}
    \pdf(2 \F | \hyp{N}) = \chi^{2}_{4}(2\F)\;.
    \label{eq:2F_chi2}
\end{equation}

\subsection{Template-bank correlations}
\label{subsec:correlations}
The statistical properties of the right-hand side of Eq.~\eqref{eq:obvious_chi2} are independent
of the specific phase-evolution template $\lambda$ under consideration.
This suggests that evaluating $2\F$ over a template bank using a single noise realization could,
under suitable conditions, 
be equivalent to evaluating $2\F$ for a single template over an ensemble of noise realizations.

Gaussian vectors $\{g_{\nu}[\lambda]\}$ are constructed from a noise stream as follows:
\begin{equation}
    g_{\nu}[\lambda] = \sum_{\nu=0}^{3} L_{\nu \mu}^{-1}(\lambda) \; \langle h_{\mu}(\lambda), n \rangle\;.
\end{equation}
The noise stream is projected onto four different deterministic functions of $\lambda$, $\{h_{\mu}(\lambda)\}$, 
and combined using a set of weights $L_{\nu \mu}^{-1}(\lambda)$, also dependent on $\lambda$.
Such a projection is a weighted average of the atomic spectrogram bins visited by the frequency-evolution 
track associated to $\lambda$.
Since the atomic spectrograms are constructed using finite time and frequency resolutions,
the number of independent Gaussian vectors constructible out of them is equivalent to the 
number of templates with \emph{non-overlapping} frequency tracks over the spectrograms
(i.e.~crossing different spectrogram bins).
This result was stated in a simpler fashion in~\cite{Wette:2009uea} 
by arguing that the typical number of bins in a narrow-banded atomic spectrogram 
is orders of magnitude smaller than the number of templates in a typical CW search crossing said spectrogram.

As discussed in \cite{Tenorio:2020cqm}, the average dissimilarity in frequency-evolution tracks
of nearby parameter-space points is related to the fractional loss in detection statistic, usually
referred to as \emph{mismatch} \cite{Prix:2006wm}
\begin{equation}
    m = 1 - \frac{2\F(\lambda + \dl)}{2\F(\lambda)} 
    \simeq \sum_{i, j} \dl_{i} \dl_{j} g_{ij} + \mathcal{O}(\dl^3)\;,
\end{equation}
where $\dl_{i}$ represents an offset in an arbitrary parameter-space dimension and $g_{ij}$ is the 
parameter-space metric \cite{Prix:2006wm, Pletsch:2010xb, Wette:2013wza, Wette:2015lfa,
Leaci:2015bka,Allen:2019vcl}. In the context of a grid-based CW search, 
the parameter-space metric can be employed to set up a template bank at a pre-specified maximum mismatch
value \cite{Prix:2007ks, Wette:2014tca, Allen:2021eju}: the higher the mismatch, the coarser the template bank. 
An ensemble of templates with non-overlapping frequency-evolution tracks, then, 
corresponds to a coarse-enough template bank in the sense of large parameter-space mismatch.

In a real search, template banks tend to be set up using a moderate mismatch (e.~g.~$m \sim 0.2$) in order 
to produce dense-enough parameter-space coverings~\cite{Prix:2007ks}. A first kind
of template-bank correlation arises, then, as a result of the template-bank construction strategy. Latest
developments on the subject \cite{Wette:2016raf, Allen:2019vcl, Allen:2021yuy, Allen:2021eju}, 
however, suggest higher mismatch values ($m \sim 1$) could actually be compatible with a successful CW search, 
potentially suppressing the effect of these correlations.

A second kind of template-bank correlations, briefly discussed in \cite{Tenorio:2021njf}, arises due to 
non-Gaussianities in the data 
(e.g.~narrow instrumental features~\cite{2018PhRvD..97h2002C} or transient ``pizza-slice'' 
disturbances \cite{piccinni2014:_thesis, Keitel:2015ova, LIGOScientific:2016ahk}).
In this case, 
it is not a matter of re-using the same data on different templates; rather, 
a region of a priori independent spectrogram bins gets correlated due to the presence of a strong disturbance. 
As a result, non-overlapping templates crossing said correlated spectrogram region become correlated as well.

This same formalism applies to the search for tCWs, as the standard strategy in such cases is either to maximize
or marginalize out any dependency on the transient parameters~\cite{Prix:2011qv},
obtaining in the end a detection statistic over an equivalent template bank to that of CW searches. Discussion
on specific tCW detection statistics is postponed to Sec.~\ref{sec:real_data}.

The presence of correlations in a template bank, thus, is a generic property of (t)CW searches, and their effects
on any newly proposed method should be properly understood before attempting to interpret results on a real setup.

\subsection{$\F$-statistic under the signal hypothesis}
\label{subsec:Fsignal}
We conclude this summary of standard CW search methods
by considering the distribution of the $\F$-statistic when there is a signal in the data.
To derive its probability distribution under the signal hypothesis $\hyp{S}$, we simply apply
\mbox{$n = x - s(\lambda, \A)$} and repeat the same reasoning up to Eq.~\eqref{eq:Gauss_Cholesky}, 
obtaining
\begin{equation}
    x_{\mu}(\lambda) = \sum_{\nu=0}^{3} L_{\mu \nu}(\lambda) 
    \left(g_{\nu}[\lambda] + m_{\nu}(\lambda, \A)\right)\;,
\end{equation}
where 
\begin{equation}
    m_{\nu}(\lambda, \A) = \sum_{\kappa = 0}^{3} L^{-1}_{\nu \kappa}(\lambda) s_{\kappa}(\lambda, \A)
    \label{eq:non_central_mean}
\end{equation}
and, consequently, $g_{\nu}[\lambda] + m_{\nu}(\lambda, \A)$ is a Gaussian random number with mean  $m_{\nu}$ and 
unit variance. 

Introducing these results in Eq.~\eqref{eq:F_M}, 
\begin{equation}
    2\F(\lambda)= \sum_{\nu=0}^{3} \left(g_{\nu}[\lambda] + m_{\nu}(\lambda, \A) \right)^2 \;,
    \label{eq:obvious_chi2}
\end{equation}
the $2\F$ under the signal hypothesis corresponds to the
norm of a four-dimensional uncorrelated Gaussian vector with identity covariance matrix and mean vector equal to 
$\{m_{\nu}(\lambda, \A)\}$. The probability distribution is, as a result, a \emph{non-central} chi-squared 
distribution with four degrees of freedom
\begin{equation}
    p(2\F|\rho^{2}\hyp{S}) = \chi^2_{4}(2\F;\rho^2) \;,
\end{equation}
where the non-centrality parameter $\rho^2$ is defined as
\begin{equation}
    \rho^2 = \sum_{\nu=0}^{3} m_{\nu}^2 
    = \sum_{\mu,\nu=0}^{3}\A_{\mu} \M_{\mu \nu} \A_{\nu}
    = \langle s, s \rangle\;.
\end{equation}
This quantity is referred to as the (squared) signal-to-noise ratio (SNR) in the literature. Concretely, $\rho^2$
is the maximum attainable SNR corresponding to the case where signal parameters are perfectly matched by a 
phase-evolution template \cite{CFSv2}.
 
\section{Loudest candidates,\\the ``effective number of templates'',\\and extreme value theory}
\label{sec:practical_applications}
The standard problem of estimating the distribution of the loudest candidate in a search is posed as follows: 
Let $\xi = \{\xi_i, i = 1, \dots, \mathcal{N}\}$ be a set of detection statistic values 
obtained by evaluating a template bank with $\N$ templates in a noise-only data stream. 
Let $f$ be the probability distribution of such a detection statistic under the noise hypothesis. 
Describe the probability distribution of the loudest candidate \mbox{$\max_{i=1, \dots, \N}{\xi_{i}}$}.

We can easily construct said distribution using the joint cumulative density function (CDF) 
of the entire template bank
\begin{equation}
    \Pro(\max_{i} \xi_{i} \leq \xi^{*} | \N) = 
    \Pro\left(\xi_{1} \leq \xi^{*}
    \;\textrm{and}\; \dots \;\textrm{and}\; 
    \xi_{\N}\leq \xi^{*} \right)
\;.
\end{equation}
For the case of an uncorrelated template bank, each template is independent and the joint CDF factors into 
the product of individual CDFs:
\begin{equation}
    \Pro(\max_{i} \xi_{i} \leq \xi^{*}| \N) = \prod_{i=1}^{\N} \Pro(\xi_{i} \leq \xi^{*}) 
    = \left[ \int^{\xi^{*}} \mathrm{d}\xi f(\xi) \right]^{\N}\;.
    \label{eq:independent_cdf}
\end{equation}
Consequently, the probability density function associated to \mbox{$\xi^{*} = \max_{i=1, \dots, \N}{\xi_{i}}$}
is simply
\begin{equation}
    \pdf(\xi^{*} | \N) = \N f(\xi^{*}) \left[ \int^{\xi^{*}} \mathrm{d}\xi f(\xi) \right]^{\N - 1}\;.
    \label{eq:p_xi_star}
\end{equation}

Template-bank correlations imply that we sample fewer independent combinations of the data
than with an uncorrelated bank of the same $\N$.
In other words, they reduce the ``trials factor'' of a search,
diminishing the expected detection statistic of the loudest candidate in a similar fashion to evaluating
a smaller template bank. 
Given a fixed false-alarm probability,
neglecting template-bank correlations and naively using Eq.~\eqref{eq:p_xi_star} would overestimate
the corresponding threshold, potentially leading to missing interesting candidates.

Extensive analyses in \cite{Wette:2009uea, Wette:2011eu} concluded the effect of template-bank correlations
on Eq.~\eqref{eq:p_xi_star} could be reproduced to an acceptable level by adjusting $\N$ to the 
``effective number of templates'' in the template bank at hand. 
Although in some cases an empirical estimate was possible~\cite{Abadie:2010hv}, most applications
obtained an effective number $\N'$ via numerical fits to search results~\cite{PhysRevD.91.064007,
LIGOScientific:2013wcb, Zhu:2016ghk, Abbott:2017pqa, 2019PhRvD.100f4058K, Papa:2020vfz}.
Further studies on this topic \cite{Dreissigacker:2018afk}, however, 
exposed a systematic discrepancy between the family of distributions spanned by Eq.~\eqref{eq:p_xi_star} 
and the actual distributions obtained due to template-bank correlations.

\begin{figure}
    \includegraphics[]{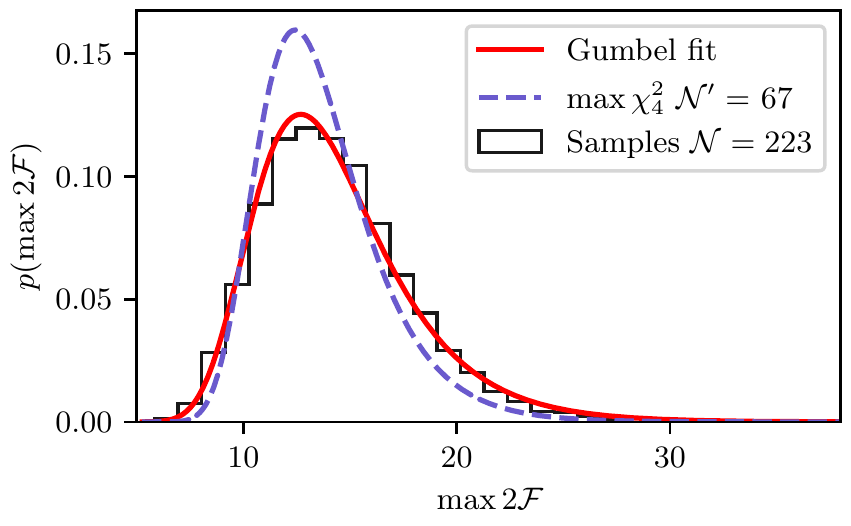}
    \includegraphics[]{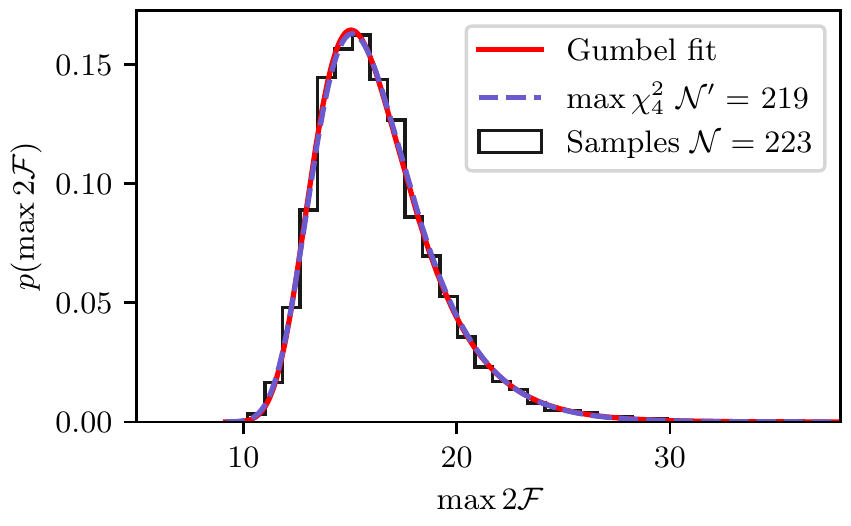}
    \caption{Distribution of the loudest $2\F$ values produced by the evaluation of a template bank on
    a Gaussian-noise data stream lasting for 7 days. The template bank was set up using the \texttt{gridType=8}
    option of \texttt{ComputeFstatistic\_v2} \cite{lalsuite} with mismatch $m = 0.2$, 
    $f_0=\SI{49.5}{\hertz}$ and $f_1=\SI{-10}{
ano\hertz/\s}$ covering bands of $\Delta f_0 = \SI{0.22}{\hertz}$ 
    and $\Delta f_1 = \SI{45}{\pico\hertz/\s}$. 
    The sky position was fixed to a fiducial value $(\alpha, \delta) = (0, 0)$ in equatorial coordinates.
    Loudest values were obtained by selecting the loudest $2\F$ over different segmentations
    of the template bank. 
    The upper panel corresponds to selecting the loudest value within every $\SI{5}{\milli\hertz}$ subband.
    The lower panel corresponds to shuffling the results and taking the loudest values from batches of the
    same size as the subbands.
    The stepped line is the histogram of the data; the dashed line is the best
    fit value $\N'$ for $\N$ in Eq.~\eqref{eq:p_xi_star}; and the solid line is the best fit of a Gumbel
    distribution.}
    \label{fig:twoF_correlations}
\end{figure}

An example of this discrepancy is illustrated in Fig.~\ref{fig:twoF_correlations}.
We evaluated a template bank containing \mbox{$2.23 \times 10^6$} CW templates over
frequency and spindown parameters \mbox{$(f_0, f_1)$} with a realistic mismatch of \mbox{$m = 0.2$}
on 7 days of simulated Gaussian noise.
We grouped the resulting \mbox{$2\F$-statistic} values into batches containing 223 templates each,
from which the loudest $2\F$-statistic value was retrieved.
The effect of parameter-space correlations was tested by either grouping templates within contiguous 
$\SI{5}{\milli\hertz}$ frequency bands or pooling an equivalent number of templates after randomly 
shuffling the results. 

Shuffling the results before retrieving the loudest value tends to break any contribution from parameter-space
correlations, as nearby templates are likely to end up in different batches. The resulting distribution can
be properly fitted assuming an uncorrelated template bank. The apparent mismatch between the obtained effective
number of templates $\N'=218$ and the actual number of independent templates $\N=223$ is consistent with the
basic claim in \cite{Abadie:2010hv} about the robustness of Eq.~\eqref{eq:p_xi_star} with respect to small
changes in $\N$.

Grouping contiguous frequency bins, on the other hand, produces a distribution out of the scope of Eq.~\eqref{eq:p_xi_star}.
This was understood in \cite{Tenorio:2021njf} using extreme value theory (EVT).\footnote{
    We acknowledge previous attempts to apply EVT to the search for CWs~\cite{Rover:2011zq, 2017PhRvD..96j2006S}.
    Ref.~\cite{Tenorio:2021njf} is the first work presenting a \emph{practical} application of 
    an EVT result improving over previous methods.
}
In the limit of $\N \rightarrow \infty$, Eq.~\eqref{eq:p_xi_star} converges to a \mbox{max-stable}
distribution~\cite{leadbetter1983extremes, coles2001introduction, beirlant2004statistics, 
de2006extreme, embrechts2013modelling}, whose functional form is determined by the behaviour of the tail of the distribution $f$
of the detection statistic.
We are primarily interested in the cases when $f$ is a $\chi^2$, $\Gamma$ or Gaussian distribution, 
for all of which $\pdf(\xi^{*} | \N)$ converges to a Gumbel distribution
\begin{equation}
    \gumbel(x ; \mu, \sigma) = 
    \frac{1}{\sigma}
    \exp{\left[ -\left(\frac{x - \mu}{\sigma}\right)
    -e^{- \left(\frac{x - \mu}{\sigma}\right)} \right] }\;
    \label{eq:gumbel_def}
\end{equation}
where $\mu$ and $\sigma$ refer to the location and scale parameters, respectively.
Analytical expressions for $\mu(\N)$ and $\sigma(\N)$ for different distributions $f$ 
are widely available in the literature~\cite{embrechts2013modelling, GASULL2015376, RePEc:spr:testjl:v:24:y:2015:i:4:p:714-733}.
As discussed in Sec.~\ref{sec:cw_searches}, the individual $2\F$ follow a $\chi^{2}_4$ distribution on Gaussian noise;
the scale parameter of the associated Gumbel distribution is, consequently, fixed to 
$\sigma=2$ \emph{regardless of} the value of $\N$ \cite{RePEc:spr:testjl:v:24:y:2015:i:4:p:714-733}. 
Naively fitting Eq.~\eqref{eq:p_xi_star} corresponds then to simply adjusting the location of the Gumbel
distribution's peak, as clearly seen in the top panel of Fig.~\ref{fig:twoF_correlations}. The apparent mismatch
is resolved if one instead tries to fit \emph{both} the location and scale parameters of the Gumbel distribution
to the data. 

This solution can be directly applied to computationally cheap searches,
such as narrow-band searches~\cite{LIGOScientific:2017ytx,LIGOScientific:2019mhs},
directed searches using the Viterbi method~\cite{Middleton:2020skz, Jones:2020htx, Beniwal:2021hvc},
or the follow-up of particular outliers~\cite{Tenorio:2021njf}, 
using the ``off-sourcing'' method~\cite{Isi:2020uxj}. 
The basic idea is that evaluating the same template bank while shifting the sky position 
away from the outlier will sample a subset of templates uncorrelated to the outlier but with a 
consistent noise background. Each off-sourced template bank is thus equivalent to a different noise realization. 
To describe the distribution of the loudest outlier, then, it suffices to evaluate \mbox{$10^{2}-10^{3}$} 
off-sourced template banks retrieving the loudest outlier of each. A Gumbel distribution can then be fitted 
to the resulting distribution~\cite{Tenorio:2021njf}. 

\begin{figure}
    \includegraphics[]{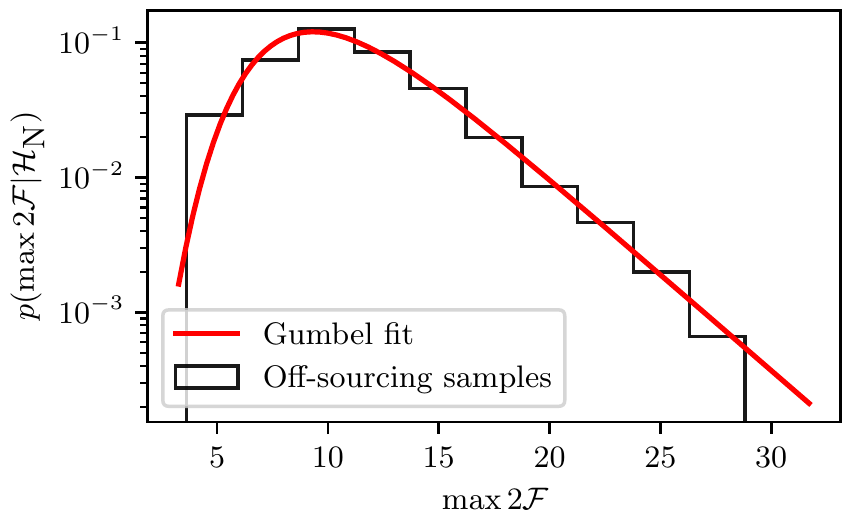}
    \caption{
        Distribution of the loudest $2\F$ of a template bank obtained using 600 off-sourcing evaluations.
        The template bank corresponds to MCMC samples from a fully-coherent follow-up of a simulated signal in
        Gaussian noise, in a similar manner to Fig.~1 in \cite{Tenorio:2021njf}.
        Each histogram entry corresponds to the loudest $2\F$ retrieved from the template bank
        evaluated at a different right ascension. 
        The solid line represents the fit of a Gumbel distribution using 
        \texttt{scipy.stats.gumbel\_r.fit}~\cite{2020SciPy-NMeth}.
    }
    \label{fig:offsourcing_example}
\end{figure}

Figure~\ref{fig:offsourcing_example} exemplifies this procedure using a template bank constructed by
MCMC sampling as implemented in \texttt{PyFstat} \cite{Ashton:2018ure,Keitel2021}.
The template bank, containing $2.5\times 10^5$ highly correlated templates across frequency,
spindown and sky positions, was shifted to 600 different sky positions excluding a $\ang{90}$ wedge around
the outlier's position. 
The resulting distribution is well described by a Gumbel distribution, 
with parameters fitted using a standard maximum-likelihood estimation.

Extreme value theory thus allows directly tackling the actual problem posed at the start of this section,
namely estimating the distribution of the loudest candidate under the noise hypothesis.
The ``effective number of independent templates'' does not play any major role,
as the parameters being fitted are the location and scale of a well-described probability distribution.
 
\section{How to estimate the distribution of the loudest outlier: an empirical approach}
\label{sec:loudest_outlier}
Estimating the loudest candidate's distribution typically entails fitting an ansatz
to a set of samples generated using a numerical procedure.
As briefly demonstrated in Sec.~\ref{sec:practical_applications}, EVT provides sensible ans\"{a}tze for this purpose;
generating samples, however, quickly becomes a burden for wide parameter-space searches, 
as template banks are orders of magnitude larger. 
In such cases, the distribution of the loudest outlier can be estimated using the search results themselves 
as a proxy for background samples~\cite{Wette:2009uea, Papa:2020vfz, Wette:2021tbv}. 

In this section, we combine the EVT ansatz described in 
Sec.~\ref{sec:practical_applications}~[Eq.~\eqref{eq:gumbel_def}]
with the proposal from~\cite{Wette:2021tbv}. 
Our new generalized method, \dmax{},
covers any sort of detection statistic whose noise-hypothesis distribution
falls into one of the three possible max-stable domains of attraction,
i.e. not only the standard $\F$-statistic, but also
\mbox{``line-robust''} statistics~\cite{Keitel:2013wga, Keitel:2015ova},
generalizations of the \mbox{$\F$-statistic} to look for tCW~\cite{Prix:2011qv}.
Other detection statistics used in the CW literature~\cite{universe}, 
such as 
Hough number-count~\cite{PhysRevD.70.082001, 2014PhRvD..90d2002A, Miller:2018rbg, Oliver:2019ksl, Covas:2019jqa},
cross-correlation~\cite{Whelan:2015dha},
or power-based statistics~\cite{Dergachev:2010tm, Goetz:2011bd, Dergachev:2019wqa},
could potentially benefit from \dmax{} as well.

\subsection{Basic formulation}
\label{subsec:basic}

We are interested in describing $\pdf(\xi^{*}|\hyp{N})$ solely using 
the available detection statistic samples from the search $\xi$,
that is, without any further evaluation of the template bank (e.g. off-sourcing).
Following the argument in Sec.~\ref{sec:practical_applications}, the evaluation of a detection statistic over
a generic template bank can be interpreted as equivalent to the evaluation of said detection statistic over 
different realizations of noise with a certain (and unknown) degree of correlation.
If correlations were negligible,
a direct application of Eq.~\eqref{eq:p_xi_star} would give us the desired answer.

The key realization of~\cite{Wette:2009uea, Wette:2021tbv} is that the loudest outlier from a template bank
$\xi^{*}$ can be obtained in two steps: estimate the distirbution of the loudest candidate
of a \emph{smaller} template bank, then extrapolate such distribution to account for the template bank reduction.
Dividing the initial template bank into smaller subsets makes multiple loudest candidates available to properly
fit a distribution.

For the first step, one splits the dataset $\xi$, containing $\N$ (possibly correlated) values, into $B$ batches, 
each of them with $n = \N / B$ elements. This partition can be done such that each batch
contains a similar subset of the overall population so that the per-batch maxima (\emph{batchmax} samples)
\mbox{$\{ \xi^{*}_{b}, \; b=1, \dots, B\}$} are independent draws
from the same unknown distribution: \mbox{$\xi^{*}_{b} \sim p_n$}.
If we choose a sufficiently high number of batches $B$, the \emph{batchmax} distribution $p_n$ can be
obtained by fitting a suitable ansatz to the data.

As the second step, the overall loudest value $\xi^{*}$ is then
\begin{equation}
    \xi^{*} = \max_{b = 1, \dots, B}  \xi^{*}_{b}\;,
    \label{eq:batch_maxima}
\end{equation}
which corresponds to the loudest of $B$ $p_n$-distributed random variables.
This operation was already described in Eq.~\eqref{eq:p_xi_star},
which here we recast as an operator in the space of probability distributions for later convenience:
given a probability distribution $f$, $\mprop_{B}{f}$ corresponds to the distribution of the loudest 
candidate over a set of $B$ independent samples of $f$:
\begin{equation}
    \mprop_{B}{f}(x) = B f(x) \left[ \int^{x} \mathrm{d}x' f(x') \right]^{B - 1}\;.
    \label{eq:max_prop}
\end{equation}
The distribution of the overall loudest value $\xi^{*}$ is then simply\footnote{
    Note that \cite{Wette:2009uea}, inserting an empirical histogram as the $f$
    in Eq.~\eqref{eq:max_prop} (to account for a bias due to implementation details in 
    the \mbox{$\F$-statistic}, see also Sec.~\ref{sec:real_data}), 
    constitutes an earlier application of this principle.}
\begin{equation}
    \pdf(\xi^{*}|\hyp{N}) = \mprop_{B}{p_{n}}(\xi^{*})\;.
    \label{eq:batch_propagation} 
\end{equation}

The initial proposal in~\cite{Wette:2021tbv} described the batchmax distribution $p_{n}$ 
using a Gaussian Kernel Density Estimation (KDE) over the set of batchmax samples
\mbox{$\{ \xi^{*}_{b}\}$}.
Eq.~\eqref{eq:batch_propagation} was then implemented as a numerical integration. 
The final decision threshold was based on the support of the resulting distribution.

\begin{figure}
    \centering
    \includegraphics[]{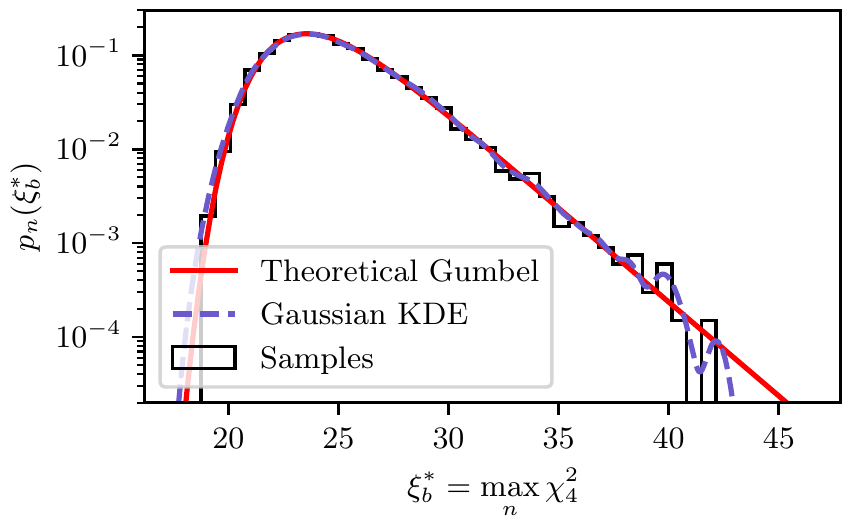}
    \includegraphics[]{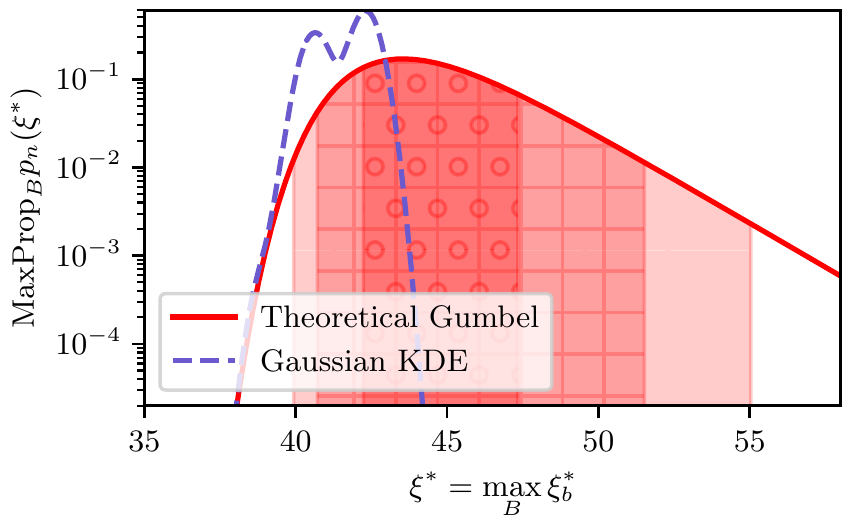}
    \caption{
    Upper panel: 
        KDE fit to a set of $B=10^4$ samples drawn from the theoretical Gumbel distribution~\cite{RePEc:spr:testjl:v:24:y:2015:i:4:p:714-733}
        of the loudest sample out of an ensemble of 
        $n=10^{4}$ $\chi^2_{4}$ random variables.
        The stepped line corresponds to the histogram of samples.
        The blue dashed line corresponds to the Gaussian KDE.
        The red solid line corresponds to the theoretical distribution. 
    Lower panel: 
        Application of the numerical $\mprop_{B}$ operator with $B=10^4$ to the KDE computed from the
        upper panel (blue dashed line).
        We compare the result to the theoretical distribution of the maximum sample over
        \mbox{$\mathcal{N} = n \times B = 10^8$} $\chi^2_4$ samples
        (red solid line).
        Shaded regions correspond to the 68\%, 95\%, and 99\% probability intervals.
        KDE bandwidths are estimated using the default method (``scott'') implemented in~\cite{2020SciPy-NMeth}.
    }
    \label{fig:kde_example}
\end{figure}

We find that the use of Gaussian KDEs introduces inaccuracies into the estimation of $\pdf(\xi^{*}|\hyp{N})$.
The reason is twofold. First, KDEs are prone to overfitting histogram artifacts which arise due to finite sample sizes.
This is illustrated in the upper panel of Fig.~\ref{fig:kde_example}. As a result, the propagated distribution in this case
displays an unintended bimodality, as shown in the lower panel of Fig.~\ref{fig:kde_example}.
Second, for the detection statistics we consider here, the batchmax distribution falls off exponentially
(see Appendix~\ref{sec:extreme_value_theory}), at a much slower pace than a Gaussian tail.
This tends to cause $\mprop_{B}$ to underestimate the variance of the resulting distribution, 
as shown in the lower panel of Fig.~\ref{fig:kde_example}.

Our main innovation with \dmax{} is to propose a cogent ansatz
to circumvent the non-parametric description of the batchmax distribution.
Our specific proposal, a \mbox{max-stable} distribution, 
corresponds to the asymptotic behaviour of the batchmax distribution in the limit of $n \rightarrow \infty$.
The max-stable property also simplifies the $\mprop_{B}$ operator into a simple algebraic operation.

\subsection{Introducing \dmax{}}
\label{subsec:evt}

Batchmax samples in Eq.~\eqref{eq:batch_maxima} are constructed so that they correspond to 
independent and identically distributed random variables from a certain underlying distribution $p_n$. 
In the case of a data stream free of loud disturbances, 
this can be simply achieved by randomly shuffling the results of a search before grouping them into batches.
(A discussion of the effects of shuffling data with loud disturbances is deferred 
to Secs.~\ref{subsec:accuracy} and~\ref{subsec:discussion} and Appendix~\ref{sec:loud_disturbances}.)
The batchmax distribution then corresponds to that of the loudest candidate over $n$ templates, which, 
as discussed in Sec.~\ref{sec:practical_applications}, tends to a Gumbel distribution as $n \rightarrow \infty$.
Hence, we propose the following ansatz for the batchmax distribution:
\begin{equation}
    p_{n}(x) = \gumbel{(x; \mu_{n}, \sigma_{n})}\;,
    \label{eq:use_gumbel}
\end{equation}
where $\mu_{n}, \sigma_{n}$ are obtained by direct fit to the batchmax samples.
This choice is similar to that of~\cite{LIGOScientific:2021ozr}, 
which directly fitted an exponential tail (upper tail of a Gumbel distribution) to the batchmax distribution.

EVT distributions, such as $\gumbel$, are \mbox{max-stable} distributions: 
the distribution of the loudest outlier from a set of EVT distributions is itself 
an EVT distribution of the same kind,
albeit with different parameter values. 
As a result, the $\mprop$ operator can be re-expressed in a closed form in terms 
of the location and scale parameters of the distribution. 
Concretely, it is straightforward to show that
\begin{equation}
    \mprop_{B} \gumbel(x;\mu_{n}, \sigma_{n}) = \gumbel(x; \mu_{*}, \sigma_{*})
    \label{eq:mprop}
\end{equation}
where
\begin{equation}
        \mu_{*} = \mu_{n} + \sigma_{n} \ln{B} \;,
        \label{eq:mu_star}
\end{equation}
\begin{equation}
        \sigma_{*} = \sigma_{n} \;.
        \label{eq:sigma_star}
\end{equation}
Thus, the target distribution is readily obtainable through a simple algebraic calculation after performing
a fit to the batch-max samples:
\begin{equation}
    \pdf(\xi^{*}|\hyp{N}) = \gumbel(\xi^{*}; \mu_{*}, \sigma_{*})\;.
\end{equation}

Summarizing, \dmax{} exploits the max-stability of the Gumbel distribution to estimate the 
distribution of the loudest candidate of a search, $p(\xi^{*}|\hyp{N})$.
To do so, 
search results are shuffled into $B$ disjoint batches from which the loudest candidates are retrieved.
These $B$ batchmax candidates, by construction, 
can be interpreted as draws from an EVT distribution $p_n$ whose parameters can be estimated
using a standard maximum-likelihood fit such as \texttt{scipy.rv\_continuous.fit}~\cite{2020SciPy-NMeth}.

In broad terms, the batch size $n$ determines how close the batchmax distribution is to an EVT one,
whereas the number of batches $B$ determines how sharply the parameters of $p_n$ can be determined.
Real searches usually contain a fixed number of templates $\N = n B$, 
meaning a trade-off is required:
On the one hand,
choosing a large $n$ (hence a small $B$) produces a small number of samples, each well consistent with an EVT distribution, 
but increases the variance of the $p_n$ fit.
On the other hand,
a large $B$ (hence a small $n$) produces a big number of samples drawn from a distribution which has not
fully converged to an EVT distribution, meaning the estimated parameters may be biased with
respect to the actual distribution.
We devote the following subsections to further discuss the role played by each of these parameters.

\subsection{The \mprop{} operator}
\label{subsec:mprop}

\begin{figure}
    \includegraphics[]{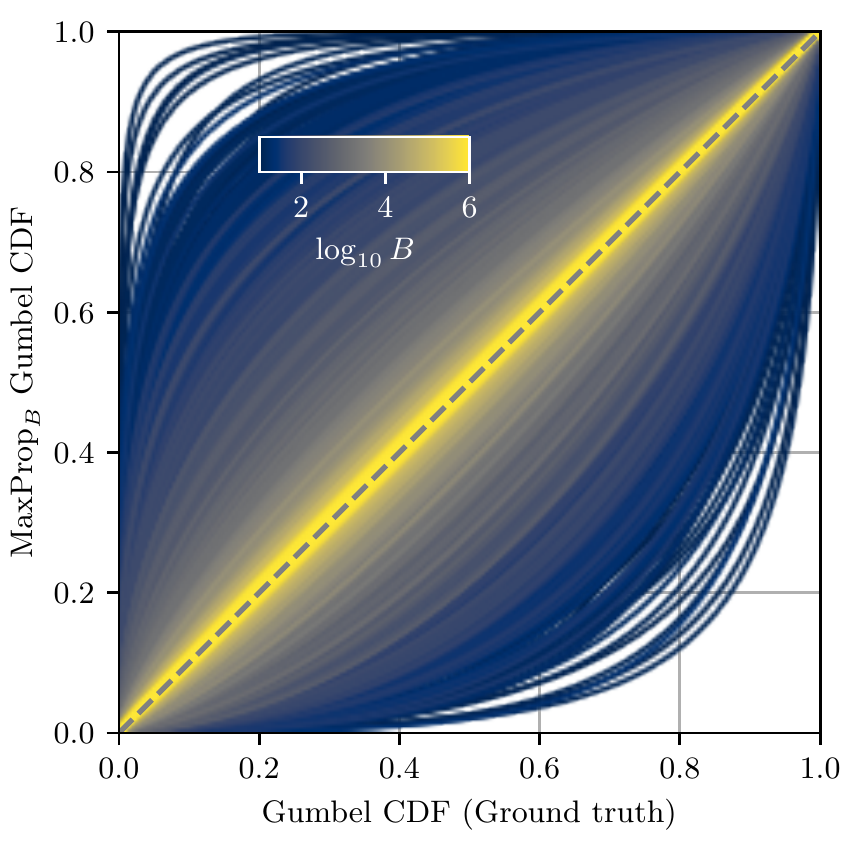}
    \caption{CDF comparison between the ground truth Gumbel distribution and the \mprop{} Gumbel 
    distributions using different numbers of batches $B$.
    Each line corresponds to a propagated Gumbel distribution obtained by drawing $B$ samples from
    the batchmax distribution (as discussed in the main text), fitting a Gumbel distribution,
    and applying the \mprop{} operator. 
    This procedure is repeated 500 times for each of the 48 selected values of $B$ between 10 and $10^6$.
    }
    \label{fig:propagated_gumbel}
\end{figure}

\begin{figure}
    \includegraphics[]{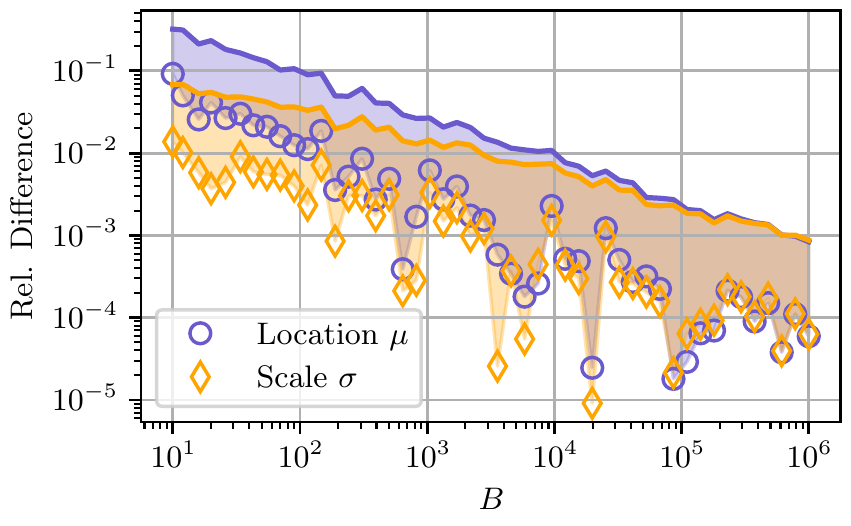}
    \caption{Relative difference in Gumbel parameters obtained by the application of \mprop{} 
    using different numbers of batches $B$. Each marker shows the average relative deviation
    over 500 realizations, corresponding to the distributions shown in Fig.~\ref{fig:propagated_gumbel}.
    The upper envelope represents sample standard deviation. 
    Lower envelopes are omitted due to the logarithmic scale.
    }
    \label{fig:loc_scale_rel_diff}
\end{figure}

We now characterize the phenomenology of the $\mprop{}$ operator on a Gumbel distribution,
which corresponds to the asymptotic distribution followed by standard (t)CW detection statistics.

Let us consider a template bank with $\N=10^6$ templates.
Given a batch size $n$, we model a batchmax distribution as a Gumbel distribution 
with $\sigma_{n}=2$ and $\mu_{n} = \sigma_{n} \ln{n}$.
This is equivalent to considering a detection statistic following a $\chi^2_{4}$ distribution
which has already converged to its corresponding EVT distribution, 
preventing finite sample-size effects from polluting the analysis.
The ground truth distribution of the loudest candidate from the template bank then 
corresponds to the propagation of said Gumbel distribution over \mbox{$B = 10^6/n$} batches,
i.e.~a Gumbel distribution with \mbox{$\sigma = \sigma_{n}$} and \mbox{$\mu = \sigma \ln{\N}$}.

Batch sizes $B \in [1, 10^6]$ are analyzed by drawing $B$ batchmax samples from the 
aforementioned batchmax distribution with $n = 10^6 / B$;
$\mu_n$ and $\sigma_n$ are fitted using \texttt{scipy.stats.gumbel\_r.fit} 
and propagated using Eq.~\eqref{eq:mprop}. 
The resulting CDF is compared against the ground truth CDF, shown in Figure~\ref{fig:propagated_gumbel}.
The relative error in the estimated location and scale parameters is shown in 
Fig.~\ref{fig:loc_scale_rel_diff}. 

As previously anticipated,
a low number of batches \mbox{$B \lesssim 10^{3}$} results in a greater dispersion of the estimated parameters.
As the number of batches reach the \mbox{$10^{3} \lesssim B \lesssim 10^{4}$} range, 
batchmax histograms become more robust and relative parameter deviations achieve sub-percent levels.

\subsection{Characterizing the \bmax{} distribution}
\label{subsec:bmax}

\begin{figure}
    \includegraphics[]{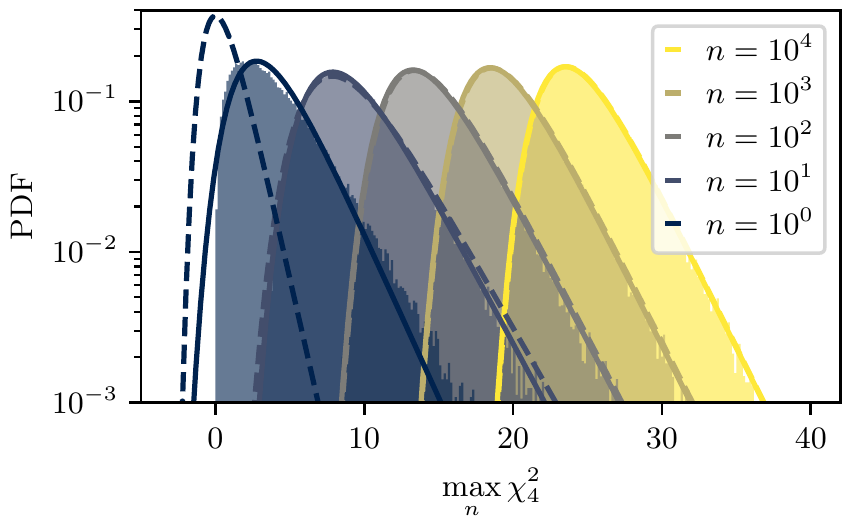}
    \caption{Batchmax distribution of an ensemble of $n$ $\chi^2_4$ random variables for different choices
    of $n$. Shaded regions represent histograms of $10^4$ batchmax values
    obtained by drawing $10^4 \times n$ values 
    and retrieving the loudest out of each group of $n$. 
    Each solid line shows a direct fit of a Gumbel distribution to the data using~
    \texttt{scipy.stats.rv\_continuous.fit}~\cite{2020SciPy-NMeth}.
    Each dashed line represents the corresponding Gumbel distribution using Eqs.~(32) 
    and (33) of~\cite{RePEc:spr:testjl:v:24:y:2015:i:4:p:714-733}, 
    implemented in the \dmax{} package~\cite{distromax}.}
    \label{fig:example_chi2_to_gumbel}
\end{figure}

On the other hand, to test convergence of \bmax{} distributions,
we take as an example the case of a $\chi^2$ distribution with 4 degrees of freedom
($\Gamma$ distribution with shape parameter \mbox{$k=2$} 
and scale parameter \mbox{$\theta=2$}), deferring to Appendix~\ref{sec:extreme_value_theory} 
other generic distributions and further references.
For a $\chi^2_4$ distribution, the limit \mbox{$n \rightarrow \infty$} corresponds to
a Gumbel distribution~\cite{embrechts2013modelling}, as shown in Fig.~\ref{fig:example_chi2_to_gumbel}.

Given a set of random variables following a specific distribution $f$, 
the convergence of the loudest draw towards an EVT distribution is driven by the behaviour of $f$'s
tail (\emph{tail-equivalence}~\cite{embrechts2013modelling}).
More specifically, the role of the batch size $n$ is related to how likely it is to draw a sample
within the tail of the distribution: the higher the number of samples $n$, 
the more likely it is to retrieve a value from the upper tail, 
hence the lower the dependency on other details of the distribution's shape.
Using a low batch size causes batchmax samples to be dominated 
by the bulk instead of the tail,
keeping the resulting distribution from properly converging to an EVT distribution.

$\chi^2$ random variables are non-negative,
as they are the sum of the squares of standard Gaussian random variables.
Gumbel distributions, on the other hand, 
present a double exponential decay in their lower tail [Eq.~\eqref{eq:gumbel_def}].
Consequently, as shown in the case $n=1$ in Fig.~\ref{fig:example_chi2_to_gumbel},
batchmax distributions with low $n$ tend not to follow a Gumbel distribution.
As the number of samples $n$ increases, the effects due to the distribution's bulk become 
milder and we find a better agreement to the expected distribution. 

\subsection{Parameter estimation accuracy\\and comparison to previous approaches}
\label{subsec:accuracy}

Finally, we present a more realistic set of results evaluating the $\F$-statistic 
over an actual template bank on 7 days of Gaussian noise using 
\texttt{ComputeFstatistic\_v2}~\cite{lalsuite}.
The template bank is constructed using the \texttt{gridType=8} option
with maximum mismatch $m = 0.2$,
for a fixed sky position $(0, 0)$ in equatorial coordinates around $f_{0} = \SI{50}{\hertz}$ 
and $f_{1} = {-10^{-8}}\; \textrm{Hz}/\textrm{s}$, containing $\N \simeq 8 \times 10^{6}$ templates. 
A ground truth distribution is numerically constructed by evaluating this template bank on 900 
realizations of Gaussian noise and retrieving the loudest $2\F$ value from each. 
\dmax{} is then applied to the individual realizations in order to test 
its accuracy.

We compare two different batching approaches: 
batching contiguous frequency bins and shuffling the results into random batches. 
To produce comparable results,
the shuffled batches contain the same number of templates $n$ as each contiguous batch.
The motivation behind these two approaches is related to the potential presence of correlated outliers in real detector data:
instrumental artifacts tend to affect
relatively well-localized frequency bands~\cite{2018PhRvD..97h2002C}.
Frequency-wise batching could thus prevent very loud outliers from polluting a high number of batches 
and overestimating the expected loudest outlier. 
Not shuffling the template bank, however, 
could require an increase in the batch size to obtain proper convergence to a Gumbel distribution,
but then the reduced number of batches would
imply an increase in the variance of the estimate.

\begin{figure}
    \includegraphics[]{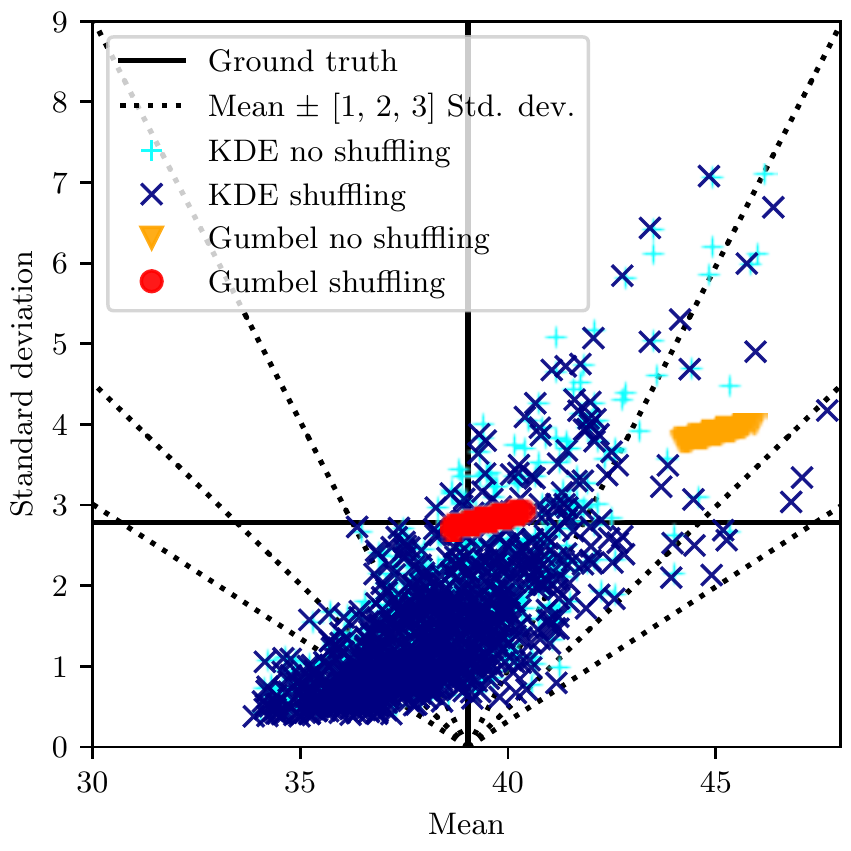}
    \caption{
        Comparison of methods to estimate the distribution of the loudest $\F$-statistic outlier 
        from a template bank.
        The data stream and template bank are constructed as explained in the main text.
        In this figure, the \mbox{$\N \simeq 8 \times 10^{6}$} templates are grouped together by
        joining 25 consecutive frequency bins or shuffled
        into \mbox{$B=7236$} batches with a batch size of \mbox{$n = 1105$}.
        Red circles and orange triangles represent \dmax{} results with and without shuffling,
        while blue crosses and light blue plus signs represent the results obtained using a Gaussian KDE
        again with and without shuffling.
        Solid lines show the mean and standard deviation of the ground truth distribution.
        Dotted lines represent one, two and three standard deviations with respect to the
        ground truth mean.
    }
    \label{fig:triangle0}
\end{figure}

\begin{figure}
    \includegraphics[]{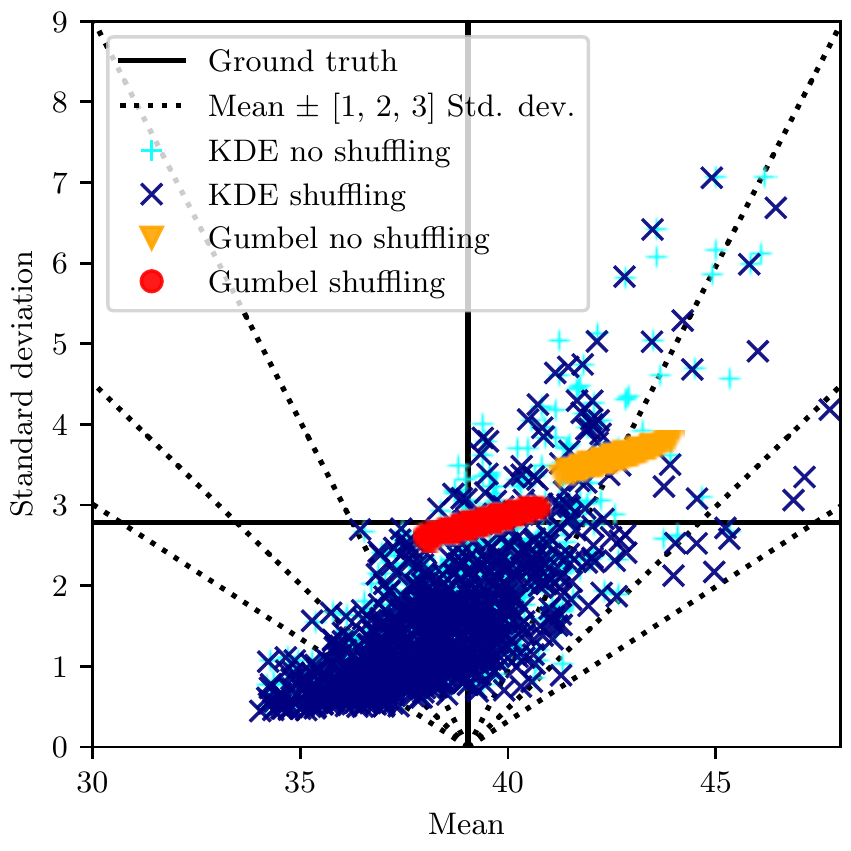}
    \caption{
        Comparison of methods to estimate the distribution of the loudest $\F$-statistic outlier 
        from a template bank using the same dataset as in Fig.~\ref{fig:triangle0}.
        In this figure, the \mbox{$\N \simeq 8 \times 10^{6}$} templates are grouped together by
        joining 100 consecutive frequency bins or shuffled
        into \mbox{$B=1801$} batches with a batch size of \mbox{$n = 4419$}.
    }
    \label{fig:triangle1}
\end{figure}

\begin{figure}
    \includegraphics[]{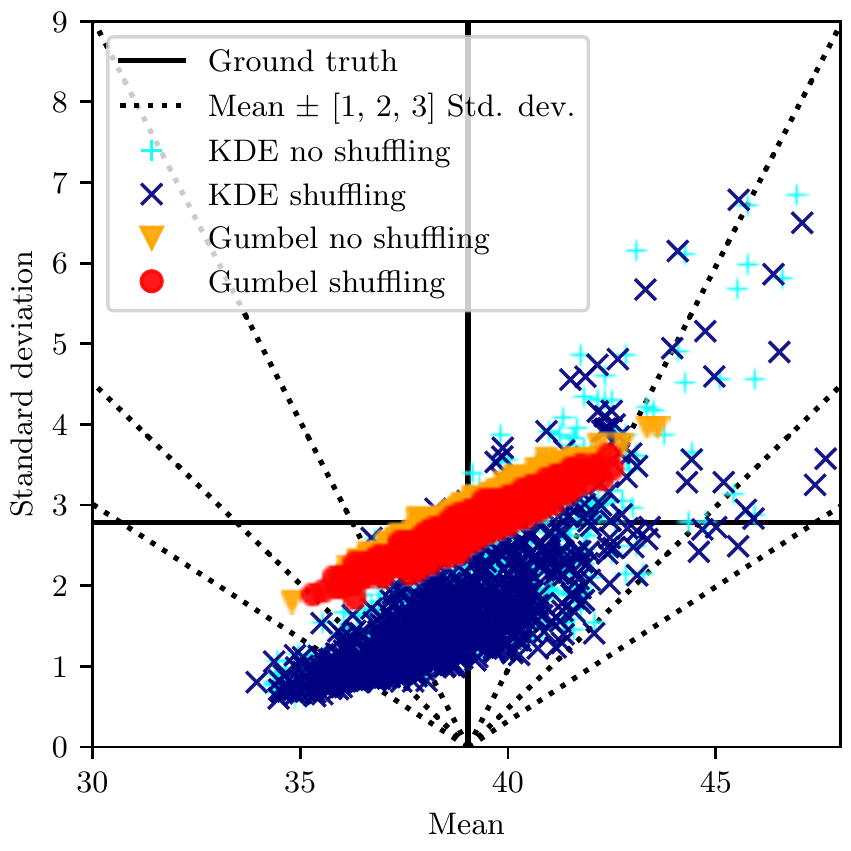}
    \caption{
        Comparison of methods to estimate the distribution of the loudest $\F$-statistic outlier 
        from a template bank using the same dataset as in Fig.~\ref{fig:triangle0}.
        In this figure, the \mbox{$\N \simeq 8 \times 10^{6}$} templates are grouped together by
        joining 3000 consecutive frequency bins or shuffled
        into \mbox{$B=60$} batches with a batch size of \mbox{$n = 132540$}.
    }
    \label{fig:triangle2}
\end{figure}

Results are shown in Figs.~\ref{fig:triangle0},~\ref{fig:triangle1}, and~\ref{fig:triangle2}
in terms of the estimated mean and standard deviation from each method
against those of the ground-truth distribution.
We also compare to the original proposal of \cite{Wette:2021tbv} by 
using a Gaussian KDE to approximate $p_{n}$ in Eq.~\eqref{eq:use_gumbel}.

We start by discussing the performance of the Gaussian KDE. 
The first significant feature is the lack in precision of the estimated parameters,
which are over-dispersed regardless of the choice of $B$ and $n$. 
We also note that the bulk of these results tend to underestimate both 
the location and scale parameters of the Gumbel distribution with respect to the ground truth. 
This is related to the shape of the kernel function being used, 
as previously discussed in Sec.~\ref{subsec:basic}:
the tails of a Gaussian distribution fall off more rapidly than those of a Gumbel distribution,
yielding a lower mean and standard deviation.
The insensitivity of these results to the choice of $B$ and $n$ and to shuffling suggests that 
this particular KDE-based ansatz does not return a reliable estimate of the batchmax distribution.

\dmax{} results, on the other hand, return a more consistent picture. 
Sample variance increases as the number of batches goes down both with and without shuffling.
Complementarily, the bias in the estimated parameters reduces as $n$ increases ($B$ decreases),
although this effect is only significant for the method without shuffling.

We observe a significant bias reduction by shuffling.
Randomly shuffling samples results in more homogeneous batches with weaker inner correlations;
as a result, batchmax samples are closer to the expected Gumbel distribution, 
improving the accuracy of the recovered parameters.
Also, as previously anticipated, 
using bigger batches in the non-shuffling case does improve accuracy,
although with a significant increment in variance due to the correspondingly lower number of batches.

\begin{figure}
    \includegraphics[]{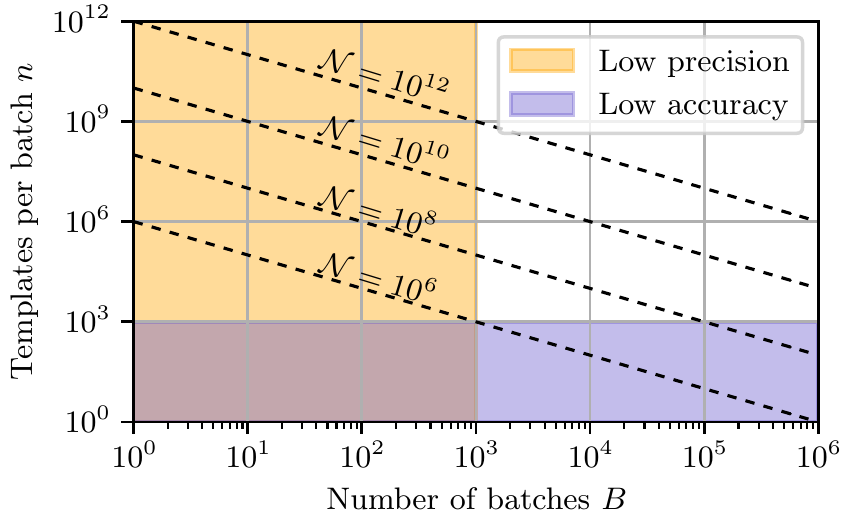}
    \caption{
        Summary of regimes in which \dmax{} operates. 
        Shaded regions represent combinations of $n$ and $B$ for which the \dmax{} results
        suffer from low precision (high variance) due to a low number of batches (Sec.~\ref{subsec:mprop})
        or low accuracy (high bias) due to a low batch size (Sec.~\ref{subsec:bmax}).
    }
    \label{fig:precision_accuracy}
\end{figure}

These features are consistent with the basic phenomenology discussed 
in Secs.~\ref{subsec:mprop} and~\ref{subsec:bmax}. 
The different regimes in which \dmax{} operates depending on $\N$, $n$, and $B$ are summarized
in Fig.~\ref{fig:precision_accuracy}. 
These values are extracted from the general behaviour of \dmax{} observed throughout the tests
performed in this section.
As a general working principle, \dmax{} requires at least $\N \simeq 10^6$ in order to return
a cogent answer; for smaller template banks, on the other hand,
off-sourcing as described in Sec.~\ref{sec:practical_applications}
requires little computational effort.

\subsection{Discussion}
\label{subsec:discussion}

The results above on simulated data demonstrate the overall performance of \dmax{} under Gaussian-noise conditions.
The concrete output of \dmax{} is a simple estimation on the probability of the loudest candidate of a search
under the noise hypothesis $\pdf{(\xi^{*}|\hyp{N})}$. However, the specific statement to be drawn from
$\pdf{(\xi^{*}|\hyp{N})}$, such as a threshold choice,
is entirely dependent upon the scope of the analysis at hand.
For the sake of completeness, we briefly review the assumptions on which the \dmax{} method relies, 
as well as possible consequences of violating them.

First, the distribution of the detection statistic at hand must belong to the domain of attraction 
of the Gumbel distribution; roughly, this means its probability distribution should be unbounded and decay 
at a slower rate than a power-law~\cite{embrechts2013modelling}. 
Nonetheless, as discussed in Appendix~\ref{sec:extreme_value_theory}, 
this method could be easily adapted to detection statistics within a different domain of attraction;
in such cases, Eqs.~\eqref{eq:mu_star} and~\eqref{eq:sigma_star} would have to be adapted
to the corresponding EVT distribution.

Second, the data at hand must be free of strong disturbances. 
As discussed in Sec.~\ref{subsec:correlations}, loud disturbances in a data stream typically
translate into parameter-space regions returning enhanced detection statistics 
with respect to a non-disturbed data stream.
The width of the affected region will depend on the characteristics of the disturbance and the template bank,
but in general there can be an extended set of templates with correlated response to the disturbance.
Attempting to construct a batchmax distribution by shuffling the samples into different batches
would result in a distribution shifted towards the right-hand side of the expected Gumbel distribution.
Not using shuffling would suppress the effect of disturbances if the resulting associated
population of templates was well localized in frequency; 
the resulting distribution, however, would be less accurate
and in particular could still overestimate the Gumbel parameters,
as discussed in Figs.~\ref{fig:triangle0},~\ref{fig:triangle1}, and~\ref{fig:triangle2}.
The robustness of \dmax{} results to the effect of mild disturbances on the data can also be tested by generating
several sets of batchmax samples and comparing the location and scale parameters
of the corresponding Gumbel distributions.
The wider the distribution over shuffling realizations, the bigger the effect of noise disturbances.

Narrow spectral features (``lines''), in particular, are common noise disturbances affecting (t)CW searches,
with excess power typically concentrated within a few frequency 
bins~\cite{2018PhRvD..97h2002C}.
CW signals themselves are another typical example of well-localized ``disturbances'':
should a (strong) CW signal be present in a datastream, 
a blind application of \dmax{} could result in overestimation of the loudest candidate's
distribution, potentially flagging the CW signal itself as a background-noise fluctuation.
The sensitivity of current interferometric detectors, however, 
makes CW searches to operate in the weak-signal regime (see Appendix~\ref{sec:injections}).
As a result, CW signals are unlikely to actually affect the estimations provided by \dmax{}.

Due to their crucial role in CW searches, we discuss in Appendix~\ref{sec:loud_disturbances}
a simple proposal to reduce the effect of narrow-band disturbances so that \dmax{} results 
can still provide a cogent answer. 
Whenever possible, we recommend the application of informed veto strategies against instrumental artifacts
(see~\cite{universe} and references therein) \emph{before} attempting to process the results using \dmax{}.
The method discussed in Appendix~\ref{sec:loud_disturbances} is just a complementary 
algorithm to prevent a specific type of strong disturbances from invalidating an analysis.
The characterization and improvement of this or similar algorithms to deal with more generic disturbances
is left for future work.

Third, in principle \dmax{} assumes the complete set of detection statistic values 
from the full template bank is available.
Several wide parameter-space searches, however, 
use \emph{toplists}~\cite{PhysRevD.70.082001, Pletsch:2009uu, Wette:2018bhc},
meaning they only keep a small fraction of detection statistic values corresponding to the louder templates.
The basic requirement is to use a toplist such that the tail of the distribution is properly represented.
Falling short (i.~e.~not reaching the bulk of the distribution) could result in inaccurate fits to 
batchmax samples.
Incidentally, the results discussed in Appendix~\ref{sec:loud_disturbances} clarify the suitability
of \dmax{} to these searches. 

Fourth, and closely related to the second point, template banks must not be too strongly correlated.
EVT ensures \dmax{} is robust to a certain degree of template-bank correlations; 
more specifically, 
answers provided by \dmax{} will be cogent as long as the dominant contribution to batchmax samples 
comes from the tails of the involved distributions. 
As discussed in e.g.~\cite{Prix:2006wm}, 
template bank setups using a small mismatch return highly correlated samples in the vicinity
of local parameter-space maxima, adding additional features to the results distribution
(see e.~g.~Fig.~2 in~\cite{Dreissigacker:2018afk}). 
This may have similar effects to the presence of loud disturbances, 
as a significant fraction of the resulting batchmax samples would come from samples around
a few local maxima and not be representative of the tail of the background distribution.
Consequently, the corresponding batchmax distribution will not be fully converged to an EVT 
distribution and the final Gumbel parameter estimation will be affected.
No simple amendment, other than using a higher mismatch~\cite{Allen:2019vcl, Allen:2021yuy}, 
is currently available to obtain robust results with \dmax{} in this situation.

This phenomenon was observed in~\cite{LIGOScientific:2021quq, narrow_band_proc},
where \dmax{} was applied to process the result of both a search for CW signals
and a search for long-duration transient GWs from glitching pulsars.
The CW search used a mismatch of $m\simeq 0.02$ combined with a toplist,
which reduced the effect of such a dense parameter-space converging on the batchmax distribution.
Preliminary studies for the long-duration transient GW search using $m=0.02$ \emph{without a toplist}, 
on the other hand, revealed a poor performance of \dmax{}; in that case, 
the solution was to increase the mismatch to $m=0.2$, using a similar setup as in~\cite{2019PhRvD.100f4058K}.

As we will discuss during Sec.~\ref{sec:real_data}, \dmax{} is suitable to be applied in 
real-data searches with typical mismatch setups.
Moreover, further detection statistics beyond $2\F$, 
such as ``line-robust''' statistics~\cite{Keitel:2013wga, Keitel:2015ova}
or tCW search statistics~\cite{Prix:2011qv}, can be processed using the same method.
This has the effect of improving the quality of \dmax{} results, 
as these statistics are designed to diminish the effect of noise disturbances on
(t)CW searches, providing a cleaner set of batchmax samples.
 
\section{Application to O2 data}
\label{sec:real_data}
\begin{figure*}[t]
    \centering
    \includegraphics[width=0.495\textwidth]{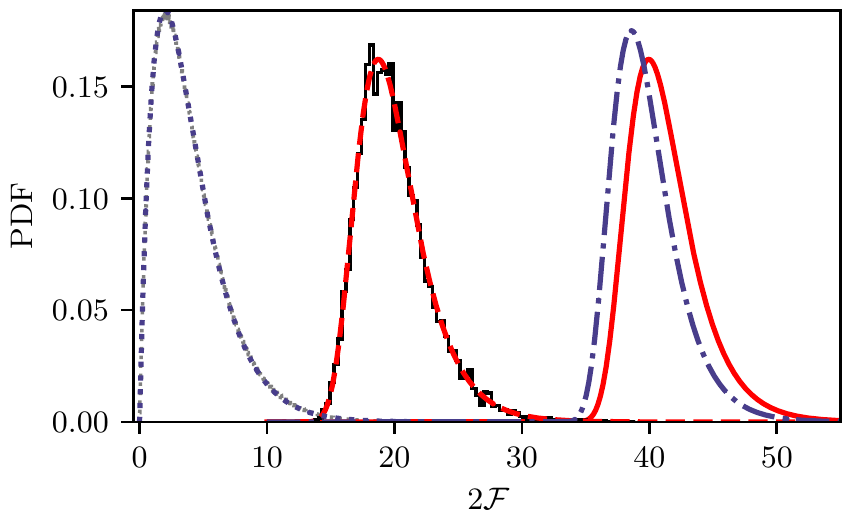}
    \includegraphics[width=0.495\textwidth]{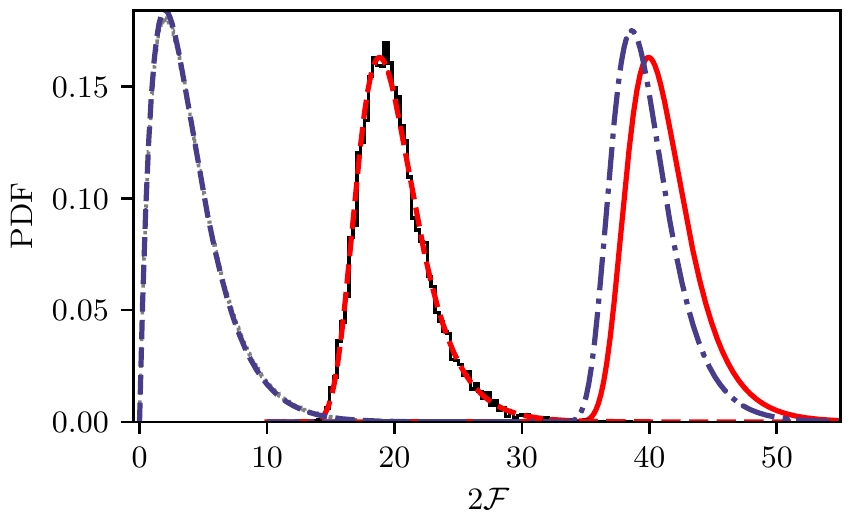}
    \caption{
         \label{fig:O2_2F_dist}
         $2\F$ values obtained in~\cite{2019PhRvD.100f4058K}
         analysing LIGO O2 data after glitches in the Vela (left panel) and Crab (right panel) pulsars.
         In each panel,
         the gray dotted line is the histogram of the samples obtained by the search. 
         The blue dotted line is the expected $\chi^2_4$ distribution for independent samples.
         This is propagated using the total number of templates $\Nlambda \approx 1.15 \times 10^7$ as $B$ in Eq.~\eqref{eq:max_prop}, 
         yielding an estimated distribution 
         for the maximum of $2\F$ shown by the dashed-dotted blue line.
         On the other hand using \dmax{} we plot the \bmax{} histogram (using $n=1000$, black solid line)
         and we fit it with a Gumbel distribution (dashed red line). 
         The propagated distribution is obtained by applying Eq.~\eqref{eq:mprop} with $B = \Nlambda/ n$ (solid red line).  
    }
\end{figure*}

\begin{figure*}[t]
    \centering
    \includegraphics[width=0.495\textwidth]{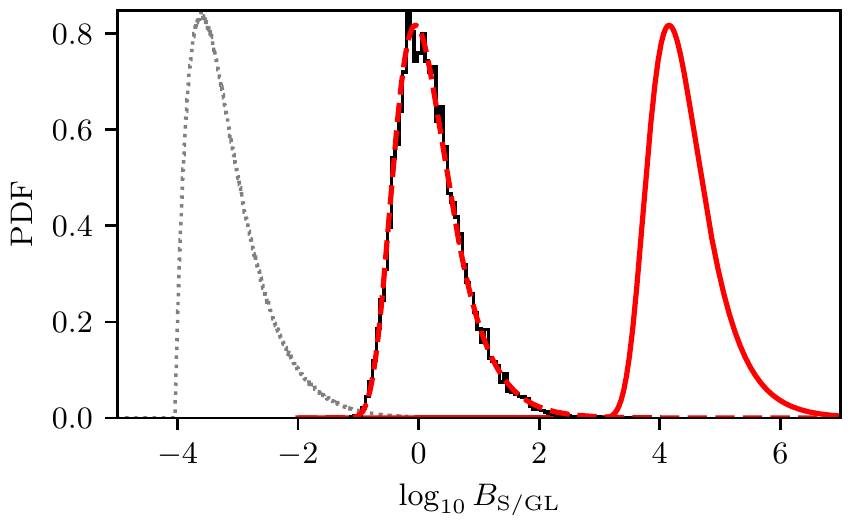}
    \includegraphics[width=0.495\textwidth]{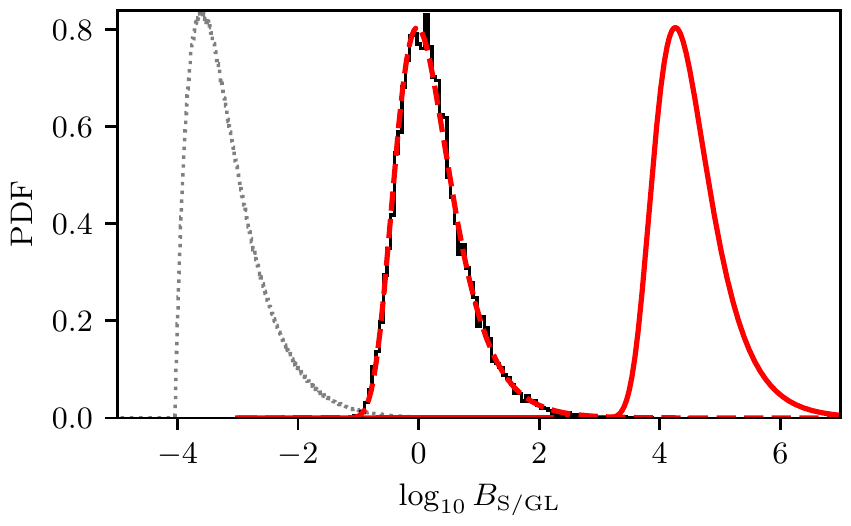}
    \caption{
        \label{fig:O2_logBSGL_dist}
        Values of the $\log_{10}\BSGL{}$ statistic obtained in the same analysis~\cite{2019PhRvD.100f4058K} as in Fig.~\ref{fig:O2_2F_dist}.
        Vela results are on the left panel, Crab results are on the right panel. 
        No closed-form expression for the distribution of $B_{\textrm{S}/{\textrm{GL}}}$ is known,
        so only the search samples (dotted gray line), the \bmax{} histogram (black solid line), 
        the Gumbel fit to it (dashed red line), and the propagated final distribution (solid red line) from \dmax{} are shown.
    }
\end{figure*}

As a demonstration, we apply \dmax{} to data of the two advanced LIGO detectors \cite{AdvancedLIGO} from the O2 observing run \cite{O2Data}
in Hanford and Livingston. We study the statistics of results obtained in \cite{2019PhRvD.100f4058K} for narrow-band searches
targeting the Vela and Crab pulsars, which experienced glitches on 12th December 2016 and on 27th March 2017 respectively.
The template bank for each target was a grid in \mbox{$\lambda = (f_0,f_1)$} of size \mbox{$\Nlambda \approx 1.15 \times 10^7$}.
This template bank was not constructed with a fixed mismatch but an estimate of $m \approx 0.2$
was given in \cite{2019PhRvD.100f4058K}. 
Here we consider the results using various $\F$-statistic based detection statistics for both CWs and tCWs.
For both cases we choose a batch size $n = 1000$, and the number of batches as $B=\Nlambda /n$.
Since this is a narrow-band search, the low number of templates places this particular application of \dmax{} at 
the border of the suitable regime described in Fig.~\ref{fig:precision_accuracy}; 
the obtained results, however, are not negatively affected by this.

\subsection{CW detection statistics}
We first consider results for CWs of duration four months corresponding to the maximum observation time in \cite{2019PhRvD.100f4058K}. 
We begin with the standard $2\mathcal{F}$ as its distribution is well known. 
As we see from Fig.~\ref{fig:O2_2F_dist}, the histogram of the full 2$\F$ results matches well with a standard $\chi^2_4$ distribution. 
One can also fit Eq.~\eqref{eq:p_xi_star} treating $\N$ as a free parameter, obtaining $\N' \approx 1$.
This is equivalent to considering each $2\F$ sample as the trivial maximum of a single draw from a $\chi^2_4$ distribution.
This $\chi^2_4$ distribution can then be propagated using the total number of templates $\Nlambda$ as $B$ in Eq.~\eqref{eq:max_prop}, 
yielding an estimated distribution for the maximum of $2\F$, 
which assumes that the template bank correlations are negligible.\footnote{As
discussed in detail in Sec.~8.7.1 of \cite{Wette:2009uea},
the result of any fit of Eq.~\eqref{eq:p_xi_star} to $\F$-statistic-based detection statistic samples
cannot be directly interpreted as an ``effective number of templates'' 
even in the absence of template-bank correlations.
This is due to a small upwards implementation bias in the $\F$-statistic computation \cite{CFSv2,CFSv2bias}.
For the $\N' \approx 1$ fit to the $\F$-statistic samples,
this seems to approximately cancel with the effect of template bank correlations.}
We then compare the resulting distribution with the one obtained by the \dmax{} method in Fig.~\ref{fig:O2_2F_dist}. 
The two resulting distributions for the maximum agree well.

\begin{figure*}[t]
    \centering
    \includegraphics[width=0.495\textwidth]{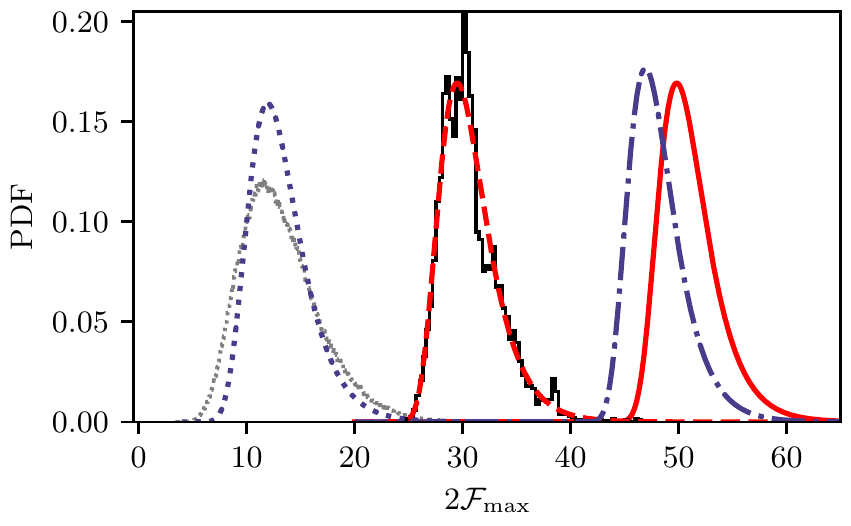}
    \includegraphics[width=0.495\textwidth]{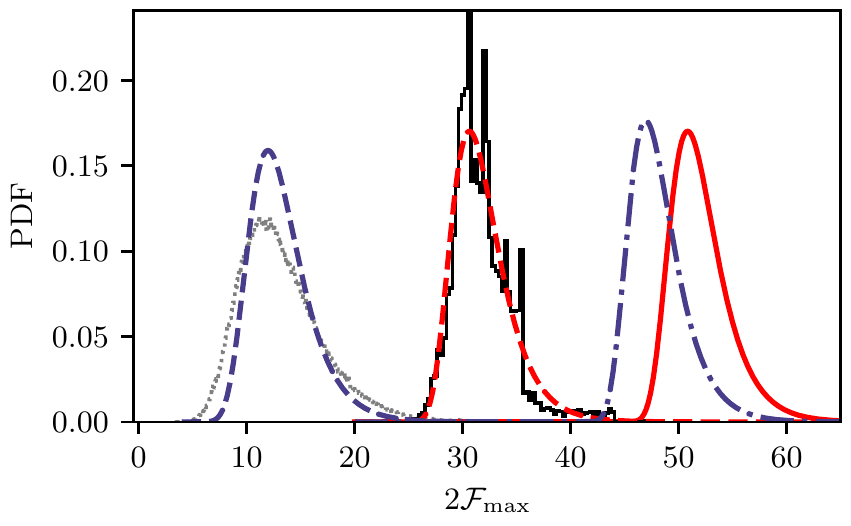}
\caption{
    \label{fig:O2_max2F_dist}
    Values of $\Ft$ obtained in the tCW analysis of~\cite{2019PhRvD.100f4058K}
    on LIGO O2 data after glitches in the Vela (left panel) and Crab (right panel) pulsars.
    The gray dotted line is the histogram of the samples obtained by the search. 
    There is no known distribution for $\Ft$, but it is a maximum over $\F$-statistics over the transient parameters
    which individually follow a $\chi^2_4$ distribution. 
    Due to the high degree of correlations in the transient parameter space,
    in \cite{2019PhRvD.100f4058K} the ``effective number of transient templates''
    was obtained by fitting Eq.~\eqref{eq:p_xi_star} to the $\Ft$ samples, obtaining $\N'\approx 55$ (dotted blue line).   
    The distribution of the overall loudest was then obtained
    by propagating the $\chi^2_4$ distribution using $B=\Nlambda \times \N'$ in Eq.~\eqref{eq:max_prop} (dash-dotted blue line).
    This corresponds to treating each sample as a batch with a single element.
    On the other hand using \dmax{} we plot the \bmax{} histogram ($n=1000$, black solid line)
    and we fit it with a Gumbel distribution (dashed red line). 
    The propagated distribution is obtained by applying Eq.~\eqref{eq:mprop} with $B = \Nlambda / n$ (solid red line).
}
\end{figure*}

\begin{figure*}[t]
    \centering
    \includegraphics[width=0.495\textwidth]{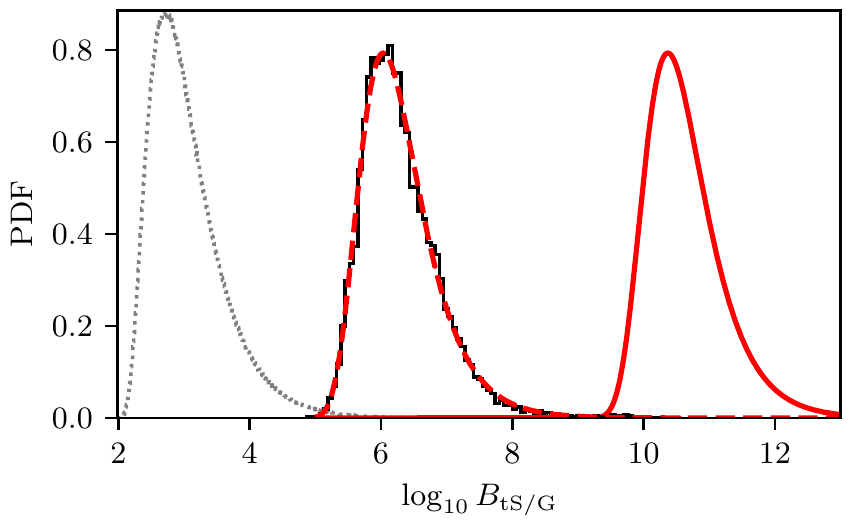}
    \includegraphics[width=0.495\textwidth]{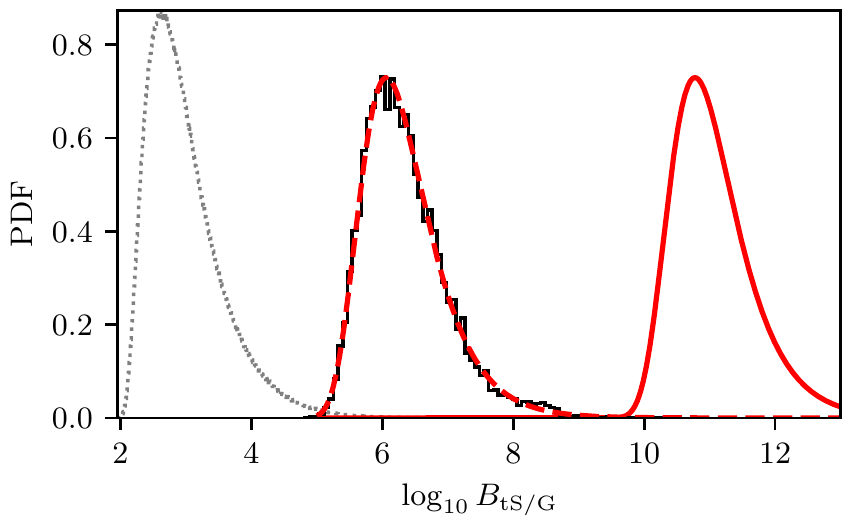}
\caption{
    \label{fig:O2_logB_dist}
    Values of $\log_{10}{\BtSG}$ obtained in the same~\cite{2019PhRvD.100f4058K} 
    tCW analysis as in Fig.~\ref{fig:O2_max2F_dist}.
    Vela results are on the left panel, Crab results are on the right panel. 
    Here there is no direct fit of Eq.~\eqref{eq:p_xi_star}
    because no closed-form expression for the distribution of $B_{\textrm{tS}/{\textrm{G}}}$ is known,
    and it cannot be easily related to the original $\F$-statistic,
    so only the search samples (dotted gray line), the \bmax{} histogram (black solid line), 
    the Gumbel fit to it (dashed red line), and the propagated final distribution (solid red line) from \dmax{} are shown. 
    }
\end{figure*}

We also apply the \dmax{} method on a different statistic for the CW search, namely the line-robust statistic $B_{\textrm{S}/{\textrm{GL}}}$ ~\cite{Keitel:2013wga}.
This is a Bayes factor derived from the likelihood ratio between the signal hypothesis and the combined hypothesis noise hypothesis of Gaussian noise and 
lines. The lines are modelled based on the assumption that they look exactly line a signal, but are present in only one detector. 
The results are shown in Fig.~\ref{fig:O2_logBSGL_dist}. Since the underlying distribution of $\BSGL{}$ is unknown, 
one cannot  do the equivalent of fitting Eq.~\eqref{eq:p_xi_star}. Nevertheless, we can still apply the \dmax{} method,
for which the only constraint is that said distribution falls off faster than a power-law (for the case here discussed involving a Gumbel distribution). 
The details of the exact distribution are not needed. 
Indeed, the \bmax{} distribution is well-fitted by a Gumbel distribution. 
In previous studies using $B_{\textrm{S}/{\textrm{GL}}}$ on real data
\cite{Steltner:2020hfd, Papa:2020vfz, Ming:2019xse, Abbott:2017pqa, Zhu:2016ghk, Papa:2016cwb, LIGOScientific:2016ahk, Ming:2021xtz},
it was used only to improve the robustness of the search against disturbances by using it as the toplist ranking statistic,
but the final significance statements were made returning to $2\F$
because no closed-form distribution was known for $B_{\textrm{S}/{\textrm{GL}}}$. Now with \dmax{} we can directly estimate thresholds from the samples 
allowing for end-to-end analysis using $B_{\textrm{S}/{\textrm{GL}}}$.

\subsection{tCW detection statistics}
We now investigate the case of tCWs, which was the main focus of \cite{2019PhRvD.100f4058K}.
As briefly mentioned in Sec.~\ref{sec:cw_searches}, 
such signals can be modelled as CWs modulated by a window function dependent on the 
transient parameters $\mathcal{T}$, namely the start time of the transient signal $t_0$
and its duration $\tau$. For this analysis a rectangular window function was used.
As a detection statistic for tCWs, the $\F$-statistic
at fixed $\lambda$ can be maximized over 
transient parameters \cite{Prix:2011qv}, thus obtaining \mbox{$\Ft = \max_{\mathcal{T}} 2\F$}.
(We use this notation instead of simply $\max2\F$
to avoid confusion with the maximum CW detection statistic $2\F$ over a full template bank.) 
We expect local correlations to have a more severe impact
when using this statistic because 
it can pick up short-duration non-Gaussianities or simple fluctuations
that the CW $\F$-statistic would not be susceptible to.

When the \dmax{} package shuffles the dataset in the \bmax{} stage, several batches can be contaminated by
the same noise fluctuation, and therefore the \bmax{} distribution
reflects this contamination.
The result is a more ragged distribution with peak-like features, as one can see in Fig.~\ref{fig:O2_max2F_dist}.

To estimate the distribution of the loudest candidate for this dataset,
\cite{2019PhRvD.100f4058K} made several simplifying assumptions.
While there is no known distribution for $\Ft$ that could be directly inserted into Eq.~\eqref{eq:p_xi_star}, 
its value at each template $\lambda$ is the maximum of $2\F$ values over the transient parameters $\mathcal{T}$,
which individually follow a $\chi^2_4$ distribution. 
However, there is a high degree of correlations in the transient parameter space.
Hence, fitting Eq.~\eqref{eq:p_xi_star} to the $\Ft$ samples, the result is an ``effective number of transient templates'' $\N'\approx 55$
(compared to a nominal number of \mbox{$N_{\mathcal{T}}\approx 2 \times 10^6$} transient templates at each $\lambda$).\footnote{Here
the difference between the fitted ``effective'' and nominal number of templates
is much larger than for the CW $\F$-statistic and hence the previously discussed bias is small enough to be ignored.
} 
It was then assumed that \mbox{$\Nlambda \times \N'$} could be interpreted as an ``effective number of templates'' over the full,
non-maximized, parameter space $(\lambda, \mathcal{T})$.
Consequently, the distribution of the overall loudest was obtained
by propagating a $\chi^2_4$ distribution using $B=\Nlambda \times \N'$ in Eq.~\eqref{eq:max_prop}.

The fits to the $\Ft$ sample histograms approximately catch the peak of the distribution, but fail
to correctly recover the overall shape. 
On the other hand, with \dmax{}, the Gumbel fits to the \bmax{} samples are 
noticeably better aligned to both the peaks and the tails of the \bmax{} histograms
than the $\N'$ based fits are to the full samples histograms.
The propagated distributions from both methods still overlap,
but their differences are larger than in the CW case.

Again, we apply \dmax{} also on an alternative detection statistic for the tCW search,
namely the $B_{\textrm{tS}/{\textrm{G}}}$ statistic, also derived in~\cite{Prix:2011qv}.
This statistic does not deal with the transient parameters by maximizing over the $\mathcal{T}$ space, but rather marginalizes
over it using a uniform prior. This results in less contamination from disturbances and noise fluctuations.
Despite its known better detection efficiency \cite{Prix:2011qv}, one reason why this detection statistic has not been
used in \cite{2019PhRvD.100f4058K} is that its distribution is not analytically known, hence no simple
fit of Eq.~\eqref{eq:p_xi_star} could be done.
With the \dmax{} method~\cite{distromax}, this is no longer a problem.
The results are shown in Fig.~\ref{fig:O2_logB_dist}.
Using the same data sets as before, 
the \bmax{} histograms are much smoother and better fit by a Gumbel distribution than their $\Ft$ counterparts.
This indicates that $B_{\textrm{tS}/{\textrm{G}}}$ is a more robust detection statistic
than $\Ft$ on real data. 
As for $\BSGL$ for CWs, with \dmax{} it can now also be used as an end-to-end detection statistic.
 
\section{Conclusion}
\label{sec:conclusion}
We have introduced \dmax{}, 
a new method to estimate the distribution of the loudest candidate in a gravitational-wave search. 
This method culminates a series of developments in the continuous gravitational-wave 
literature aimed at re-cycling wide parameter-space search results into
a proxy distribution for the expectation over different background noise realizations.
An implementation of the method is freely available as a Python package~\cite{distromax}.

Our specific proposal uses max-stable distributions from extreme value theory 
to provide a generic approach,
applicable to any detection statistic displaying a light-tailed distribution under
the noise hypothesis (that is, unbounded and decaying faster than a power-law).
This is in contrast with previous approaches based on the \mbox{$\F$-statistic}, 
whose very specific assumptions prevented a successful generalization.

Although we have focused on the case of detection statistics with light-tailed
distributions, as that is the standard encountered in CW searches, 
extensions to other kinds of distributions are possible by using a different
family of max-stable distributions.

We have demonstrated the general applicability of \dmax{} using both synthetic
Gaussian-noise data and the results of a real search on Advanced LIGO O2 data
for (transient) continuous gravitational-wave signals from the Vela and Crab pulsars.
Results show a significant improvement with respect to previous estimation methods
due to the robustness of \dmax{} to realistic template-bank correlations. 

Additionally, the possibility of using further detection statistics suppressing the effect of
lines ($\BSGL, \BSGLtL$) or transient instrumental artifacts ($\BtSG$) presents two further
advantages for (transient) continuous gravitational-wave searches: 
first,
\dmax{} allows us to process the results directly in terms of these more informative statistics;
second, the built-in suppression of instrumental features in these statistics itself improves
the convergence of batchmax samples to a max-stable distribution, 
improving the quality of the results provided by \dmax{}. 
This last point also makes plots of the batchmax distribution a useful tool to diagnose the data quality of a
specific frequency band using its deviation with respect to the expected max-stable distribution.

\section*{Acknowledgements}
We gratefully thank 
Patrick M. Meyers for testing and discussing improvements to the \texttt{distromax} package,
Karl Wette for sharing code for the KDE version of estimating loudest-candidate distributions and comments on the manuscript,
John T. Whelan for valuable comments on the manuscript,
and the LIGO--Virgo--KAGRA Continuous Waves working group for fruitful discussions.
This work was supported by European Union FEDER funds, 
the Spanish Ministerio de Ciencia e Innovaci\'on,
and the Spanish Agencia Estatal de Investigaci\'on grants
PID2019-106416GB-I00/AEI/MCIN/10.13039/501100011033,
RED2018-102661-T,
RED2018-102573-E,
Comunitat Aut\`onoma de les Illes Balears through the Conselleria de Fons Europeus, Universitat i Cultura
and the Direcci\'o General de Pol\'itica Universitaria i Recerca with funds from the Tourist Stay Tax Law ITS 2017-006 (PRD2018/24, PRD2020/11),
Generalitat Valenciana (PROMETEO/2019/071),  
EU COST Actions CA18108, CA17137, CA16214, and CA16104.
R.~T.~is supported by the Spanish Ministerio de Universidades (ref.~FPU 18/00694).
L.~M.~is supported by the Universitat de les Illes Balears.
D.~K.~is supported by the Spanish Ministerio de Ciencia, Innovaci\'on y Universidades (ref.~BEAGAL 18/00148)
and cofinanced by the Universitat de les Illes Balears.
 This research has made use of data, software and/or web tools obtained from the Gravitational Wave Open Science Center (https://www.gw-openscience.org/ ), a service of LIGO Laboratory, the LIGO Scientific Collaboration and the Virgo Collaboration. LIGO Laboratory and Advanced LIGO are funded by the United States National Science Foundation (NSF) as well as the Science and Technology Facilities Council (STFC) of the United Kingdom, the Max-Planck-Society (MPS), and the State of Niedersachsen/Germany for support of the construction of Advanced LIGO and construction and operation of the GEO600 detector. Additional support for Advanced LIGO was provided by the Australian Research Council. Virgo is funded, through the European Gravitational Observatory (EGO), by the French Centre National de Recherche Scientifique (CNRS), the Italian Istituto Nazionale di Fisica Nucleare (INFN) and the Dutch Nikhef, with contributions by institutions from Belgium, Germany, Greece, Hungary, Ireland, Japan, Monaco, Poland, Portugal, Spain.
 The authors are grateful for computational resources provided by the LIGO Laboratory and supported by  	
National Science Foundation Grants PHY-0757058 and PHY-0823459 
  This paper has been assigned document number LIGO-P2100277.
 
\appendix
\section{Basic results of extreme value theory}
\label{sec:extreme_value_theory}
Let us consider a set of $n$ independent and identically distributed random variables $\{x_{1}, \dots, x_{n}\}$
each following a probability distribution $f$.
These variables can be identified with a detection statistic evaluated on a set of parameter-space templates,
with $f$ corresponding to the detection statistic's distribution under the noise hypothesis. 
We are interested in describing the probability distribution of the highest detection statistic
value (usually referred to as the largest order statistic~\cite{leadbetter1983extremes})
\mbox{$\starx = \max\{x_{1}, \dots, x_{n}\}$}, 
denoted as $\starf$, in order to evaluate the significance of outliers resulting from a CW search.

A first ansatz for $\starf$ can be constructed by considering the probability of drawing $x_{i} = \starx$
for a single $i \in [1, n]$ and $x_{i} < \starx$ for the remaining $n-1$ values, taking into account all possible
sortings:
\begin{equation}
    \starf_n(x) = n \; f(x) \; \left[ \int^{x} \mathrm{d}x' \; f(x')\right]^{n-1}\;.
    \label{eq:f_star}
\end{equation}
(An alternative derivation of this results is presented in Sec.~\ref{sec:practical_applications}.)
This approach is sufficient if the probability distribution $f$ is well understood and different
random variables $x_i$ are independent from one another so that the joint distribution factors into
the product of individual distributions.
In the case of searches on real data, however, 
parameter-space correlations cause $f^{*}_n$ to deviate from Eq.~\eqref{eq:f_star}~\cite{Wette:2011eu, 
Dreissigacker:2018afk, Tenorio:2021njf}.

\begin{figure*}
    \centering
    \includegraphics[]{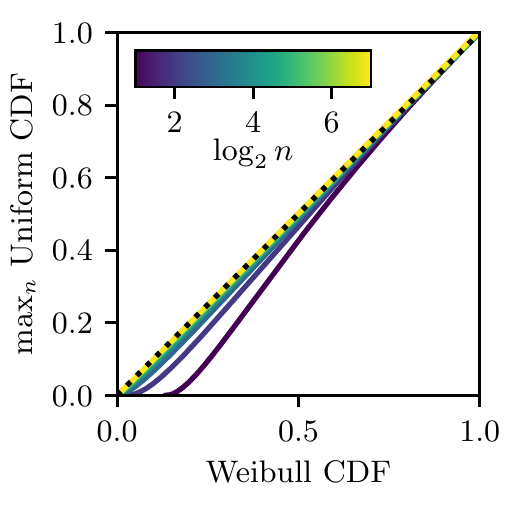}
    \includegraphics[]{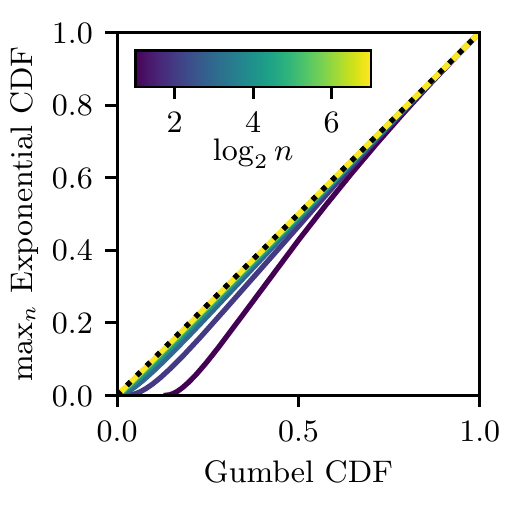}
    \includegraphics[]{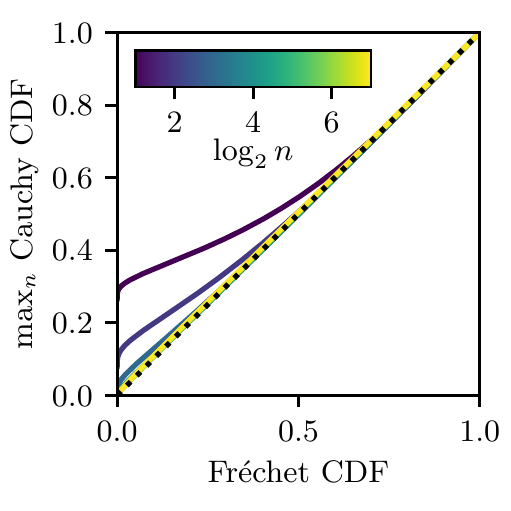}
    \caption{
        CDF comparison of the loudest sample out of $n$ draws from a uniform (upper panel), 
        exponential (middle panel),
        and Cauchy (lower panel) distribution to their corresponding generalized extreme value distribution.
        Different line colors represent different numbers of draws $n$ over which the maximization
        was performed.
    }
    \label{fig:max_example}
\end{figure*}

\begin{figure*}
    \centering
    \includegraphics[]{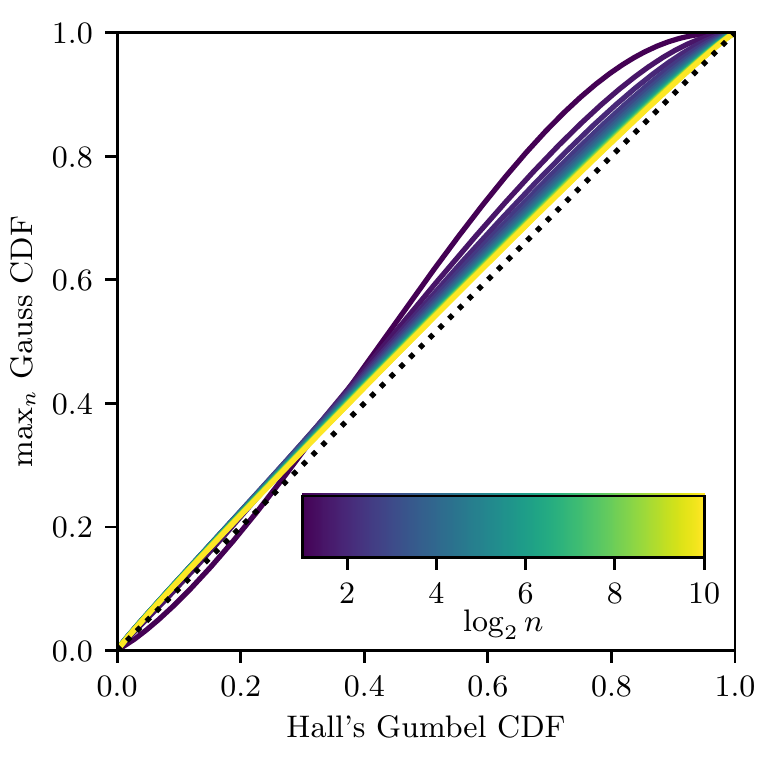}
    \includegraphics[]{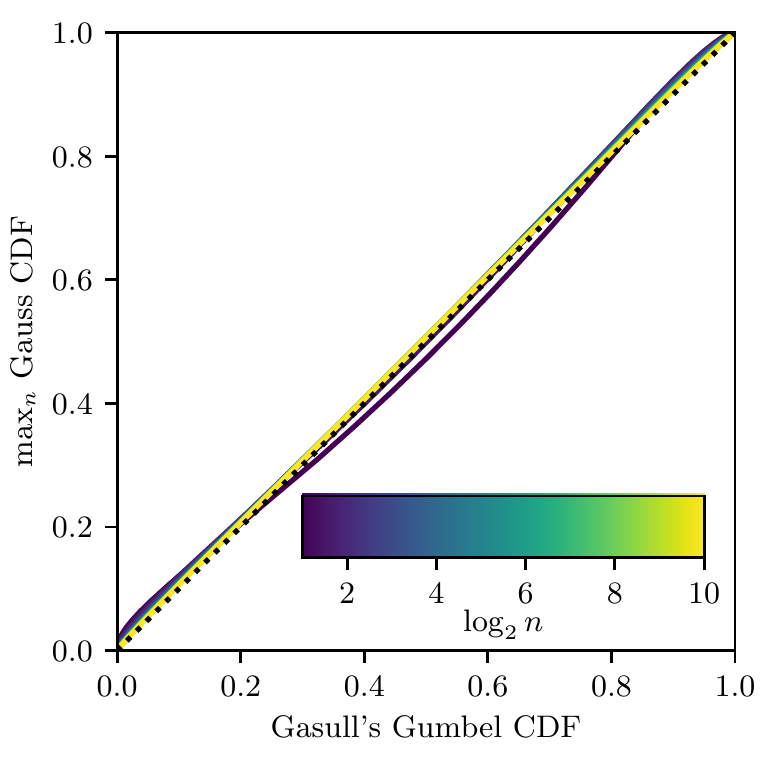}
    \caption{
        Comparison of the speed of convergence of the distribution of the loudest sample out of $n$ draws
        from a standard Gaussian distribution. 
        The upper panel shows the classical result, derived in~\cite{10.2307/3212912}.
        The lower panel shows an improvement later presented in~\cite{GASULL2015376}, 
        which achieves a lower level of discrepancy than the previous one at the same number of draws.
        Different line colors represent different numbers of draws $n$ over which the maximization
        was performed.
    }
    \label{fig:max_normal}
\end{figure*}

\begin{figure*}
    \centering
    \includegraphics[]{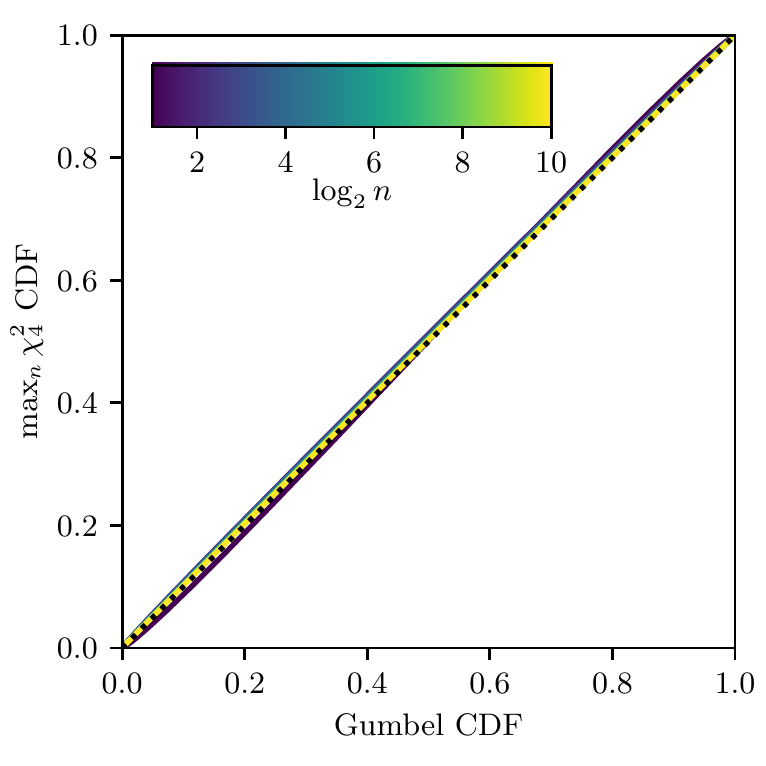}
    \includegraphics[]{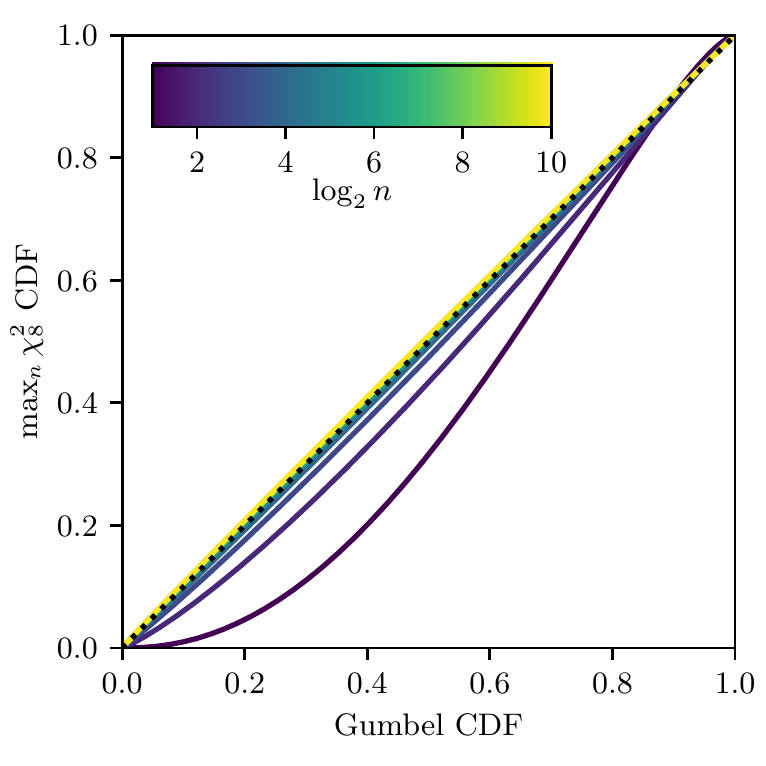}
    \includegraphics[]{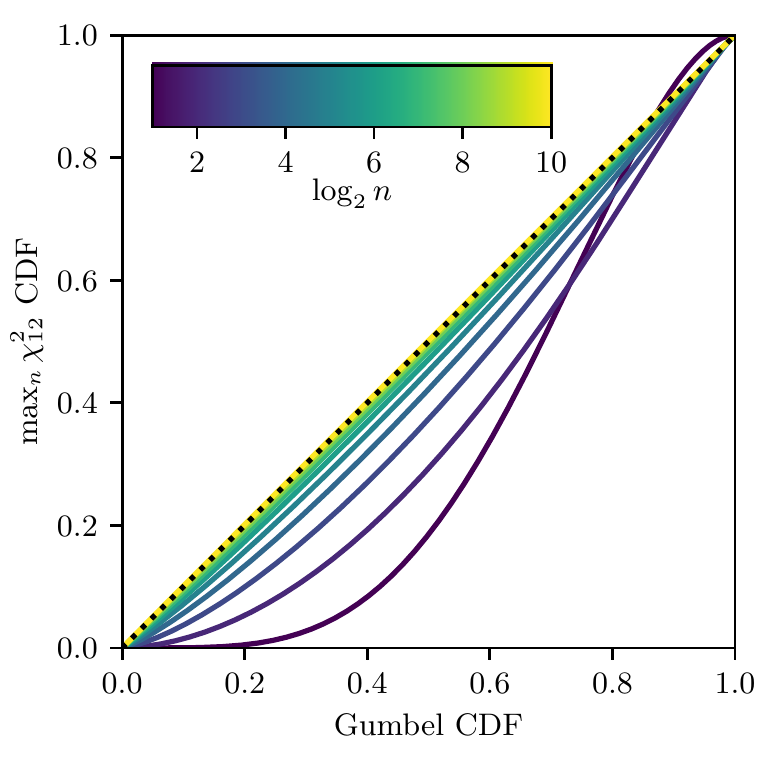}
    \includegraphics[]{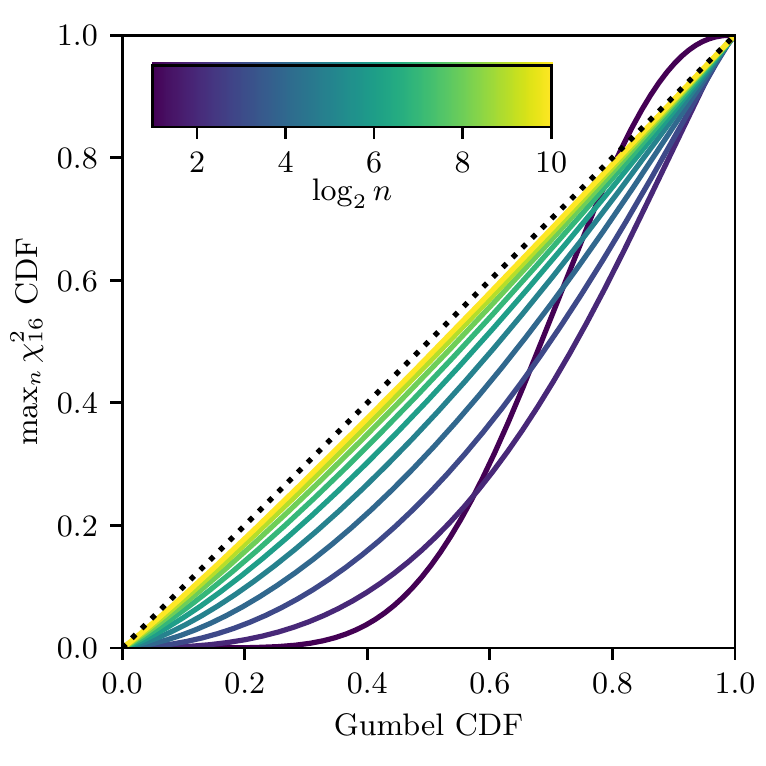}
    \caption{
        CDF comparison of the maximum sample out of $n$ draws from a $\chi^2_k$-distributed random variable
        with different number of degrees of freedom $k$ to their asymptotic Gumbel distribution. 
        Different line colors represent different numbers of draws $n$ over which the maximization
        was performed.}
    \label{fig:chi2_to_gumbel}
\end{figure*}

Extreme Value Theory (EVT) provides asymptotic closed forms for Eq.~\eqref{eq:f_star} in the limit of 
$n \to \infty$
\begin{equation}
    \starf_{n}
    \xrightarrow{n \to \infty} \textrm{GEV}(\gamma)\;,
    \label{eq:starrr}
\end{equation}
where $\textrm{GEV}$ refers to the generalized extreme value distribution
and $\gamma \in \mathbb{R}$ is referred to as the
extreme value index~\cite{leadbetter1983extremes, coles2001introduction, de2006extreme, embrechts2013modelling}.
According to the specific properties of the random variables at hand, 
$\textrm{GEV}$ distributions can be shifted and rescaled by location and scale parameters, 
$\mu_n$ and $\sigma_n$. 
The general dependency on the number of random variables being drawn $n$ is due to the increased chances of
drawing an extreme value as the number of independent trials increases. This is usually referred to as the 
\emph{trials factor}.

In a practical case, assuming $n$ so that the convergence is suitable for the application at hand,
Eq.~\eqref{eq:starrr} can be recast into a closed form
\begin{equation}
    \starf_{n}(x) = \textrm{GEV}(\mu_{n} + x \;\sigma_{n}; \gamma) \;.
    \label{eq:starrr}
\end{equation}
Typical prescriptions for $\mu_{n}$ and $\sigma_{n}$~\cite{GASULL2015376, 
RePEc:spr:testjl:v:24:y:2015:i:4:p:714-733}
tend to be valid for~\mbox{$n \gtrsim 10^4$}.
The $\textrm{GEV}$ distribution has the specific property of being max-stable,
meaning that the distribution of the maximum sample out of $n$ draws from a GEV distribution
is again a GEV of the same kind (same $\gamma$).

The value of $\gamma$ depends on the right-hand tail behaviour of $f$, and determines the functional form of 
$\starf$ out of three possibilities.
We follow the definitions given in~\cite{embrechts2013modelling} (where $\gamma$ is referred to as $\xi$):
finite tails with power-law behaviour correspond to $\gamma < 0$ (Weibull distribution), light tails correspond 
to $\gamma = 0$ (Gumbel distribution), and power-law tails correspond to $\gamma > 0$ 
(Fr\'echet distribution)\footnote{
    The \texttt{scipy} Python package~\cite{2020SciPy-NMeth}
    implements these three distributions under the \texttt{stats} module, 
    although it uses a different sign criterion for the extreme value index,
    therein referred to as $c$. 
    Setting \mbox{$c=\gamma$}, 
    the Gumbel distribution is~\texttt{gumbel\_r}, 
    the Fr\'echet distribution is~\texttt{invweibull}, 
    and the Weibull distribution is~\texttt{weibull\_max}.
}.
Fig.~\ref{fig:max_example} illustrates the convergence towards each of these families using paradigmatic
probability distributions, namely a uniform distribution in [0, 1], a standard exponential distribution, and
a standard Cauchy distribution.

We focus our attention on the location and scale parameters, as they are relevant in terms of convergence speed.
EVT imposes very loose conditions on them, so the choice of $\mu_{n}$ and $\sigma_{n}$ as functions of $n$
is not unique for a given distribution, 
and the main difference across different choices is the speed with which the resulting distribution
will approach the \textrm{GEV} one.
We illustrate this using a Gaussian distribution, which is in the domain of
attraction of the Gumbel distribution and is famous for being quite slow to converge. 
Figure~\ref{fig:max_normal} compares the prescription of location and scale parameters originally proposed by 
Hall~\cite{10.2307/3212912} to the improvement proposed by Gasull~\cite{GASULL2015376}.

However, in this paper we are mainly interested in $\chi^2_k$ distributions, where $k \in \mathbb{N}$ denotes the degrees of
freedom of the distribution, as CW statistics are quite frequently constructed as the norm of a Gaussian vector
and hence follow $\chi^2_k$ distributions.
A significant improvement over the classical literature was 
presented in~\cite{RePEc:spr:testjl:v:24:y:2015:i:4:p:714-733}, where closed expressions for $\mu$ and $\sigma$
for a generic $\Gamma$ distribution were obtained. 
Figure~\ref{fig:chi2_to_gumbel} shows the convergence of different $\chi^2_k$ distributions towards a Gumbel distribution.

We provide an implementation of the corresponding expressions discussed in~\cite{RePEc:spr:testjl:v:24:y:2015:i:4:p:714-733}
within the \dmax{} Python package~\cite{distromax}:
Although the \dmax{} method itself does not use any of these results (since $\mu$ and $\sigma$ are estimated from the data),
they can still be used to produce theoretical estimates.
 
\section{Addressing disturbed data}
\label{sec:loud_disturbances}
The intended output of \dmax{} is an empirical estimation of the distribution of the loudest
candidate produced by noise-only data in a CW search $\pdf(\xi^{*}|\hyp{N})$. 
To do so, the basic assumption is that the output of a search $\xi$ corresponds mostly to
samples of the detection statistic from a single well-behaved distribution.
In practice, this generally means Gaussian noise, plus only a small number of
samples coming from another population such as a non-Gaussianity in the data or a CW signal.
If the number of such samples is negligible compared to the number of batches used in
\dmax{}, so would be their effect in the batchmax distribution.

CW searches in real data, however, are populated by various kinds of noise disturbances.
Concretely, a prominent type are narrow-band instrumental features (``lines''), 
which tend to concentrate their effect within a few frequency bins,
but especially for higher-dimensional searches
(several spin-down terms and/or all-sky searches)
can still affect a large number of templates.
Probability theory provides the right tools to deal with this situation.
Specifically, as discussed in Chapter 21 of~\cite{jaynes_2003}, 
one should describe the results of a search $\xi$ as a mixture of two populations, 
namely a population of samples belonging to the background and another one
belonging to the noise disturbance.
Further populations describing additional effects, such as the presence of a CW signal,
can also be included in the analysis.
The distribution of the loudest candidate produced by the background, $\pdf(\xi^{*}|\hyp{N})$,
would then be obtained by marginalizing out all but the background component of the mixture.
Part of the idea of using multiple candidate populations was implemented in~\cite{Bennett:2013nea}.
In this Appendix, however, 
we concentrate on the typical case of loud noise disturbances polluting a small number of frequency bins, 
for which the ad hoc approach of excising or \emph{notching} 
the disturbed frequency band returns a similar result to a proper Bayesian analysis. 
A full treatment of the mixture model problem is left for future work.

As previously discussed in Sec.~\ref{sec:loudest_outlier}, 
this approach is conceived to deal with a specific set of common noise disturbances so that 
\dmax{} can be applied on a larger range of real-data results.
However, users are encouraged to understand and curate their search results using standard CW vetoes 
(see \cite{universe} and references therein) before falling back to this specific notching algorithm.

Leveraging thresholding algorithms from the image-processing literature \cite{sezgin2004survey}
we propose a simple algorithm capable of notching frequency bands containing prominent disturbances.
Since we focus on noise disturbances within well-localized frequency bands, 
we attempt to flag their corresponding candidates focusing on
the loudest detection statistic in each frequency bin $f_0$
\begin{equation}
    \xi^{*}(f_0) = \max_{\slambda} \xi(f_0, \slambda)\;,
\end{equation}
where $\slambda$ contains any other relevant parameter-space dimension. 
The resulting envelope can be thought of as a one-dimensional gray-scale image in which 
we are interested to discern the background from an object (the polluted band); 
the distinction is made by properly selecting a gray-value (detection-statistic) threshold such 
that object pixels (polluted-band samples) lie above it, leaving nothing but background below. 
\begin{figure}
    \includegraphics[]{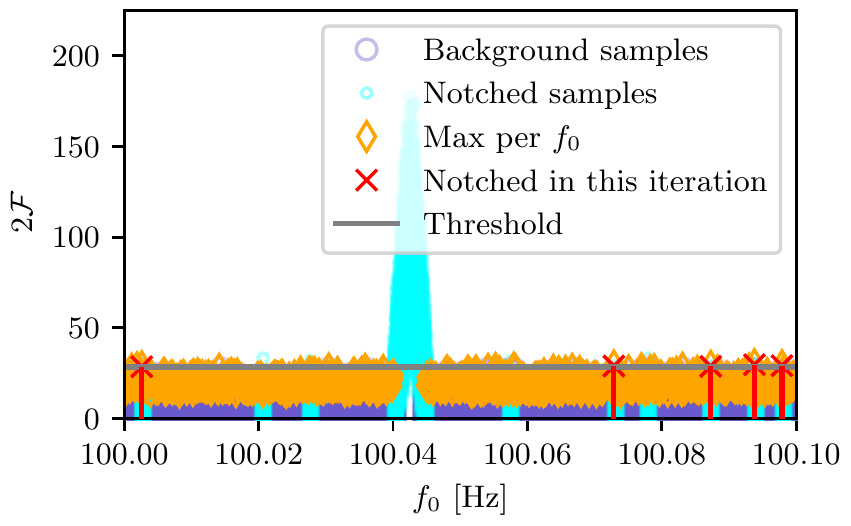}
    \caption{
        Simulated samples corresponding to $10^6$ templates in a square grid across $(f_0, f_1)$.
        $2\F$ background samples correspond to draws from a $\chi^2_{4}$ distribution.
        $2\F$ outlier samples are drawn from a non-central $\chi^{2}_{4}$ distribution 
        with non-centrality parameter $\rho^2 = 25$. 
        This figure shows the third notching iteration.
        Orange diamonds correspond to the loudest outlier per frequency bin $\xi^{*}(f_0)$.
        Cyan dots mark samples notched in a previous iteration.
        Red crosses and the corresponding vertical lines denote the frequency bins being
        notched in the present iteration.
        The solid horizontal line corresponds to the threshold computed 
        on the maximum samples using \texttt{skimage.filters.threshold\_minimum.}
    }
    \label{fig:outlier_samples}
\end{figure}

We illustrate the effects of our notching algorithm using 
a synthetic template bank containing a narrow-band disturbance.
The template bank contains $1000 \times 1000$ templates spanning the $(f_0, f_1)$ parameter space over the 
\mbox{$[100, 100.1] \;\textrm{Hz}$} frequency band. 
The corresponding $2\F$ is drawn from a $\chi^2_{4}$ distribution for each template in the bank. 
We refer to this $\chi^2_{4}$-drawn set of samples as the ground truth. 
An outlier is introduced by replacing samples in the $[100.04, 100.05]\;\textrm{Hz}$ sub-band with an equal
amount of samples drawn from a non-central $\chi^2_{4}$ distribution with non-centrality parameter 
$\rho^2 = 25$.
The sample projection over the $f_0$ subspace is shown in Fig.~\ref{fig:outlier_samples}.

We tested different thresholding techniques,
including standard approaches such as the Otsu threshold~\cite{4310076}, 
the minimum cross-entropy threshold~\cite{LI1993617, LI1998771}, 
and the minimum method threshold~\cite{min_threshold},
using the implementations available in the \texttt{skimage} package \cite{scikit-image}.
We find the minimum method threshold~\texttt{skimage.filters.threshold\_minimum} 
performs best in our specific study, 
noting that the implementation of the notching procedure in~\cite{distromax}
allows for a flexible selection of thresholding strategies.

Once an appropriate threshold $\xit$ has been established, 
we proceed to \emph{notch} any frequency bin containing at least one sample above threshold. 
Specifically, we remove \emph{all} the samples from the frequency bins $f_0$ where
\mbox{$\xi^{*}(f_0) > \xit$}.
This step can be applied multiple times in order to take care of multiple lines in a band with 
very different amplitudes, ``shoulders'' of broad lines, or features such as spectral leakage.
The specific implementation provided in~\cite{distromax} implements a simple stopping criterion:
notching iterations stop whenever the threshold $\xit$ falls below a pre-specified quantile of 
$\xi^{*}(f_0)$. 
The default value, which performs well for our specific example, stops whenever $\xit$ is lower
than the top $20\%$ values of $\xi^{*}(f_0)$.
\begin{figure}
    \includegraphics[]{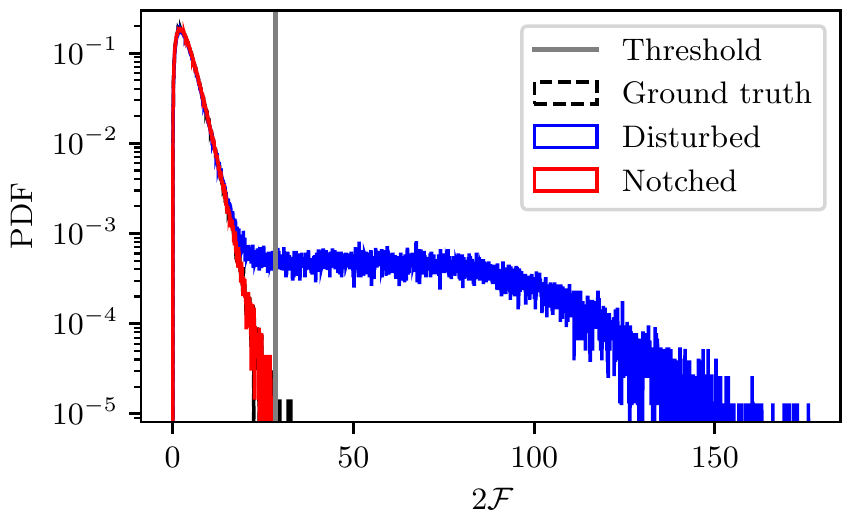}
    \caption{
        Effects of notching disturbed frequency bins on the background distribution.
        Each histogram shows a set of $2\F$ samples akin to that in Fig.~\ref{fig:outlier_samples}. 
        The black dashed histogram represents ground truth samples in which no disturbance has been included. 
        The blue histogram represents the ground truth samples plus a loud narrow-band disturbance.
        The red histogram represents the result of notching the blue histogram as discussed in the main text
        and the caption of Fig.~\ref{fig:outlier_samples}.
    }
    \label{fig:outlier_histograms}
\end{figure}

The result of notching, as opposed to simply removing samples over threshold, 
is illustrated in Fig.~\ref{fig:outlier_histograms}, 
where the distribution being notched is shown as a blue histogram. 
Simply removing samples above threshold would be equivalent to \emph{cutting} the tail of
the histogram while leaving the bulk untouched. 
While such an approach would be relatively harmless in the case of a disturbance strong enough 
to be cleanly separated from the background, 
it is rendered ineffective in the case of a relatively mild disturbance,
as the polluted-band samples tend to overpopulate the tail of the distribution itself.
Completely notching the band, on the other hand (red histogram), 
properly deals with the overpopulation of outliers and 
returns a distribution consistent with an undisturbed background.

\begin{figure}
    \includegraphics[]{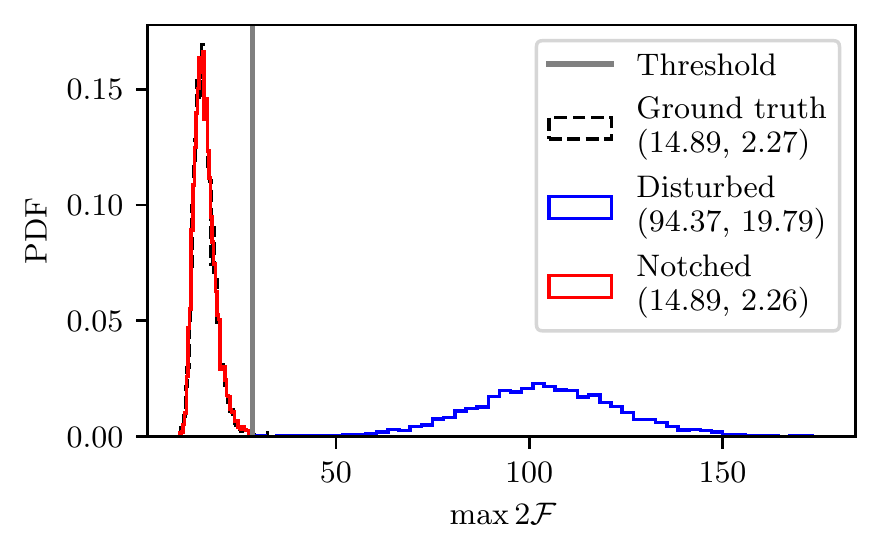}
    \caption{
        Distribution of batchmax samples. 
        Each histogram contains the maxima of 5000 batches generated by randomly shuffling samples from 
        Fig.~\ref{fig:outlier_histograms}. 
        Numbers in the legend indicate the location and scale parameters of a Gumbel distribution 
        fitted to each histogram.
    }
    \label{fig:outlier_max_histograms}
\end{figure}

\begin{figure}
    \includegraphics[]{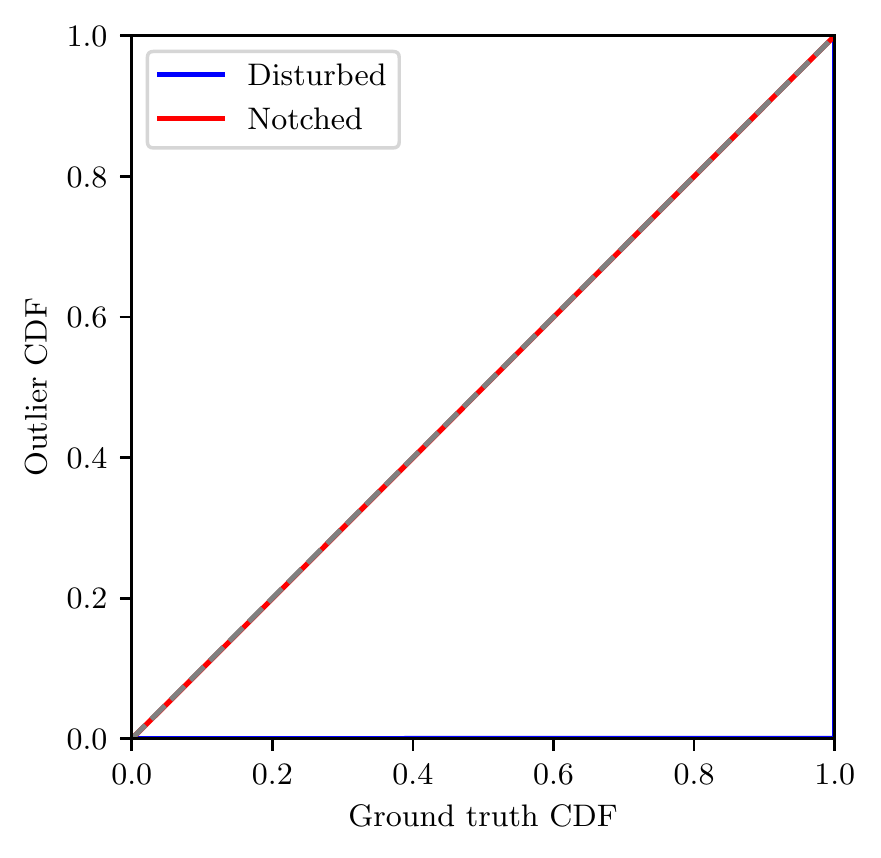}
    \caption{
        CDF comparison of the batchmax distributions obtained with (red) and without (blue)
        the application of notching to the ground truth CDF.
        Each line compares the corresponding CDF of the distribution shown in Fig.~\ref{fig:outlier_max_histograms}
        to the CDF of the ground truth (black histogram).
        The dashed gray line represents equality.
    }
    \label{fig:pp_plot_notching}
\end{figure}

After notching disturbed bands, we can simply apply the shuffling and batching procedure described in 
Sec.~\ref{subsec:evt}. The resulting batchmax distributions, including the unnotched and ground truth 
distributions, are shown in Fig.~\ref{fig:outlier_max_histograms}. Location and scale parameters of the 
best fitting Gumbel distributions are shown in the legend and compared to the ground truth distribution 
in Fig.~\ref{fig:pp_plot_notching}. 

In this example, the estimated parameters using notching show only $1\%$ relative difference with respect to 
ground truth parameters,
as opposed to the strong bias suffered by the unnotched estimates.
Moreover, we remark the robustness of the method to a mild overnotching of non-polluted frequency bins:
the convergence to a Gumbel distribution is mainly related to the properties of the ``bulk of the tail'';
trimming the most extreme events from the background distribution does not affect significantly 
the fitting of a Gumbel distribution.
This is clearly seen in Fig.~\ref{fig:outlier_histograms}, where the notched distribution differs from the
ground truth by a few samples. 
These samples correspond to the background samples over threshold in Fig.~\ref{fig:outlier_samples}, 
which belong to the tail of the non-disturbed distribution.
As briefly commented in Sec.~\ref{subsec:discussion}, 
this result justifies the extent up to which \dmax{} is applicable to toplist-based searches:
so long as the toplist reaches the bulk of the distribution, 
\dmax{} should be capable of returning a cogent answer. 

The application of more notching iterations than strictly required, however, 
could result in an underestimation of the Gumbel parameters due to the removal 
of too many samples in the tail of the distribution.
This is particularly important for the scale parameter:
underestimations around $5\%$ are often obtained across several realizations of the example setup discussed here
when using the notching procedure with the stopping criterion as described above.
For a typical batch size of $B \simeq 10^3 - 10^4$ ($\ln{B} \simeq 10$) and fiducial
values of $\mu_{*} = 50$ and $\sigma_{n} = 2$ [Eq.~\eqref{eq:mu_star}]
(similar to the values encountered in a CW search using the $2\F$ statistic),
a 5\% underestimation in $\sigma_{n}$ implies about 20\% of underestimation in $\mu_{*}$.
The main consequence of this is a shift of the resulting  $\pdf(\xi^{*}|\hyp{N})$ towards lower values,
potentially resulting in an increased number of candidates scoring over the specified threshold.
 
\section{Robustness against injections}
\label{sec:injections}
\begin{figure*}[t]
    \centering
    \includegraphics[width=0.495\textwidth]{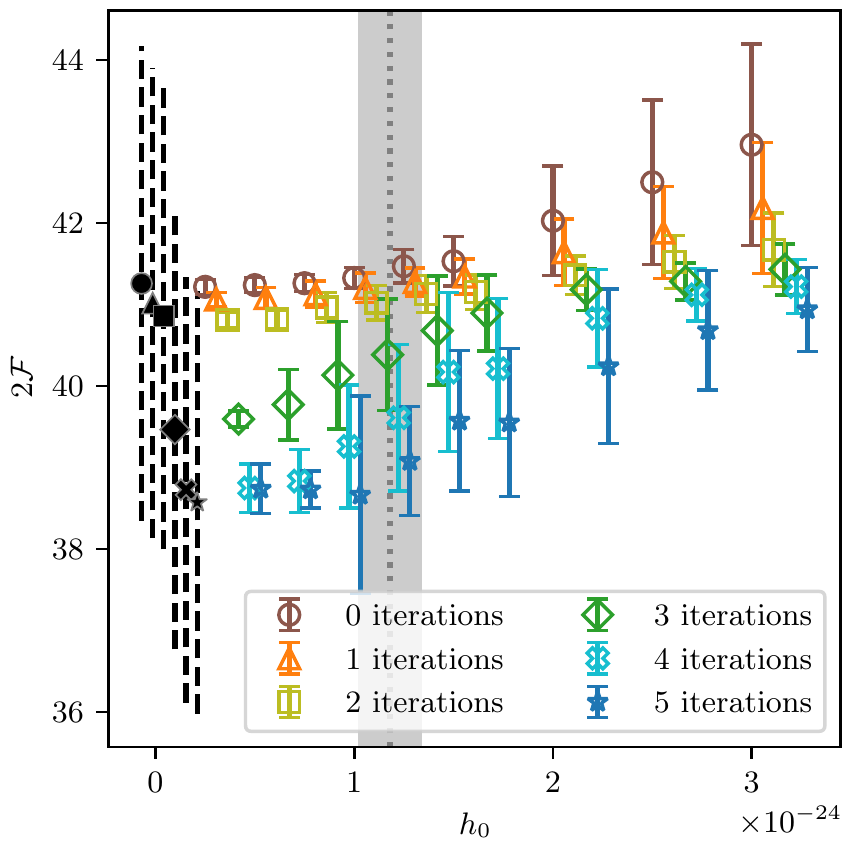}
    \includegraphics[width=0.495\textwidth]{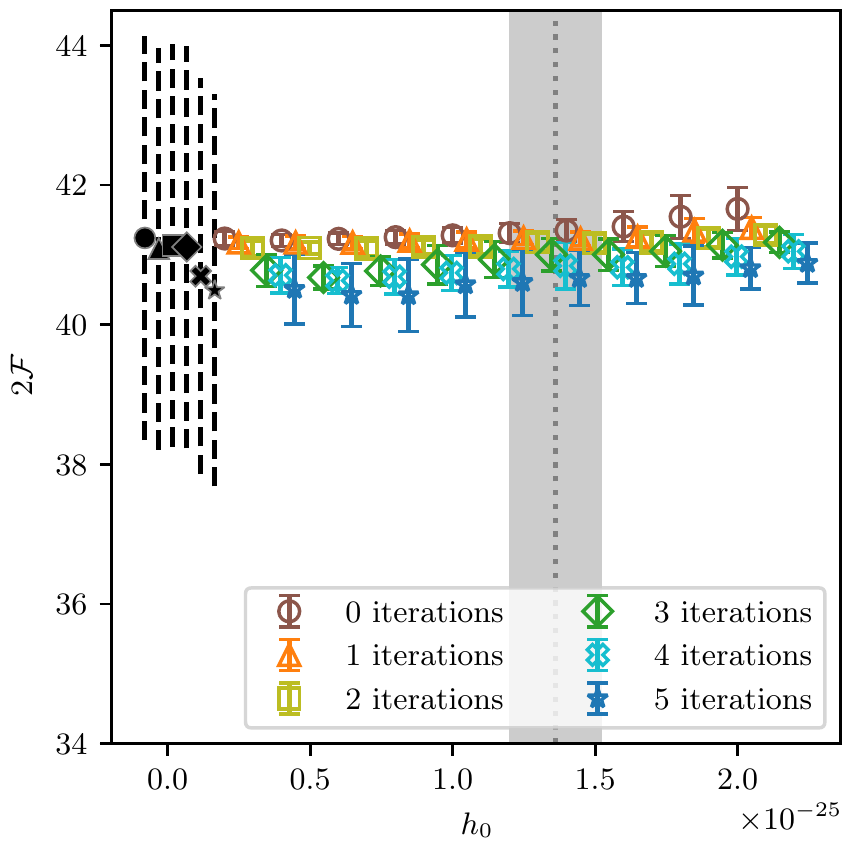}
\caption{
    \label{fig:2F_injs}
        Results of testing \dmax{} on $2\F$ values from simulated signal injections
        for the Vela (left) and Crab (right) O2 search parameter spaces,
        matching the setup from \cite{2019PhRvD.100f4058K}, plotted as a function of 
        injected amplitude $h_0$.
        The black data points corresponding to $h_0=0$
        are the means of the estimated Gumbel distribution on the original data
        without injected signals and their dashed vertical lines 
        correspond to the standard deviation of the same distribution. 
        The different markers correspond to different choices of notching iterations.
        For each injected amplitude, the colored data points with different markers are 
        the average distribution means over 50 injected signals. 
        Their error bars show the standard deviations of these 50 means.
        A small horizontal shift between the data points
        belonging to the same $h_0$ has been inserted for readability. 
        The gray highlight areas are the 90\% upper limits taken from \cite{2019PhRvD.100f4058K} for Vela and Crab, 
        considering their uncertainty.
    }
\end{figure*}

\begin{figure*}[t]
    \centering
    \includegraphics[width=0.495\textwidth]{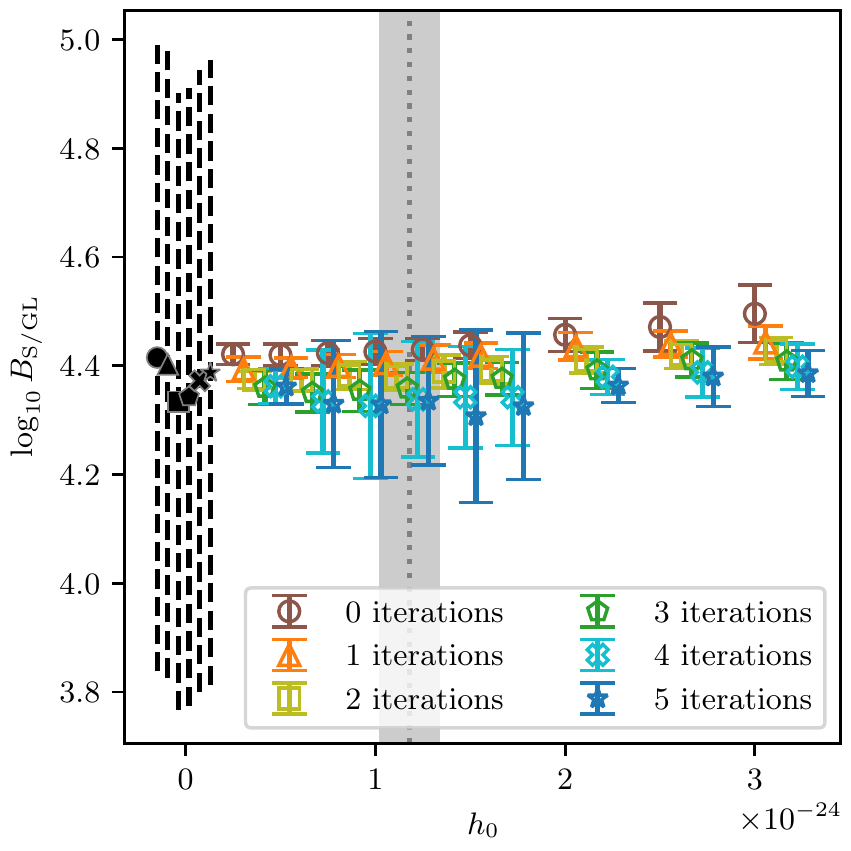}
    \includegraphics[width=0.495\textwidth]{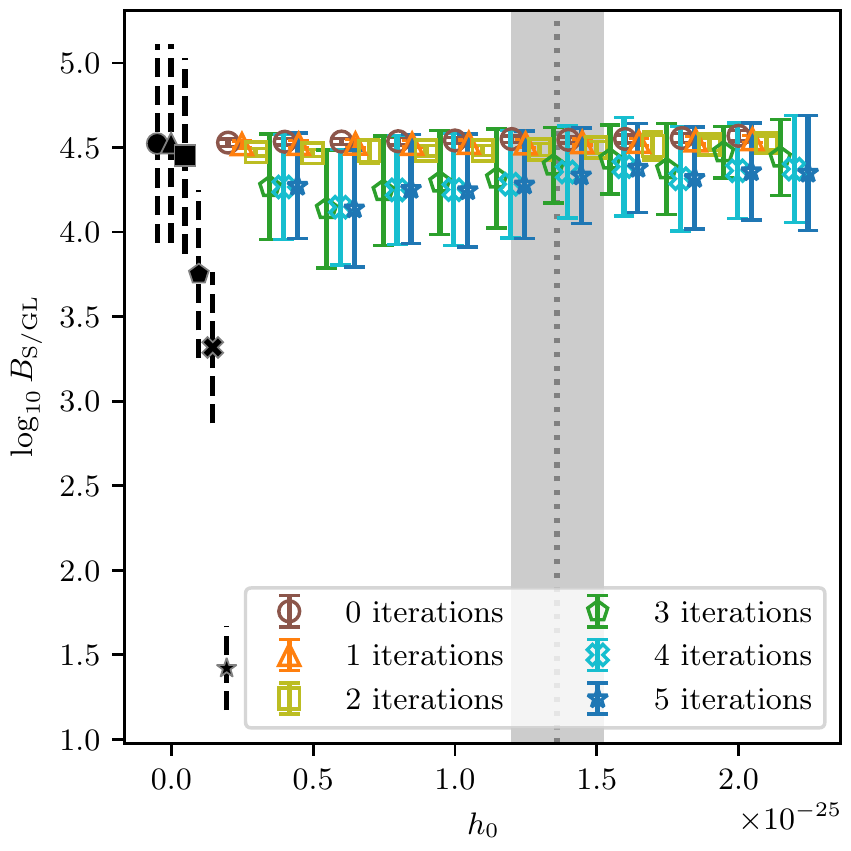}
\caption{
    \label{fig:logBSGL_injs}
    Results of testing \dmax{} on $\log_{10}B_{\textrm{S}/{\textrm{GL}}}$ values from simulated signal injections
    for the Vela (left) and Crab (right) O2 search parameter spaces,
    with all details as in Fig.~\ref{fig:2F_injs}.
    }
\end{figure*}

\begin{figure*}[t]
    \centering
    \includegraphics[width=0.495\textwidth]{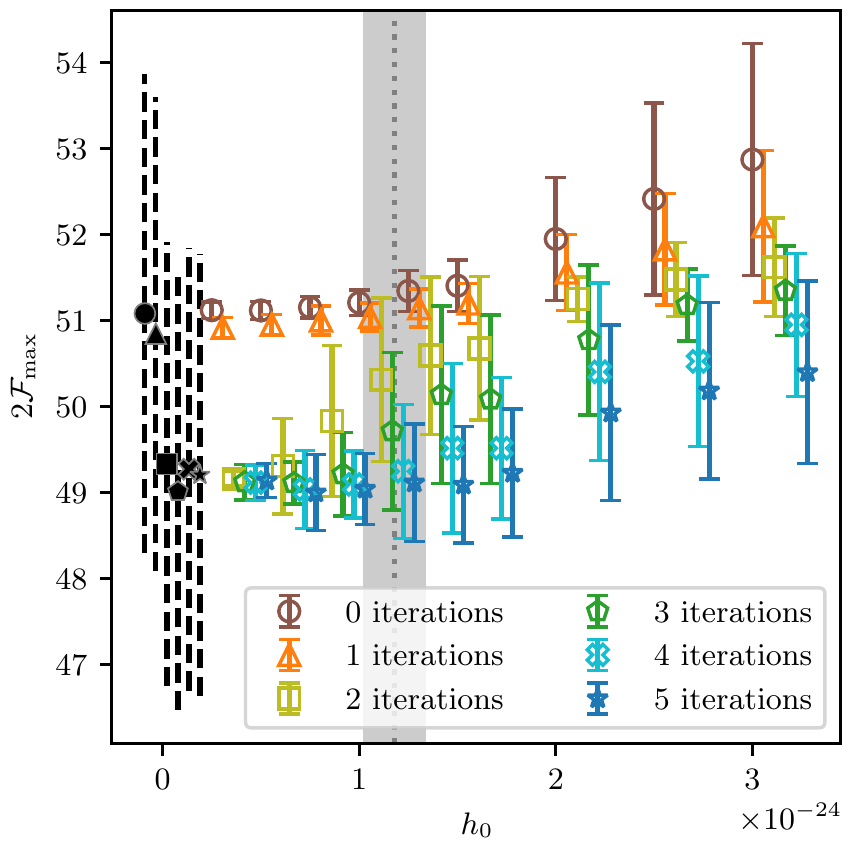}
    \includegraphics[width=0.495\textwidth]{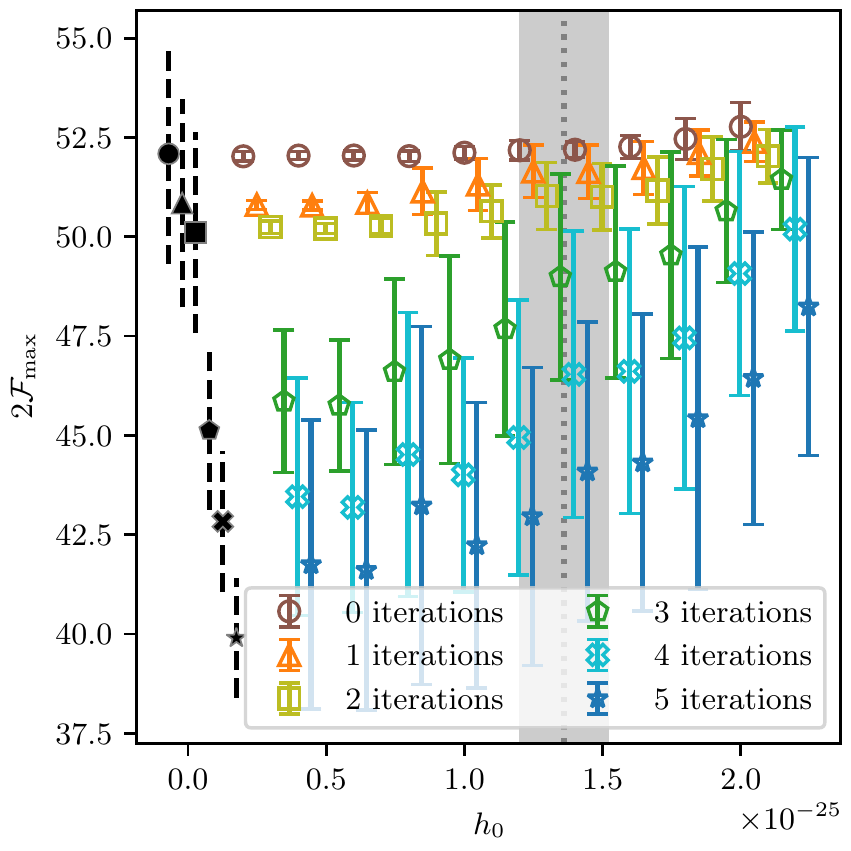}
\caption{
    \label{fig:max2F_injs}
    Results of testing \dmax{} on $2\F_{\max}$ values from simulated signal injections
    for the Vela (left) and Crab (right) O2 search parameter spaces,
    with all details as in Fig.~\ref{fig:2F_injs}.
    }
\end{figure*}

\begin{figure*}[t]
    \centering
    \includegraphics[width=0.495\textwidth]{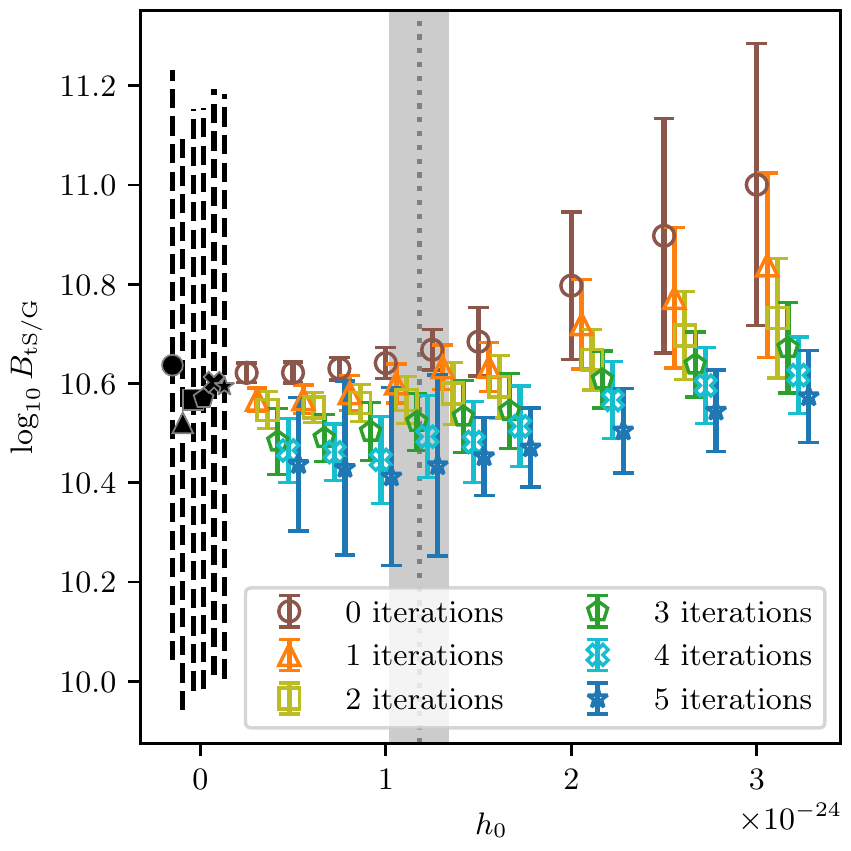}
    \includegraphics[width=0.495\textwidth]{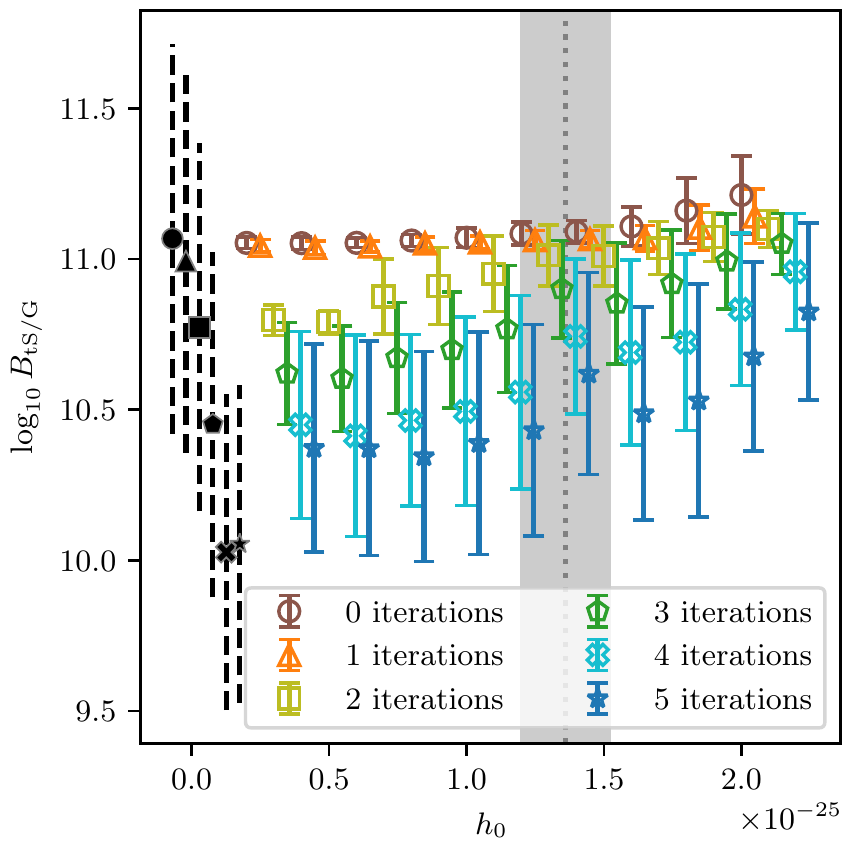}
\caption{
    \label{fig:logB_injs}
    Results of testing \dmax{} on $\log_{10}B_{\textrm{tS}/{\textrm{G}}}$ values from simulated signal injections
    for the Vela (left) and Crab (right) O2 search parameter spaces, with all details as in Fig.~\ref{fig:2F_injs}.
    }
\end{figure*}

As discussed in Sec.~\ref{subsec:discussion},
(t)CW signals themselves can be considered as disturbances
when trying to estimate a background distribution.
We therefore need to test that the \dmax{} method is robust to the presence
of (t)CW signals and will not be biased upwards,
which would lead to picking too high a threshold and missing those signal-related candidates.
Such a bias is not entirely avoidable,
but should ideally only appear for signals stronger than expected in any practical real-data search situation.

To test this, we re-use the upper limits injections in O2 data from 
the same analysis \cite{2019PhRvD.100f4058K} as in Sec.~\ref{sec:real_data}.
Simulated signals of increasing amplitude $h_0$ were added to the original short Fourier Transform (SFT) data (``injections'').
For each chosen $h_0$, there are 50 data sets
with parameters $\{f_0, f_1, t_0\}$ uniformly distributed over their respective search ranges (see \cite{2019PhRvD.100f4058K})
and the remaining amplitude parameters $\{\cos{\iota}, \psi, \phi_0\}$ randomized over their natural ranges.
\texttt{ComputeFstatistic\_v2} \cite{lalsuite} was used to reanalyze a small range around the injection point
and we combine these results with the original detection statistic samples for the rest of the search band.
We then apply \dmax{} as in Sec.~\ref{sec:real_data},
checking the resulting propagated distribution for the maximum of each detection statistic.
We also test the notching procedure introduced in Appendix~\ref{sec:loud_disturbances} by gradually increasing 
the \texttt{num\_iterations} option, from 0 to 5.
The means and standard deviations of the propagated distributions for the four statistics 
discussed in Sec.~\ref{sec:real_data} are shown in 
Fig.~\ref{fig:2F_injs}, Fig.~\ref{fig:logBSGL_injs}, Fig.~\ref{fig:max2F_injs} and Fig.~\ref{fig:logB_injs}.

As one increases the amplitude $h_0$ of the injected signal, an increasing number of templates will produce elevated values of the detection statistic.
In Sec.~\ref{subsec:accuracy}, we found that shuffling of batches in the \bmax{} step is generally preferred.
However, with a strong signal present that affects multiple templates,
the shuffled \bmax{} distribution will inevitably become contaminated,
leading to an overestimation of the final distribution parameters.
Such a trend is indeed visible in the results for all detection statistics,
more clearly for the Vela analyses (left-side panels).
However, the effect is small compared to the actual increase of the detection statistic at the templates with injections.
For the highest amplitudes tested, the detection statistic can reach values $\sim 20$ times 
above the expected loudest background sample.
Furthermore, the upwards shift in the estimated distribution is generally mild,
with the mean shifting by less than one standard deviation of the original injection-free mean,
at least as long as $h_0$ does not reach significantly above the 90\% upper limits set in \cite{2019PhRvD.100f4058K}.

In addition, the notching feature can be useful in limiting the rise of the estimated means
in the presence of signals with large $h_0$,
by treating the templates with elevated detection statistic as disturbances for the purpose of background estimation
and removing them before applying the \bmax{} procedure.
Generally 1--2 iterations of notching have little influence on the estimated 
Gumbel distribution mean of the original data without injections
while helping to reduce the rise of the estimated distribution mean with injection $h_0$.
With more iterations of notching, results become more robust towards strong injections,
while for the original data the means in some cases are estimated lower.
This would always be conservative in the sense that one would retain more candidates for follow-up even when ``over-notching'' clean data,
however it would lead to additional human and computing effort to follow up candidates that are clearly noise fluctuations.

Hence, as already discussed in Appendix~\ref{sec:loud_disturbances}, we recommend using the notching feature only if required.
If data is clean -- i.e. no unusual features in the \bmax{} histograms --
notching is not necessary, and more likely leads to underestimated distribution parameters which would correspond to an overly conservative threshold choice.
If, on the other hand, data exhibits strong and numerous spikes in the statistic,
the plain \dmax{} method may lead to overestimated distribution parameters,
and notching can be a useful tool in such situations.
Regarding the potential presence of (t)CW signals in the data,
\dmax{} seems robust to these, with or even without notching,
for the typical target signals of current (t)CW searches
(not standing out far above the noise background);
and if in doubt, notching can still help to provide more conservative thresholds.
 
\bibliography{references}

\begin{thebibliography}{105}%
\makeatletter
\providecommand \@ifxundefined [1]{%
 \@ifx{#1\undefined}
}%
\providecommand \@ifnum [1]{%
 \ifnum #1\expandafter \@firstoftwo
 \else \expandafter \@secondoftwo
 \fi
}%
\providecommand \@ifx [1]{%
 \ifx #1\expandafter \@firstoftwo
 \else \expandafter \@secondoftwo
 \fi
}%
\providecommand \natexlab [1]{#1}%
\providecommand \enquote  [1]{``#1''}%
\providecommand \bibnamefont  [1]{#1}%
\providecommand \bibfnamefont [1]{#1}%
\providecommand \citenamefont [1]{#1}%
\providecommand \href@noop [0]{\@secondoftwo}%
\providecommand \href [0]{\begingroup \@sanitize@url \@href}%
\providecommand \@href[1]{\@@startlink{#1}\@@href}%
\providecommand \@@href[1]{\endgroup#1\@@endlink}%
\providecommand \@sanitize@url [0]{\catcode `\\12\catcode `\$12\catcode
  `\&12\catcode `\#12\catcode `\^12\catcode `\_12\catcode `\%12\relax}%
\providecommand \@@startlink[1]{}%
\providecommand \@@endlink[0]{}%
\providecommand \url  [0]{\begingroup\@sanitize@url \@url }%
\providecommand \@url [1]{\endgroup\@href {#1}{\urlprefix }}%
\providecommand \urlprefix  [0]{URL }%
\providecommand \Eprint [0]{\href }%
\providecommand \doibase [0]{https://doi.org/}%
\providecommand \selectlanguage [0]{\@gobble}%
\providecommand \bibinfo  [0]{\@secondoftwo}%
\providecommand \bibfield  [0]{\@secondoftwo}%
\providecommand \translation [1]{[#1]}%
\providecommand \BibitemOpen [0]{}%
\providecommand \bibitemStop [0]{}%
\providecommand \bibitemNoStop [0]{.\EOS\space}%
\providecommand \EOS [0]{\spacefactor3000\relax}%
\providecommand \BibitemShut  [1]{\csname bibitem#1\endcsname}%
\let\auto@bib@innerbib\@empty
\bibitem [{\citenamefont {Jaynes}(2003)}]{jaynes_2003}%
  \BibitemOpen
  \bibfield  {author} {\bibinfo {author} {\bibfnamefont {E.~T.}\ \bibnamefont
  {Jaynes}},\ }\href {https://doi.org/https://doi.org/10.1017/CBO9780511790423}
  {\emph {\bibinfo {title} {Probability Theory: The Logic of Science}}},\
  edited by\ \bibinfo {editor} {\bibfnamefont {G.~L.}\ \bibnamefont
  {Bretthorst}}\ (\bibinfo  {publisher} {Cambridge University Press},\ \bibinfo
  {year} {2003})\BibitemShut {NoStop}%
\bibitem [{\citenamefont {Owen}(1996)}]{Owen:1995tm}%
  \BibitemOpen
  \bibfield  {author} {\bibinfo {author} {\bibfnamefont {B.~J.}\ \bibnamefont
  {Owen}},\ }\bibfield  {title} {\bibinfo {title} {{Search templates for
  gravitational waves from inspiraling binaries: Choice of template spacing}},\
  }\href {https://doi.org/10.1103/PhysRevD.53.6749} {\bibfield  {journal}
  {\bibinfo  {journal} {Phys. Rev. D}\ }\textbf {\bibinfo {volume} {53}},\
  \bibinfo {pages} {6749} (\bibinfo {year} {1996})},\ \Eprint
  {https://arxiv.org/abs/gr-qc/9511032} {arXiv:gr-qc/9511032} \BibitemShut
  {NoStop}%
\bibitem [{\citenamefont {Brady}\ and\ \citenamefont
  {Creighton}(2000)}]{Brady:1998nj}%
  \BibitemOpen
  \bibfield  {author} {\bibinfo {author} {\bibfnamefont {P.~R.}\ \bibnamefont
  {Brady}}\ and\ \bibinfo {author} {\bibfnamefont {T.}~\bibnamefont
  {Creighton}},\ }\bibfield  {title} {\bibinfo {title} {{Searching for periodic
  sources with LIGO. 2. Hierarchical searches}},\ }\href
  {https://doi.org/10.1103/PhysRevD.61.082001} {\bibfield  {journal} {\bibinfo
  {journal} {Phys. Rev. D}\ }\textbf {\bibinfo {volume} {61}},\ \bibinfo
  {pages} {082001} (\bibinfo {year} {2000})},\ \Eprint
  {https://arxiv.org/abs/gr-qc/9812014} {arXiv:gr-qc/9812014} \BibitemShut
  {NoStop}%
\bibitem [{\citenamefont {Sathyaprakash}\ and\ \citenamefont
  {Dhurandhar}(1991)}]{Sathyaprakash:1991mt}%
  \BibitemOpen
  \bibfield  {author} {\bibinfo {author} {\bibfnamefont {B.~S.}\ \bibnamefont
  {Sathyaprakash}}\ and\ \bibinfo {author} {\bibfnamefont {S.~V.}\ \bibnamefont
  {Dhurandhar}},\ }\bibfield  {title} {\bibinfo {title} {{Choice of filters for
  the detection of gravitational waves from coalescing binaries}},\ }\href
  {https://doi.org/10.1103/PhysRevD.44.3819} {\bibfield  {journal} {\bibinfo
  {journal} {Phys. Rev. D}\ }\textbf {\bibinfo {volume} {44}},\ \bibinfo
  {pages} {3819} (\bibinfo {year} {1991})}\BibitemShut {NoStop}%
\bibitem [{\citenamefont {Owen}\ and\ \citenamefont
  {Sathyaprakash}(1999)}]{Owen:1998dk}%
  \BibitemOpen
  \bibfield  {author} {\bibinfo {author} {\bibfnamefont {B.~J.}\ \bibnamefont
  {Owen}}\ and\ \bibinfo {author} {\bibfnamefont {B.~S.}\ \bibnamefont
  {Sathyaprakash}},\ }\bibfield  {title} {\bibinfo {title} {{Matched filtering
  of gravitational waves from inspiraling compact binaries: Computational cost
  and template placement}},\ }\href
  {https://doi.org/10.1103/PhysRevD.60.022002} {\bibfield  {journal} {\bibinfo
  {journal} {Phys. Rev. D}\ }\textbf {\bibinfo {volume} {60}},\ \bibinfo
  {pages} {022002} (\bibinfo {year} {1999})},\ \Eprint
  {https://arxiv.org/abs/gr-qc/9808076} {arXiv:gr-qc/9808076} \BibitemShut
  {NoStop}%
\bibitem [{\citenamefont {Allen}\ \emph {et~al.}(2012)\citenamefont {Allen},
  \citenamefont {Anderson}, \citenamefont {Brady}, \citenamefont {Brown},\ and\
  \citenamefont {Creighton}}]{Allen:2005fk}%
  \BibitemOpen
  \bibfield  {author} {\bibinfo {author} {\bibfnamefont {B.}~\bibnamefont
  {Allen}}, \bibinfo {author} {\bibfnamefont {W.~G.}\ \bibnamefont {Anderson}},
  \bibinfo {author} {\bibfnamefont {P.~R.}\ \bibnamefont {Brady}}, \bibinfo
  {author} {\bibfnamefont {D.~A.}\ \bibnamefont {Brown}},\ and\ \bibinfo
  {author} {\bibfnamefont {J.~D.~E.}\ \bibnamefont {Creighton}},\ }\bibfield
  {title} {\bibinfo {title} {{FINDCHIRP: An Algorithm for detection of
  gravitational waves from inspiraling compact binaries}},\ }\href
  {https://doi.org/10.1103/PhysRevD.85.122006} {\bibfield  {journal} {\bibinfo
  {journal} {Phys. Rev. D}\ }\textbf {\bibinfo {volume} {85}},\ \bibinfo
  {pages} {122006} (\bibinfo {year} {2012})},\ \Eprint
  {https://arxiv.org/abs/gr-qc/0509116} {arXiv:gr-qc/0509116} \BibitemShut
  {NoStop}%
\bibitem [{\citenamefont {Prix}(2007{\natexlab{a}})}]{Prix:2007ks}%
  \BibitemOpen
  \bibfield  {author} {\bibinfo {author} {\bibfnamefont {R.}~\bibnamefont
  {Prix}},\ }\bibfield  {title} {\bibinfo {title} {{Template-based searches for
  gravitational waves: Efficient lattice covering of flat parameter spaces}},\
  }\href {https://doi.org/10.1088/0264-9381/24/19/S11} {\bibfield  {journal}
  {\bibinfo  {journal} {Class. Quant. Grav.}\ }\textbf {\bibinfo {volume}
  {24}},\ \bibinfo {pages} {S481} (\bibinfo {year} {2007}{\natexlab{a}})},\
  \Eprint {https://arxiv.org/abs/0707.0428} {arXiv:0707.0428 [gr-qc]}
  \BibitemShut {NoStop}%
\bibitem [{\citenamefont {Harry}\ \emph {et~al.}(2009)\citenamefont {Harry},
  \citenamefont {Allen},\ and\ \citenamefont {Sathyaprakash}}]{Harry:2009ea}%
  \BibitemOpen
  \bibfield  {author} {\bibinfo {author} {\bibfnamefont {I.~W.}\ \bibnamefont
  {Harry}}, \bibinfo {author} {\bibfnamefont {B.}~\bibnamefont {Allen}},\ and\
  \bibinfo {author} {\bibfnamefont {B.~S.}\ \bibnamefont {Sathyaprakash}},\
  }\bibfield  {title} {\bibinfo {title} {{A Stochastic template placement
  algorithm for gravitational wave data analysis}},\ }\href
  {https://doi.org/10.1103/PhysRevD.80.104014} {\bibfield  {journal} {\bibinfo
  {journal} {Phys. Rev. D}\ }\textbf {\bibinfo {volume} {80}},\ \bibinfo
  {pages} {104014} (\bibinfo {year} {2009})},\ \Eprint
  {https://arxiv.org/abs/0908.2090} {arXiv:0908.2090 [gr-qc]} \BibitemShut
  {NoStop}%
\bibitem [{\citenamefont {Messenger}\ \emph {et~al.}(2009)\citenamefont
  {Messenger}, \citenamefont {Prix},\ and\ \citenamefont
  {Papa}}]{Messenger:2008ta}%
  \BibitemOpen
  \bibfield  {author} {\bibinfo {author} {\bibfnamefont {C.}~\bibnamefont
  {Messenger}}, \bibinfo {author} {\bibfnamefont {R.}~\bibnamefont {Prix}},\
  and\ \bibinfo {author} {\bibfnamefont {M.~A.}\ \bibnamefont {Papa}},\
  }\bibfield  {title} {\bibinfo {title} {{Random template banks and relaxed
  lattice coverings}},\ }\href {https://doi.org/10.1103/PhysRevD.79.104017}
  {\bibfield  {journal} {\bibinfo  {journal} {Phys. Rev. D}\ }\textbf {\bibinfo
  {volume} {79}},\ \bibinfo {pages} {104017} (\bibinfo {year} {2009})},\
  \Eprint {https://arxiv.org/abs/0809.5223} {arXiv:0809.5223 [gr-qc]}
  \BibitemShut {NoStop}%
\bibitem [{\citenamefont {Allen}(2021)}]{Allen:2021yuy}%
  \BibitemOpen
  \bibfield  {author} {\bibinfo {author} {\bibfnamefont {B.}~\bibnamefont
  {Allen}},\ }\bibfield  {title} {\bibinfo {title} {{Optimal template banks}},\
  }\href {https://doi.org/10.1103/PhysRevD.104.042005} {\bibfield  {journal}
  {\bibinfo  {journal} {Phys. Rev. D}\ }\textbf {\bibinfo {volume} {104}},\
  \bibinfo {pages} {042005} (\bibinfo {year} {2021})},\ \Eprint
  {https://arxiv.org/abs/2102.11254} {arXiv:2102.11254 [astro-ph.IM]}
  \BibitemShut {NoStop}%
\bibitem [{\citenamefont {Wagner}\ \emph {et~al.}(2021)\citenamefont {Wagner},
  \citenamefont {Whelan}, \citenamefont {Wofford},\ and\ \citenamefont
  {Wette}}]{Wagner:2021hgv}%
  \BibitemOpen
  \bibfield  {author} {\bibinfo {author} {\bibfnamefont {K.~J.}\ \bibnamefont
  {Wagner}}, \bibinfo {author} {\bibfnamefont {J.~T.}\ \bibnamefont {Whelan}},
  \bibinfo {author} {\bibfnamefont {J.~K.}\ \bibnamefont {Wofford}},\ and\
  \bibinfo {author} {\bibfnamefont {K.}~\bibnamefont {Wette}},\ }\bibfield
  {title} {\bibinfo {title} {{Template Lattices for a Cross-Correlation Search
  for Gravitational Waves from Scorpius X-1}},\ }\href@noop {} {\bibfield
  {journal} {\bibinfo  {journal} {arXiv e-print}\ } (\bibinfo {year} {2021})},\
  \Eprint {https://arxiv.org/abs/2106.16142} {arXiv:2106.16142 [gr-qc]}
  \BibitemShut {NoStop}%
\bibitem [{\citenamefont {Prix}\ and\ \citenamefont
  {Krishnan}(2009)}]{Prix:2009tq}%
  \BibitemOpen
  \bibfield  {author} {\bibinfo {author} {\bibfnamefont {R.}~\bibnamefont
  {Prix}}\ and\ \bibinfo {author} {\bibfnamefont {B.}~\bibnamefont
  {Krishnan}},\ }\bibfield  {title} {\bibinfo {title} {{Targeted search for
  continuous gravitational waves: Bayesian versus maximum-likelihood
  statistics}},\ }\href {https://doi.org/10.1088/0264-9381/26/20/204013}
  {\bibfield  {journal} {\bibinfo  {journal} {Class. Quant. Grav.}\ }\textbf
  {\bibinfo {volume} {26}},\ \bibinfo {pages} {204013} (\bibinfo {year}
  {2009})},\ \Eprint {https://arxiv.org/abs/0907.2569} {arXiv:0907.2569
  [gr-qc]} \BibitemShut {NoStop}%
\bibitem [{\citenamefont {{Covas}}\ \emph {et~al.}(2018)\citenamefont {{Covas}}
  \emph {et~al.}}]{2018PhRvD..97h2002C}%
  \BibitemOpen
  \bibfield  {author} {\bibinfo {author} {\bibfnamefont {P.~B.}\ \bibnamefont
  {{Covas}}} \emph {et~al.},\ }\bibfield  {title} {\bibinfo {title}
  {{Identification and mitigation of narrow spectral artifacts that degrade
  searches for persistent gravitational waves in the first two observing runs
  of Advanced LIGO}},\ }\href {https://doi.org/10.1103/PhysRevD.97.082002}
  {\bibfield  {journal} {\bibinfo  {journal} {Phys. Rev. D}\ }\textbf {\bibinfo
  {volume} {97}},\ \bibinfo {eid} {082002} (\bibinfo {year} {2018})},\ \Eprint
  {https://arxiv.org/abs/1801.07204} {arXiv:1801.07204 [astro-ph.IM]}
  \BibitemShut {NoStop}%
\bibitem [{\citenamefont {Davis}\ \emph {et~al.}(2020)\citenamefont {Davis},
  \citenamefont {White},\ and\ \citenamefont {Saulson}}]{Davis:2020nyf}%
  \BibitemOpen
  \bibfield  {author} {\bibinfo {author} {\bibfnamefont {D.}~\bibnamefont
  {Davis}}, \bibinfo {author} {\bibfnamefont {L.~V.}\ \bibnamefont {White}},\
  and\ \bibinfo {author} {\bibfnamefont {P.~R.}\ \bibnamefont {Saulson}},\
  }\bibfield  {title} {\bibinfo {title} {{Utilizing aLIGO Glitch
  Classifications to Validate Gravitational-Wave Candidates}},\ }\href
  {https://doi.org/10.1088/1361-6382/ab91e6} {\bibfield  {journal} {\bibinfo
  {journal} {Class. Quant. Grav.}\ }\textbf {\bibinfo {volume} {37}},\ \bibinfo
  {pages} {145001} (\bibinfo {year} {2020})},\ \Eprint
  {https://arxiv.org/abs/2002.09429} {arXiv:2002.09429 [gr-qc]} \BibitemShut
  {NoStop}%
\bibitem [{\citenamefont {Abbott}\ \emph
  {et~al.}(2016{\natexlab{a}})\citenamefont {Abbott} \emph
  {et~al.}}]{LIGOScientific:2016aoc}%
  \BibitemOpen
  \bibfield  {author} {\bibinfo {author} {\bibfnamefont {B.~P.}\ \bibnamefont
  {Abbott}} \emph {et~al.} (\bibinfo {collaboration} {LIGO Scientific,
  Virgo}),\ }\bibfield  {title} {\bibinfo {title} {{Observation of
  Gravitational Waves from a Binary Black Hole Merger}},\ }\href
  {https://doi.org/10.1103/PhysRevLett.116.061102} {\bibfield  {journal}
  {\bibinfo  {journal} {Phys. Rev. Lett.}\ }\textbf {\bibinfo {volume} {116}},\
  \bibinfo {pages} {061102} (\bibinfo {year} {2016}{\natexlab{a}})},\ \Eprint
  {https://arxiv.org/abs/1602.03837} {arXiv:1602.03837 [gr-qc]} \BibitemShut
  {NoStop}%
\bibitem [{\citenamefont {Tenorio}\ \emph
  {et~al.}(2021{\natexlab{a}})\citenamefont {Tenorio}, \citenamefont {Keitel},\
  and\ \citenamefont {Sintes}}]{universe}%
  \BibitemOpen
  \bibfield  {author} {\bibinfo {author} {\bibfnamefont {R.}~\bibnamefont
  {Tenorio}}, \bibinfo {author} {\bibfnamefont {D.}~\bibnamefont {Keitel}},\
  and\ \bibinfo {author} {\bibfnamefont {A.~M.}\ \bibnamefont {Sintes}},\
  }\bibfield  {title} {\bibinfo {title} {{Search Methods for Continuous
  Gravitational-Wave Signals from Unknown Sources in the Advanced-Detector
  Era}},\ }\href {https://doi.org/10.3390/universe7120474} {\bibfield
  {journal} {\bibinfo  {journal} {Universe}\ }\textbf {\bibinfo {volume} {7}},\
  \bibinfo {pages} {474} (\bibinfo {year} {2021}{\natexlab{a}})},\ \Eprint
  {https://arxiv.org/abs/2111.12575} {arXiv:2111.12575 [gr-qc]} \BibitemShut
  {NoStop}%
\bibitem [{\citenamefont {Sieniawska}\ and\ \citenamefont
  {Bejger}(2019)}]{Sieniawska_2019}%
  \BibitemOpen
  \bibfield  {author} {\bibinfo {author} {\bibfnamefont {M.}~\bibnamefont
  {Sieniawska}}\ and\ \bibinfo {author} {\bibfnamefont {M.}~\bibnamefont
  {Bejger}},\ }\bibfield  {title} {\bibinfo {title} {{Continuous gravitational
  waves from neutron stars: current status and prospects}},\ }\href
  {https://doi.org/10.3390/universe5110217} {\bibfield  {journal} {\bibinfo
  {journal} {Universe}\ }\textbf {\bibinfo {volume} {5}},\ \bibinfo {pages}
  {217} (\bibinfo {year} {2019})},\ \Eprint {https://arxiv.org/abs/1909.12600}
  {arXiv:1909.12600 [astro-ph.HE]} \BibitemShut {NoStop}%
\bibitem [{\citenamefont {Aasi}\ \emph {et~al.}(2015)\citenamefont {Aasi} \emph
  {et~al.}}]{AdvancedLIGO}%
  \BibitemOpen
  \bibfield  {author} {\bibinfo {author} {\bibfnamefont {J.}~\bibnamefont
  {Aasi}} \emph {et~al.},\ }\bibfield  {title} {\bibinfo {title} {Advanced
  {LIGO}},\ }\href {https://doi.org/10.1088/0264-9381/32/7/074001} {\bibfield
  {journal} {\bibinfo  {journal} {Class. Quant. Grav.}\ }\textbf {\bibinfo
  {volume} {32}},\ \bibinfo {pages} {074001} (\bibinfo {year} {2015})},\
  \Eprint {https://arxiv.org/abs/1411.4547} {arXiv:1411.4547 [gr-qc]}
  \BibitemShut {NoStop}%
\bibitem [{\citenamefont {Acernese}\ \emph {et~al.}(2014)\citenamefont
  {Acernese} \emph {et~al.}}]{AdvancedVirgo}%
  \BibitemOpen
  \bibfield  {author} {\bibinfo {author} {\bibfnamefont {F.}~\bibnamefont
  {Acernese}} \emph {et~al.},\ }\bibfield  {title} {\bibinfo {title} {{Advanced
  Virgo: a second-generation interferometric gravitational wave detector}},\
  }\href {https://doi.org/10.1088/0264-9381/32/2/024001} {\bibfield  {journal}
  {\bibinfo  {journal} {Class. Quant. Grav.}\ }\textbf {\bibinfo {volume}
  {32}},\ \bibinfo {pages} {024001} (\bibinfo {year} {2014})},\ \Eprint
  {https://arxiv.org/abs/1408.3978} {arXiv:1408.3978 [gr-qc]} \BibitemShut
  {NoStop}%
\bibitem [{\citenamefont {Akutsu}\ \emph {et~al.}(2019)\citenamefont {Akutsu}
  \emph {et~al.}}]{KAGRA:2018plz}%
  \BibitemOpen
  \bibfield  {author} {\bibinfo {author} {\bibfnamefont {T.}~\bibnamefont
  {Akutsu}} \emph {et~al.} (\bibinfo {collaboration} {KAGRA}),\ }\bibfield
  {title} {\bibinfo {title} {{KAGRA: 2.5 Generation Interferometric
  Gravitational Wave Detector}},\ }\href
  {https://doi.org/10.1038/s41550-018-0658-y} {\bibfield  {journal} {\bibinfo
  {journal} {Nat. Astron.}\ }\textbf {\bibinfo {volume} {3}},\ \bibinfo {pages}
  {35} (\bibinfo {year} {2019})},\ \Eprint {https://arxiv.org/abs/1811.08079}
  {arXiv:1811.08079 [gr-qc]} \BibitemShut {NoStop}%
\bibitem [{\citenamefont {Jaranowski}\ \emph {et~al.}(1998)\citenamefont
  {Jaranowski}, \citenamefont {Kr\'olak},\ and\ \citenamefont
  {Schutz}}]{JKS1998}%
  \BibitemOpen
  \bibfield  {author} {\bibinfo {author} {\bibfnamefont {P.}~\bibnamefont
  {Jaranowski}}, \bibinfo {author} {\bibfnamefont {A.}~\bibnamefont
  {Kr\'olak}},\ and\ \bibinfo {author} {\bibfnamefont {B.~F.}\ \bibnamefont
  {Schutz}},\ }\bibfield  {title} {\bibinfo {title} {Data analysis of
  gravitational-wave signals from spinning neutron stars: The signal and its
  detection},\ }\href {https://doi.org/10.1103/PhysRevD.58.063001} {\bibfield
  {journal} {\bibinfo  {journal} {Phys. Rev. D}\ }\textbf {\bibinfo {volume}
  {58}},\ \bibinfo {pages} {063001} (\bibinfo {year} {1998})},\ \Eprint
  {https://arxiv.org/abs/gr-qc/9804014} {arXiv:gr-qc/9804014} \BibitemShut
  {NoStop}%
\bibitem [{\citenamefont {Cutler}\ and\ \citenamefont
  {Schutz}(2005)}]{PhysRevD.72.063006}%
  \BibitemOpen
  \bibfield  {author} {\bibinfo {author} {\bibfnamefont {C.}~\bibnamefont
  {Cutler}}\ and\ \bibinfo {author} {\bibfnamefont {B.~F.}\ \bibnamefont
  {Schutz}},\ }\bibfield  {title} {\bibinfo {title} {Generalized
  $\mathcal{F}$-statistic: Multiple detectors and multiple gravitational wave
  pulsars},\ }\href {https://doi.org/10.1103/PhysRevD.72.063006} {\bibfield
  {journal} {\bibinfo  {journal} {Phys. Rev. D}\ }\textbf {\bibinfo {volume}
  {72}},\ \bibinfo {pages} {063006} (\bibinfo {year} {2005})},\ \Eprint
  {https://arxiv.org/abs/gr-qc/0504011} {arXiv:gr-qc/0504011} \BibitemShut
  {NoStop}%
\bibitem [{\citenamefont {Wette}(2009)}]{Wette:2009uea}%
  \BibitemOpen
  \bibfield  {author} {\bibinfo {author} {\bibfnamefont {K.~W.}\ \bibnamefont
  {Wette}},\ }\emph {\bibinfo {title} {{Gravitational waves from accreting
  neutron stars and Cassiopeia A}}},\ \href@noop {} {Ph.D. thesis},\ \bibinfo
  {school} {Australian Natl. U., Canberra} (\bibinfo {year} {2009})\BibitemShut
  {NoStop}%
\bibitem [{\citenamefont {Papa}\ \emph {et~al.}(2020)\citenamefont {Papa},
  \citenamefont {Ming}, \citenamefont {Gotthelf}, \citenamefont {Allen},
  \citenamefont {Prix}, \citenamefont {Dergachev}, \citenamefont {Eggenstein},
  \citenamefont {Singh},\ and\ \citenamefont {Zhu}}]{Papa:2020vfz}%
  \BibitemOpen
  \bibfield  {author} {\bibinfo {author} {\bibfnamefont {M.~A.}\ \bibnamefont
  {Papa}}, \bibinfo {author} {\bibfnamefont {J.}~\bibnamefont {Ming}}, \bibinfo
  {author} {\bibfnamefont {E.~V.}\ \bibnamefont {Gotthelf}}, \bibinfo {author}
  {\bibfnamefont {B.}~\bibnamefont {Allen}}, \bibinfo {author} {\bibfnamefont
  {R.}~\bibnamefont {Prix}}, \bibinfo {author} {\bibfnamefont {V.}~\bibnamefont
  {Dergachev}}, \bibinfo {author} {\bibfnamefont {H.-B.}\ \bibnamefont
  {Eggenstein}}, \bibinfo {author} {\bibfnamefont {A.}~\bibnamefont {Singh}},\
  and\ \bibinfo {author} {\bibfnamefont {S.~J.}\ \bibnamefont {Zhu}},\
  }\bibfield  {title} {\bibinfo {title} {{Search for Continuous Gravitational
  Waves from the Central Compact Objects in Supernova Remnants Cassiopeia A,
  Vela Jr., and G347.3\textendash{}0.5}},\ }\href
  {https://doi.org/10.3847/1538-4357/ab92a6} {\bibfield  {journal} {\bibinfo
  {journal} {Astrophys. J.}\ }\textbf {\bibinfo {volume} {897}},\ \bibinfo
  {pages} {22} (\bibinfo {year} {2020})},\ \Eprint
  {https://arxiv.org/abs/2005.06544} {arXiv:2005.06544 [astro-ph.HE]}
  \BibitemShut {NoStop}%
\bibitem [{\citenamefont {Wette}\ \emph {et~al.}(2021)\citenamefont {Wette},
  \citenamefont {Dunn}, \citenamefont {Clearwater},\ and\ \citenamefont
  {Melatos}}]{Wette:2021tbv}%
  \BibitemOpen
  \bibfield  {author} {\bibinfo {author} {\bibfnamefont {K.}~\bibnamefont
  {Wette}}, \bibinfo {author} {\bibfnamefont {L.}~\bibnamefont {Dunn}},
  \bibinfo {author} {\bibfnamefont {P.}~\bibnamefont {Clearwater}},\ and\
  \bibinfo {author} {\bibfnamefont {A.}~\bibnamefont {Melatos}},\ }\bibfield
  {title} {\bibinfo {title} {{Deep exploration for continuous gravitational
  waves at 171\textendash{}172 Hz in LIGO second observing run data}},\ }\href
  {https://doi.org/10.1103/PhysRevD.103.083020} {\bibfield  {journal} {\bibinfo
   {journal} {Phys. Rev. D}\ }\textbf {\bibinfo {volume} {103}},\ \bibinfo
  {pages} {083020} (\bibinfo {year} {2021})},\ \Eprint
  {https://arxiv.org/abs/2103.12976} {arXiv:2103.12976 [gr-qc]} \BibitemShut
  {NoStop}%
\bibitem [{\citenamefont {Abadie}\ \emph {et~al.}(2010)\citenamefont {Abadie}
  \emph {et~al.}}]{Abadie:2010hv}%
  \BibitemOpen
  \bibfield  {author} {\bibinfo {author} {\bibfnamefont {J.}~\bibnamefont
  {Abadie}} \emph {et~al.} (\bibinfo {collaboration} {LIGO Scientific}),\
  }\bibfield  {title} {\bibinfo {title} {{First search for gravitational waves
  from the youngest known neutron star}},\ }\href
  {https://doi.org/10.1088/0004-637X/722/2/1504} {\bibfield  {journal}
  {\bibinfo  {journal} {Astrophys. J.}\ }\textbf {\bibinfo {volume} {722}},\
  \bibinfo {pages} {1504} (\bibinfo {year} {2010})},\ \Eprint
  {https://arxiv.org/abs/1006.2535} {arXiv:1006.2535 [gr-qc]} \BibitemShut
  {NoStop}%
\bibitem [{\citenamefont {Aasi}\ \emph {et~al.}(2013)\citenamefont {Aasi} \emph
  {et~al.}}]{LIGOScientific:2013wcb}%
  \BibitemOpen
  \bibfield  {author} {\bibinfo {author} {\bibfnamefont {J.}~\bibnamefont
  {Aasi}} \emph {et~al.} (\bibinfo {collaboration} {LIGO Scientific, VIRGO}),\
  }\bibfield  {title} {\bibinfo {title} {{Directed search for continuous
  gravitational waves from the Galactic center}},\ }\href
  {https://doi.org/10.1103/PhysRevD.88.102002} {\bibfield  {journal} {\bibinfo
  {journal} {Phys. Rev. D}\ }\textbf {\bibinfo {volume} {88}},\ \bibinfo
  {pages} {102002} (\bibinfo {year} {2013})},\ \Eprint
  {https://arxiv.org/abs/1309.6221} {arXiv:1309.6221 [gr-qc]} \BibitemShut
  {NoStop}%
\bibitem [{\citenamefont {Behnke}\ \emph {et~al.}(2015)\citenamefont {Behnke},
  \citenamefont {Papa},\ and\ \citenamefont {Prix}}]{PhysRevD.91.064007}%
  \BibitemOpen
  \bibfield  {author} {\bibinfo {author} {\bibfnamefont {B.}~\bibnamefont
  {Behnke}}, \bibinfo {author} {\bibfnamefont {M.~A.}\ \bibnamefont {Papa}},\
  and\ \bibinfo {author} {\bibfnamefont {R.}~\bibnamefont {Prix}},\ }\bibfield
  {title} {\bibinfo {title} {Postprocessing methods used in the search for
  continuous gravitational-wave signals from the galactic center},\ }\href
  {https://doi.org/10.1103/PhysRevD.91.064007} {\bibfield  {journal} {\bibinfo
  {journal} {Phys. Rev. D}\ }\textbf {\bibinfo {volume} {91}},\ \bibinfo
  {pages} {064007} (\bibinfo {year} {2015})},\ \Eprint
  {https://arxiv.org/abs/1410.5997} {arXiv:1410.5997 [gr-qc]} \BibitemShut
  {NoStop}%
\bibitem [{\citenamefont {Abbott}\ \emph
  {et~al.}(2016{\natexlab{b}})\citenamefont {Abbott} \emph
  {et~al.}}]{LIGOScientific:2016ahk}%
  \BibitemOpen
  \bibfield  {author} {\bibinfo {author} {\bibfnamefont {B.~P.}\ \bibnamefont
  {Abbott}} \emph {et~al.} (\bibinfo {collaboration} {LIGO Scientific,
  Virgo}),\ }\bibfield  {title} {\bibinfo {title} {{Results of the deepest
  all-sky survey for continuous gravitational waves on LIGO S6 data running on
  the Einstein@Home volunteer distributed computing project}},\ }\href
  {https://doi.org/10.1103/PhysRevD.94.102002} {\bibfield  {journal} {\bibinfo
  {journal} {Phys. Rev. D}\ }\textbf {\bibinfo {volume} {94}},\ \bibinfo
  {pages} {102002} (\bibinfo {year} {2016}{\natexlab{b}})},\ \Eprint
  {https://arxiv.org/abs/1606.09619} {arXiv:1606.09619 [gr-qc]} \BibitemShut
  {NoStop}%
\bibitem [{\citenamefont {{Zhu}}\ \emph {et~al.}(2016)\citenamefont {{Zhu}},
  \citenamefont {{Papa}}, \citenamefont {{Eggenstein}}, \citenamefont {{Prix}},
  \citenamefont {{Wette}}, \citenamefont {{Allen}}, \citenamefont {{Bock}},
  \citenamefont {{Keitel}}, \citenamefont {{Krishnan}}, \citenamefont
  {{Machenschalk}}, \citenamefont {{Shaltev}},\ and\ \citenamefont
  {{Siemens}}}]{Zhu:2016ghk}%
  \BibitemOpen
  \bibfield  {author} {\bibinfo {author} {\bibfnamefont {S.~J.}\ \bibnamefont
  {{Zhu}}}, \bibinfo {author} {\bibfnamefont {M.~A.}\ \bibnamefont {{Papa}}},
  \bibinfo {author} {\bibfnamefont {H.-B.}\ \bibnamefont {{Eggenstein}}},
  \bibinfo {author} {\bibfnamefont {R.}~\bibnamefont {{Prix}}}, \bibinfo
  {author} {\bibfnamefont {K.}~\bibnamefont {{Wette}}}, \bibinfo {author}
  {\bibfnamefont {B.}~\bibnamefont {{Allen}}}, \bibinfo {author} {\bibfnamefont
  {O.}~\bibnamefont {{Bock}}}, \bibinfo {author} {\bibfnamefont
  {D.}~\bibnamefont {{Keitel}}}, \bibinfo {author} {\bibfnamefont
  {B.}~\bibnamefont {{Krishnan}}}, \bibinfo {author} {\bibfnamefont
  {B.}~\bibnamefont {{Machenschalk}}}, \bibinfo {author} {\bibfnamefont
  {M.}~\bibnamefont {{Shaltev}}},\ and\ \bibinfo {author} {\bibfnamefont
  {X.}~\bibnamefont {{Siemens}}},\ }\bibfield  {title} {\bibinfo {title} {{An
  Einstein@home search for continuous gravitational waves from Cassiopeia A}},\
  }\href {https://doi.org/10.1103/PhysRevD.94.082008} {\bibfield  {journal}
  {\bibinfo  {journal} {Phys. Rev. D}\ }\textbf {\bibinfo {volume} {94}},\
  \bibinfo {pages} {082008} (\bibinfo {year} {2016})},\ \Eprint
  {https://arxiv.org/abs/1608.07589} {arXiv:1608.07589 [gr-qc]} \BibitemShut
  {NoStop}%
\bibitem [{\citenamefont {Abbott}\ \emph
  {et~al.}(2017{\natexlab{a}})\citenamefont {Abbott} \emph
  {et~al.}}]{Abbott:2017pqa}%
  \BibitemOpen
  \bibfield  {author} {\bibinfo {author} {\bibfnamefont {B.~P.}\ \bibnamefont
  {Abbott}} \emph {et~al.} (\bibinfo {collaboration} {LIGO Scientific,
  Virgo}),\ }\bibfield  {title} {\bibinfo {title} {{First low-frequency
  Einstein@Home all-sky search for continuous gravitational waves in Advanced
  LIGO data}},\ }\href {https://doi.org/10.1103/PhysRevD.96.122004} {\bibfield
  {journal} {\bibinfo  {journal} {Phys. Rev. D}\ }\textbf {\bibinfo {volume}
  {96}},\ \bibinfo {pages} {122004} (\bibinfo {year} {2017}{\natexlab{a}})},\
  \Eprint {https://arxiv.org/abs/1707.02669} {arXiv:1707.02669 [gr-qc]}
  \BibitemShut {NoStop}%
\bibitem [{\citenamefont {{Keitel}}\ \emph {et~al.}(2019)\citenamefont
  {{Keitel}}, \citenamefont {{Woan}}, \citenamefont {{Pitkin}}, \citenamefont
  {{Schumacher}}, \citenamefont {{Pearlstone}}, \citenamefont {{Riles}},
  \citenamefont {{Lyne}}, \citenamefont {{Palfreyman}}, \citenamefont
  {{Stappers}},\ and\ \citenamefont {{Weltevrede}}}]{2019PhRvD.100f4058K}%
  \BibitemOpen
  \bibfield  {author} {\bibinfo {author} {\bibfnamefont {D.}~\bibnamefont
  {{Keitel}}}, \bibinfo {author} {\bibfnamefont {G.}~\bibnamefont {{Woan}}},
  \bibinfo {author} {\bibfnamefont {M.}~\bibnamefont {{Pitkin}}}, \bibinfo
  {author} {\bibfnamefont {C.}~\bibnamefont {{Schumacher}}}, \bibinfo {author}
  {\bibfnamefont {B.}~\bibnamefont {{Pearlstone}}}, \bibinfo {author}
  {\bibfnamefont {K.}~\bibnamefont {{Riles}}}, \bibinfo {author} {\bibfnamefont
  {A.~G.}\ \bibnamefont {{Lyne}}}, \bibinfo {author} {\bibfnamefont
  {J.}~\bibnamefont {{Palfreyman}}}, \bibinfo {author} {\bibfnamefont
  {B.}~\bibnamefont {{Stappers}}},\ and\ \bibinfo {author} {\bibfnamefont
  {P.}~\bibnamefont {{Weltevrede}}},\ }\bibfield  {title} {\bibinfo {title}
  {{First search for long-duration transient gravitational waves after glitches
  in the Vela and Crab pulsars}},\ }\href
  {https://doi.org/10.1103/PhysRevD.100.064058} {\bibfield  {journal} {\bibinfo
   {journal} {Phys. Rev. D}\ }\textbf {\bibinfo {volume} {100}},\ \bibinfo
  {eid} {064058} (\bibinfo {year} {2019})},\ \Eprint
  {https://arxiv.org/abs/1907.04717} {arXiv:1907.04717 [gr-qc]} \BibitemShut
  {NoStop}%
\bibitem [{\citenamefont {Dreissigacker}\ \emph {et~al.}(2018)\citenamefont
  {Dreissigacker}, \citenamefont {Prix},\ and\ \citenamefont
  {Wette}}]{Dreissigacker:2018afk}%
  \BibitemOpen
  \bibfield  {author} {\bibinfo {author} {\bibfnamefont {C.}~\bibnamefont
  {Dreissigacker}}, \bibinfo {author} {\bibfnamefont {R.}~\bibnamefont
  {Prix}},\ and\ \bibinfo {author} {\bibfnamefont {K.}~\bibnamefont {Wette}},\
  }\bibfield  {title} {\bibinfo {title} {{Fast and Accurate Sensitivity
  Estimation for Continuous-Gravitational-Wave Searches}},\ }\href
  {https://doi.org/10.1103/PhysRevD.98.084058} {\bibfield  {journal} {\bibinfo
  {journal} {Phys. Rev. D}\ }\textbf {\bibinfo {volume} {98}},\ \bibinfo
  {pages} {084058} (\bibinfo {year} {2018})},\ \Eprint
  {https://arxiv.org/abs/1808.02459} {arXiv:1808.02459 [gr-qc]} \BibitemShut
  {NoStop}%
\bibitem [{\citenamefont {Tenorio}\ \emph
  {et~al.}(2021{\natexlab{b}})\citenamefont {Tenorio}, \citenamefont {Keitel},\
  and\ \citenamefont {Sintes}}]{Tenorio:2021njf}%
  \BibitemOpen
  \bibfield  {author} {\bibinfo {author} {\bibfnamefont {R.}~\bibnamefont
  {Tenorio}}, \bibinfo {author} {\bibfnamefont {D.}~\bibnamefont {Keitel}},\
  and\ \bibinfo {author} {\bibfnamefont {A.~M.}\ \bibnamefont {Sintes}},\
  }\bibfield  {title} {\bibinfo {title} {{Application of a hierarchical MCMC
  follow-up to Advanced LIGO continuous gravitational-wave candidates}},\
  }\href {https://doi.org/10.1103/PhysRevD.104.084012} {\bibfield  {journal}
  {\bibinfo  {journal} {Phys. Rev. D}\ }\textbf {\bibinfo {volume} {104}},\
  \bibinfo {pages} {084012} (\bibinfo {year} {2021}{\natexlab{b}})},\ \Eprint
  {https://arxiv.org/abs/2105.13860} {arXiv:2105.13860 [gr-qc]} \BibitemShut
  {NoStop}%
\bibitem [{\citenamefont {Keitel}\ \emph {et~al.}(2014)\citenamefont {Keitel},
  \citenamefont {Prix}, \citenamefont {Papa}, \citenamefont {Leaci},\ and\
  \citenamefont {Siddiqi}}]{Keitel:2013wga}%
  \BibitemOpen
  \bibfield  {author} {\bibinfo {author} {\bibfnamefont {D.}~\bibnamefont
  {Keitel}}, \bibinfo {author} {\bibfnamefont {R.}~\bibnamefont {Prix}},
  \bibinfo {author} {\bibfnamefont {M.~A.}\ \bibnamefont {Papa}}, \bibinfo
  {author} {\bibfnamefont {P.}~\bibnamefont {Leaci}},\ and\ \bibinfo {author}
  {\bibfnamefont {M.}~\bibnamefont {Siddiqi}},\ }\bibfield  {title} {\bibinfo
  {title} {{Search for continuous gravitational waves: Improving robustness
  versus instrumental artifacts}},\ }\href
  {https://doi.org/10.1103/PhysRevD.89.064023} {\bibfield  {journal} {\bibinfo
  {journal} {Phys. Rev. D}\ }\textbf {\bibinfo {volume} {89}},\ \bibinfo
  {pages} {064023} (\bibinfo {year} {2014})},\ \Eprint
  {https://arxiv.org/abs/1311.5738} {arXiv:1311.5738 [gr-qc]} \BibitemShut
  {NoStop}%
\bibitem [{\citenamefont {Keitel}(2016)}]{Keitel:2015ova}%
  \BibitemOpen
  \bibfield  {author} {\bibinfo {author} {\bibfnamefont {D.}~\bibnamefont
  {Keitel}},\ }\bibfield  {title} {\bibinfo {title} {{Robust semicoherent
  searches for continuous gravitational waves with noise and signal models
  including hours to days long transients}},\ }\href
  {https://doi.org/10.1103/PhysRevD.93.084024} {\bibfield  {journal} {\bibinfo
  {journal} {Phys. Rev. D}\ }\textbf {\bibinfo {volume} {93}},\ \bibinfo
  {pages} {084024} (\bibinfo {year} {2016})},\ \Eprint
  {https://arxiv.org/abs/1509.02398} {arXiv:1509.02398 [gr-qc]} \BibitemShut
  {NoStop}%
\bibitem [{\citenamefont {Prix}\ \emph {et~al.}(2011)\citenamefont {Prix},
  \citenamefont {Giampanis},\ and\ \citenamefont {Messenger}}]{Prix:2011qv}%
  \BibitemOpen
  \bibfield  {author} {\bibinfo {author} {\bibfnamefont {R.}~\bibnamefont
  {Prix}}, \bibinfo {author} {\bibfnamefont {S.}~\bibnamefont {Giampanis}},\
  and\ \bibinfo {author} {\bibfnamefont {C.}~\bibnamefont {Messenger}},\
  }\bibfield  {title} {\bibinfo {title} {{Search method for long-duration
  gravitational-wave transients from neutron stars}},\ }\href
  {https://doi.org/10.1103/PhysRevD.84.023007} {\bibfield  {journal} {\bibinfo
  {journal} {Phys. Rev. D}\ }\textbf {\bibinfo {volume} {84}},\ \bibinfo
  {pages} {023007} (\bibinfo {year} {2011})},\ \Eprint
  {https://arxiv.org/abs/1104.1704} {arXiv:1104.1704 [gr-qc]} \BibitemShut
  {NoStop}%
\bibitem [{\citenamefont {Tenorio}\ \emph
  {et~al.}(2021{\natexlab{c}})\citenamefont {Tenorio}, \citenamefont
  {Modafferi}, \citenamefont {Keitel},\ and\ \citenamefont
  {Sintes}}]{distromax}%
  \BibitemOpen
  \bibfield  {author} {\bibinfo {author} {\bibfnamefont {R.}~\bibnamefont
  {Tenorio}}, \bibinfo {author} {\bibfnamefont {L.~M.}\ \bibnamefont
  {Modafferi}}, \bibinfo {author} {\bibfnamefont {D.}~\bibnamefont {Keitel}},\
  and\ \bibinfo {author} {\bibfnamefont {A.~M.}\ \bibnamefont {Sintes}},\
  }\href@noop {} {\bibinfo {title} {\texttt{distromax}: empirically estimating
  the distribution of the loudest candidate from a gravitational-wave
  search}},\ \bibinfo {howpublished}
  {\url{https://github.com/Rodrigo-Tenorio/distromax}} (\bibinfo {year}
  {2021}{\natexlab{c}})\BibitemShut {NoStop}%
\bibitem [{\citenamefont {Sancho de~la Jordana}\ and\ \citenamefont
  {Sintes}(2008)}]{SanchodelaJordana:2008dc}%
  \BibitemOpen
  \bibfield  {author} {\bibinfo {author} {\bibfnamefont {L.}~\bibnamefont
  {Sancho de~la Jordana}}\ and\ \bibinfo {author} {\bibfnamefont {A.~M.}\
  \bibnamefont {Sintes}},\ }\bibfield  {title} {\bibinfo {title} {{A $\chi^2$
  veto for continuous wave searches}},\ }\href
  {https://doi.org/10.1088/0264-9381/25/18/184014} {\bibfield  {journal}
  {\bibinfo  {journal} {Class. Quant. Grav.}\ }\textbf {\bibinfo {volume}
  {25}},\ \bibinfo {pages} {184014} (\bibinfo {year} {2008})},\ \Eprint
  {https://arxiv.org/abs/0804.1007} {arXiv:0804.1007 [gr-qc]} \BibitemShut
  {NoStop}%
\bibitem [{\citenamefont {Leaci}(2015)}]{Leaci:2015iuc}%
  \BibitemOpen
  \bibfield  {author} {\bibinfo {author} {\bibfnamefont {P.}~\bibnamefont
  {Leaci}},\ }\bibfield  {title} {\bibinfo {title} {{Methods to filter out
  spurious disturbances in continuous-wave searches from gravitational-wave
  detectors}},\ }\href {https://doi.org/10.1088/0031-8949/90/12/125001}
  {\bibfield  {journal} {\bibinfo  {journal} {Phys. Scr.}\ }\textbf {\bibinfo
  {volume} {90}},\ \bibinfo {pages} {125001} (\bibinfo {year}
  {2015})}\BibitemShut {NoStop}%
\bibitem [{\citenamefont {Zhu}\ \emph {et~al.}(2017)\citenamefont {Zhu},
  \citenamefont {Papa},\ and\ \citenamefont {Walsh}}]{Zhu:2017ujz}%
  \BibitemOpen
  \bibfield  {author} {\bibinfo {author} {\bibfnamefont {S.~J.}\ \bibnamefont
  {Zhu}}, \bibinfo {author} {\bibfnamefont {M.~A.}\ \bibnamefont {Papa}},\ and\
  \bibinfo {author} {\bibfnamefont {S.}~\bibnamefont {Walsh}},\ }\bibfield
  {title} {\bibinfo {title} {{New veto for continuous gravitational wave
  searches}},\ }\href {https://doi.org/10.1103/PhysRevD.96.124007} {\bibfield
  {journal} {\bibinfo  {journal} {Phys. Rev. D}\ }\textbf {\bibinfo {volume}
  {96}},\ \bibinfo {pages} {124007} (\bibinfo {year} {2017})},\ \Eprint
  {https://arxiv.org/abs/1707.05268} {arXiv:1707.05268 [gr-qc]} \BibitemShut
  {NoStop}%
\bibitem [{\citenamefont {Whelan}\ \emph {et~al.}(2014)\citenamefont {Whelan},
  \citenamefont {Prix}, \citenamefont {Cutler},\ and\ \citenamefont
  {Willis}}]{Whelan:2013xka}%
  \BibitemOpen
  \bibfield  {author} {\bibinfo {author} {\bibfnamefont {J.~T.}\ \bibnamefont
  {Whelan}}, \bibinfo {author} {\bibfnamefont {R.}~\bibnamefont {Prix}},
  \bibinfo {author} {\bibfnamefont {C.~J.}\ \bibnamefont {Cutler}},\ and\
  \bibinfo {author} {\bibfnamefont {J.~L.}\ \bibnamefont {Willis}},\ }\bibfield
   {title} {\bibinfo {title} {{New Coordinates for the Amplitude Parameter
  Space of Continuous Gravitational Waves}},\ }\href
  {https://doi.org/10.1088/0264-9381/31/6/065002} {\bibfield  {journal}
  {\bibinfo  {journal} {Class. Quant. Grav.}\ }\textbf {\bibinfo {volume}
  {31}},\ \bibinfo {pages} {065002} (\bibinfo {year} {2014})},\ \Eprint
  {https://arxiv.org/abs/1311.0065} {arXiv:1311.0065 [gr-qc]} \BibitemShut
  {NoStop}%
\bibitem [{\citenamefont {Dhurandhar}\ \emph {et~al.}(2017)\citenamefont
  {Dhurandhar}, \citenamefont {Krishnan},\ and\ \citenamefont
  {Willis}}]{Dhurandhar:2017rlr}%
  \BibitemOpen
  \bibfield  {author} {\bibinfo {author} {\bibfnamefont {S.}~\bibnamefont
  {Dhurandhar}}, \bibinfo {author} {\bibfnamefont {B.}~\bibnamefont
  {Krishnan}},\ and\ \bibinfo {author} {\bibfnamefont {J.~L.}\ \bibnamefont
  {Willis}},\ }\bibfield  {title} {\bibinfo {title} {{Marginalizing the
  likelihood function for modeled gravitational wave searches}},\ }\href@noop
  {} {\bibfield  {journal} {\bibinfo  {journal} {arXiv e-print}\ } (\bibinfo
  {year} {2017})},\ \Eprint {https://arxiv.org/abs/1707.08163}
  {arXiv:1707.08163 [gr-qc]} \BibitemShut {NoStop}%
\bibitem [{\citenamefont {Bero}\ and\ \citenamefont
  {Whelan}(2019)}]{Bero:2018xyq}%
  \BibitemOpen
  \bibfield  {author} {\bibinfo {author} {\bibfnamefont {J.~J.}\ \bibnamefont
  {Bero}}\ and\ \bibinfo {author} {\bibfnamefont {J.~T.}\ \bibnamefont
  {Whelan}},\ }\bibfield  {title} {\bibinfo {title} {{An Analytic Approximation
  to the Bayesian Detection Statistic for Continuous Gravitational Waves}},\
  }\href {https://doi.org/10.1088/1361-6382/aaed6a} {\bibfield  {journal}
  {\bibinfo  {journal} {Class. Quant. Grav.}\ }\textbf {\bibinfo {volume}
  {36}},\ \bibinfo {pages} {015013} (\bibinfo {year} {2019})},\ \bibinfo {note}
  {[Erratum: Class.Quant.Grav. 36, 049601 (2019)]},\ \Eprint
  {https://arxiv.org/abs/1808.05453} {arXiv:1808.05453 [gr-qc]} \BibitemShut
  {NoStop}%
\bibitem [{\citenamefont {{Prix, Reinhard}}(2018)}]{CFSv2}%
  \BibitemOpen
  \bibfield  {author} {\bibinfo {author} {\bibnamefont {{Prix, Reinhard}}},\
  }\href@noop {} {\bibinfo {title} {{The F-statistic and its implementation in
  ComputeFstatistic\_v2}}},\ \bibinfo {howpublished}
  {\url{https://dcc.ligo.org/LIGO-T0900149/public}} (\bibinfo {year}
  {2018})\BibitemShut {NoStop}%
\bibitem [{\citenamefont {Finn}(1992)}]{Finn:1992wt}%
  \BibitemOpen
  \bibfield  {author} {\bibinfo {author} {\bibfnamefont {L.~S.}\ \bibnamefont
  {Finn}},\ }\bibfield  {title} {\bibinfo {title} {{Detection, measurement and
  gravitational radiation}},\ }\href {https://doi.org/10.1103/PhysRevD.46.5236}
  {\bibfield  {journal} {\bibinfo  {journal} {Phys. Rev. D}\ }\textbf {\bibinfo
  {volume} {46}},\ \bibinfo {pages} {5236} (\bibinfo {year} {1992})},\ \Eprint
  {https://arxiv.org/abs/gr-qc/9209010} {arXiv:gr-qc/9209010} \BibitemShut
  {NoStop}%
\bibitem [{\citenamefont {Tenorio}\ \emph
  {et~al.}(2021{\natexlab{d}})\citenamefont {Tenorio}, \citenamefont {Keitel},\
  and\ \citenamefont {Sintes}}]{Tenorio:2020cqm}%
  \BibitemOpen
  \bibfield  {author} {\bibinfo {author} {\bibfnamefont {R.}~\bibnamefont
  {Tenorio}}, \bibinfo {author} {\bibfnamefont {D.}~\bibnamefont {Keitel}},\
  and\ \bibinfo {author} {\bibfnamefont {A.~M.}\ \bibnamefont {Sintes}},\
  }\bibfield  {title} {\bibinfo {title} {{Time-frequency track distance for
  comparing continuous gravitational wave signals}},\ }\href
  {https://doi.org/10.1103/PhysRevD.103.064053} {\bibfield  {journal} {\bibinfo
   {journal} {Phys. Rev. D}\ }\textbf {\bibinfo {volume} {103}},\ \bibinfo
  {pages} {064053} (\bibinfo {year} {2021}{\natexlab{d}})},\ \Eprint
  {https://arxiv.org/abs/2012.05752} {arXiv:2012.05752 [gr-qc]} \BibitemShut
  {NoStop}%
\bibitem [{\citenamefont {Prix}(2007{\natexlab{b}})}]{Prix:2006wm}%
  \BibitemOpen
  \bibfield  {author} {\bibinfo {author} {\bibfnamefont {R.}~\bibnamefont
  {Prix}},\ }\bibfield  {title} {\bibinfo {title} {{Search for continuous
  gravitational waves: Metric of the multi-detector F-statistic}},\ }\href
  {https://doi.org/10.1103/PhysRevD.75.023004} {\bibfield  {journal} {\bibinfo
  {journal} {Phys. Rev. D}\ }\textbf {\bibinfo {volume} {75}},\ \bibinfo
  {pages} {023004} (\bibinfo {year} {2007}{\natexlab{b}})},\ \bibinfo {note}
  {[Erratum: Phys. Rev. D 75, 069901(E) (2007)]},\ \Eprint
  {https://arxiv.org/abs/gr-qc/0606088} {arXiv:gr-qc/0606088} \BibitemShut
  {NoStop}%
\bibitem [{\citenamefont {Pletsch}(2010)}]{Pletsch:2010xb}%
  \BibitemOpen
  \bibfield  {author} {\bibinfo {author} {\bibfnamefont {H.~J.}\ \bibnamefont
  {Pletsch}},\ }\bibfield  {title} {\bibinfo {title} {{Parameter-space metric
  of semicoherent searches for continuous gravitational waves}},\ }\href
  {https://doi.org/10.1103/PhysRevD.82.042002} {\bibfield  {journal} {\bibinfo
  {journal} {Phys. Rev. D}\ }\textbf {\bibinfo {volume} {82}},\ \bibinfo
  {pages} {042002} (\bibinfo {year} {2010})},\ \Eprint
  {https://arxiv.org/abs/1005.0395} {arXiv:1005.0395 [gr-qc]} \BibitemShut
  {NoStop}%
\bibitem [{\citenamefont {Wette}\ and\ \citenamefont
  {Prix}(2013)}]{Wette:2013wza}%
  \BibitemOpen
  \bibfield  {author} {\bibinfo {author} {\bibfnamefont {K.}~\bibnamefont
  {Wette}}\ and\ \bibinfo {author} {\bibfnamefont {R.}~\bibnamefont {Prix}},\
  }\bibfield  {title} {\bibinfo {title} {{Flat parameter-space metric for
  all-sky searches for gravitational-wave pulsars}},\ }\href
  {https://doi.org/10.1103/PhysRevD.88.123005} {\bibfield  {journal} {\bibinfo
  {journal} {Phys. Rev. D}\ }\textbf {\bibinfo {volume} {88}},\ \bibinfo
  {pages} {123005} (\bibinfo {year} {2013})},\ \Eprint
  {https://arxiv.org/abs/1310.5587} {arXiv:1310.5587 [gr-qc]} \BibitemShut
  {NoStop}%
\bibitem [{\citenamefont {Wette}(2015)}]{Wette:2015lfa}%
  \BibitemOpen
  \bibfield  {author} {\bibinfo {author} {\bibfnamefont {K.}~\bibnamefont
  {Wette}},\ }\bibfield  {title} {\bibinfo {title} {{Parameter-space metric for
  all-sky semicoherent searches for gravitational-wave pulsars}},\ }\href
  {https://doi.org/10.1103/PhysRevD.92.082003} {\bibfield  {journal} {\bibinfo
  {journal} {Phys. Rev. D}\ }\textbf {\bibinfo {volume} {92}},\ \bibinfo
  {pages} {082003} (\bibinfo {year} {2015})},\ \Eprint
  {https://arxiv.org/abs/1508.02372} {arXiv:1508.02372 [gr-qc]} \BibitemShut
  {NoStop}%
\bibitem [{\citenamefont {Leaci}\ and\ \citenamefont
  {Prix}(2015)}]{Leaci:2015bka}%
  \BibitemOpen
  \bibfield  {author} {\bibinfo {author} {\bibfnamefont {P.}~\bibnamefont
  {Leaci}}\ and\ \bibinfo {author} {\bibfnamefont {R.}~\bibnamefont {Prix}},\
  }\bibfield  {title} {\bibinfo {title} {{Directed searches for continuous
  gravitational waves from binary systems: parameter-space metrics and optimal
  Scorpius X-1 sensitivity}},\ }\href
  {https://doi.org/10.1103/PhysRevD.91.102003} {\bibfield  {journal} {\bibinfo
  {journal} {Phys. Rev. D}\ }\textbf {\bibinfo {volume} {91}},\ \bibinfo
  {pages} {102003} (\bibinfo {year} {2015})},\ \Eprint
  {https://arxiv.org/abs/1502.00914} {arXiv:1502.00914 [gr-qc]} \BibitemShut
  {NoStop}%
\bibitem [{\citenamefont {Allen}(2019)}]{Allen:2019vcl}%
  \BibitemOpen
  \bibfield  {author} {\bibinfo {author} {\bibfnamefont {B.}~\bibnamefont
  {Allen}},\ }\bibfield  {title} {\bibinfo {title} {{Spherical ansatz for
  parameter-space metrics}},\ }\href
  {https://doi.org/10.1103/PhysRevD.100.124004} {\bibfield  {journal} {\bibinfo
   {journal} {Phys. Rev. D}\ }\textbf {\bibinfo {volume} {100}},\ \bibinfo
  {pages} {124004} (\bibinfo {year} {2019})},\ \Eprint
  {https://arxiv.org/abs/1906.01352} {arXiv:1906.01352 [gr-qc]} \BibitemShut
  {NoStop}%
\bibitem [{\citenamefont {Wette}(2014)}]{Wette:2014tca}%
  \BibitemOpen
  \bibfield  {author} {\bibinfo {author} {\bibfnamefont {K.}~\bibnamefont
  {Wette}},\ }\bibfield  {title} {\bibinfo {title} {{Lattice template placement
  for coherent all-sky searches for gravitational-wave pulsars}},\ }\href
  {https://doi.org/10.1103/PhysRevD.90.122010} {\bibfield  {journal} {\bibinfo
  {journal} {Phys. Rev. D}\ }\textbf {\bibinfo {volume} {90}},\ \bibinfo
  {pages} {122010} (\bibinfo {year} {2014})},\ \Eprint
  {https://arxiv.org/abs/1410.6882} {arXiv:1410.6882 [gr-qc]} \BibitemShut
  {NoStop}%
\bibitem [{\citenamefont {Allen}\ and\ \citenamefont
  {Shoom}(2021)}]{Allen:2021eju}%
  \BibitemOpen
  \bibfield  {author} {\bibinfo {author} {\bibfnamefont {B.}~\bibnamefont
  {Allen}}\ and\ \bibinfo {author} {\bibfnamefont {A.~A.}\ \bibnamefont
  {Shoom}},\ }\bibfield  {title} {\bibinfo {title} {{Template banks based on Zn
  and An* lattices}},\ }\href {https://doi.org/10.1103/PhysRevD.104.122007}
  {\bibfield  {journal} {\bibinfo  {journal} {Phys. Rev. D}\ }\textbf {\bibinfo
  {volume} {104}},\ \bibinfo {pages} {122007} (\bibinfo {year} {2021})},\
  \Eprint {https://arxiv.org/abs/2102.11631} {arXiv:2102.11631 [astro-ph.IM]}
  \BibitemShut {NoStop}%
\bibitem [{\citenamefont {Wette}(2016)}]{Wette:2016raf}%
  \BibitemOpen
  \bibfield  {author} {\bibinfo {author} {\bibfnamefont {K.}~\bibnamefont
  {Wette}},\ }\bibfield  {title} {\bibinfo {title} {{Empirically extending the
  range of validity of parameter-space metrics for all-sky searches for
  gravitational-wave pulsars}},\ }\href
  {https://doi.org/10.1103/PhysRevD.94.122002} {\bibfield  {journal} {\bibinfo
  {journal} {Phys. Rev. D}\ }\textbf {\bibinfo {volume} {94}},\ \bibinfo
  {pages} {122002} (\bibinfo {year} {2016})},\ \Eprint
  {https://arxiv.org/abs/1607.00241} {arXiv:1607.00241 [gr-qc]} \BibitemShut
  {NoStop}%
\bibitem [{\citenamefont {{Piccinni}}(2014)}]{piccinni2014:_thesis}%
  \BibitemOpen
  \bibfield  {author} {\bibinfo {author} {\bibfnamefont {O.}~\bibnamefont
  {{Piccinni}}},\ }\emph {\bibinfo {title} {Mitigation of transient
  disturbances in wide parameter space searches for continuous gravitational
  wave signals}},\ \href@noop {} {Master's thesis},\ \bibinfo  {school}
  {Universit\`a di Roma La Sapienza} (\bibinfo {year} {2014})\BibitemShut
  {NoStop}%
\bibitem [{\citenamefont {Wette}(2012)}]{Wette:2011eu}%
  \BibitemOpen
  \bibfield  {author} {\bibinfo {author} {\bibfnamefont {K.}~\bibnamefont
  {Wette}},\ }\bibfield  {title} {\bibinfo {title} {{Estimating the sensitivity
  of wide-parameter-space searches for gravitational-wave pulsars}},\ }\href
  {https://doi.org/10.1103/PhysRevD.85.042003} {\bibfield  {journal} {\bibinfo
  {journal} {Phys. Rev. D}\ }\textbf {\bibinfo {volume} {85}},\ \bibinfo
  {pages} {042003} (\bibinfo {year} {2012})},\ \Eprint
  {https://arxiv.org/abs/1111.5650} {arXiv:1111.5650 [gr-qc]} \BibitemShut
  {NoStop}%
\bibitem [{\citenamefont {{LIGO Scientific Collaboration}}(2018)}]{lalsuite}%
  \BibitemOpen
  \bibfield  {author} {\bibinfo {author} {\bibnamefont {{LIGO Scientific
  Collaboration}}},\ }\href {https://doi.org/10.7935/GT1W-FZ16} {\bibinfo
  {title} {{LIGO} {A}lgorithm {L}ibrary - {LALS}uite}},\ \bibinfo
  {howpublished} {free software (GPL)} (\bibinfo {year} {2018})\BibitemShut
  {NoStop}%
\bibitem [{\citenamefont {Rover}\ \emph {et~al.}(2011)\citenamefont {Rover},
  \citenamefont {Messenger},\ and\ \citenamefont {Prix}}]{Rover:2011zq}%
  \BibitemOpen
  \bibfield  {author} {\bibinfo {author} {\bibfnamefont {C.}~\bibnamefont
  {Rover}}, \bibinfo {author} {\bibfnamefont {C.}~\bibnamefont {Messenger}},\
  and\ \bibinfo {author} {\bibfnamefont {R.}~\bibnamefont {Prix}},\ }\bibfield
  {title} {\bibinfo {title} {{Bayesian versus frequentist upper limits}},\ }in\
  \href {https://doi.org/https://doi.org/10.5170/CERN-2011-006.158} {\emph
  {\bibinfo {booktitle} {{PHYSTAT 2011}}}}\ (\bibinfo  {publisher} {CERN},\
  \bibinfo {address} {Geneva},\ \bibinfo {year} {2011})\ pp.\ \bibinfo {pages}
  {158--163},\ \Eprint {https://arxiv.org/abs/1103.2987} {arXiv:1103.2987
  [physics.data-an]} \BibitemShut {NoStop}%
\bibitem [{\citenamefont {{Suvorova}}\ \emph {et~al.}(2017)\citenamefont
  {{Suvorova}}, \citenamefont {{Clearwater}}, \citenamefont {{Melatos}},
  \citenamefont {{Sun}}, \citenamefont {{Moran}},\ and\ \citenamefont
  {{Evans}}}]{2017PhRvD..96j2006S}%
  \BibitemOpen
  \bibfield  {author} {\bibinfo {author} {\bibfnamefont {S.}~\bibnamefont
  {{Suvorova}}}, \bibinfo {author} {\bibfnamefont {P.}~\bibnamefont
  {{Clearwater}}}, \bibinfo {author} {\bibfnamefont {A.}~\bibnamefont
  {{Melatos}}}, \bibinfo {author} {\bibfnamefont {L.}~\bibnamefont {{Sun}}},
  \bibinfo {author} {\bibfnamefont {W.}~\bibnamefont {{Moran}}},\ and\ \bibinfo
  {author} {\bibfnamefont {R.~J.}\ \bibnamefont {{Evans}}},\ }\bibfield
  {title} {\bibinfo {title} {{Hidden Markov model tracking of continuous
  gravitational waves from a binary neutron star with wandering spin. II.
  Binary orbital phase tracking}},\ }\href
  {https://doi.org/10.1103/PhysRevD.96.102006} {\bibfield  {journal} {\bibinfo
  {journal} {\prd}\ }\textbf {\bibinfo {volume} {96}},\ \bibinfo {eid} {102006}
  (\bibinfo {year} {2017})},\ \Eprint {https://arxiv.org/abs/1710.07092}
  {arXiv:1710.07092 [astro-ph.IM]} \BibitemShut {NoStop}%
\bibitem [{\citenamefont {Leadbetter}\ \emph {et~al.}(1983)\citenamefont
  {Leadbetter}, \citenamefont {Lindgren},\ and\ \citenamefont
  {Rootzen}}]{leadbetter1983extremes}%
  \BibitemOpen
  \bibfield  {author} {\bibinfo {author} {\bibfnamefont {M.}~\bibnamefont
  {Leadbetter}}, \bibinfo {author} {\bibfnamefont {G.}~\bibnamefont
  {Lindgren}},\ and\ \bibinfo {author} {\bibfnamefont {H.}~\bibnamefont
  {Rootzen}},\ }\href
  {https://doi.org/https://doi.org/10.1007/978-1-4612-5449-2} {\emph {\bibinfo
  {title} {Extremes and Related Properties of Random Sequences and
  Processes}}},\ Springer Series in Statistics\ (\bibinfo  {publisher}
  {Springer New York},\ \bibinfo {year} {1983})\BibitemShut {NoStop}%
\bibitem [{\citenamefont {Coles}(2001)}]{coles2001introduction}%
  \BibitemOpen
  \bibfield  {author} {\bibinfo {author} {\bibfnamefont {S.}~\bibnamefont
  {Coles}},\ }\href {https://doi.org/https://doi.org/10.1007/978-1-4471-3675-0}
  {\emph {\bibinfo {title} {An introduction to statistical modeling of extreme
  values}}},\ Springer Series in Statistics\ (\bibinfo  {publisher}
  {Springer-Verlag},\ \bibinfo {address} {London},\ \bibinfo {year}
  {2001})\BibitemShut {NoStop}%
\bibitem [{\citenamefont {Beirlant}\ \emph {et~al.}(2004)\citenamefont
  {Beirlant}, \citenamefont {Goegebeur}, \citenamefont {Segers}, \citenamefont
  {Teugels}, \citenamefont {De~Waal},\ and\ \citenamefont
  {Ferro}}]{beirlant2004statistics}%
  \BibitemOpen
  \bibfield  {author} {\bibinfo {author} {\bibfnamefont {J.}~\bibnamefont
  {Beirlant}}, \bibinfo {author} {\bibfnamefont {Y.}~\bibnamefont {Goegebeur}},
  \bibinfo {author} {\bibfnamefont {J.}~\bibnamefont {Segers}}, \bibinfo
  {author} {\bibfnamefont {J.}~\bibnamefont {Teugels}}, \bibinfo {author}
  {\bibfnamefont {D.}~\bibnamefont {De~Waal}},\ and\ \bibinfo {author}
  {\bibfnamefont {C.}~\bibnamefont {Ferro}},\ }\href
  {https://doi.org/https://doi.org/10.1002/0470012382} {\emph {\bibinfo {title}
  {Statistics of Extremes: Theory and Applications}}},\ Wiley Series in
  Probability and Statistics\ (\bibinfo  {publisher} {Wiley},\ \bibinfo {year}
  {2004})\BibitemShut {NoStop}%
\bibitem [{\citenamefont {de~Haan}\ and\ \citenamefont
  {Ferreira}(2006)}]{de2006extreme}%
  \BibitemOpen
  \bibfield  {author} {\bibinfo {author} {\bibfnamefont {L.}~\bibnamefont
  {de~Haan}}\ and\ \bibinfo {author} {\bibfnamefont {A.}~\bibnamefont
  {Ferreira}},\ }\href {https://doi.org/https://doi.org/10.1007/0-387-34471-3}
  {\emph {\bibinfo {title} {Extreme Value Theory: An Introduction}}},\ Springer
  Series in Operations Research and Financial Engineering\ (\bibinfo
  {publisher} {Springer New York},\ \bibinfo {year} {2006})\BibitemShut
  {NoStop}%
\bibitem [{\citenamefont {Embrechts}\ \emph {et~al.}(2013)\citenamefont
  {Embrechts}, \citenamefont {Kl{\"u}ppelberg},\ and\ \citenamefont
  {Mikosch}}]{embrechts2013modelling}%
  \BibitemOpen
  \bibfield  {author} {\bibinfo {author} {\bibfnamefont {P.}~\bibnamefont
  {Embrechts}}, \bibinfo {author} {\bibfnamefont {C.}~\bibnamefont
  {Kl{\"u}ppelberg}},\ and\ \bibinfo {author} {\bibfnamefont {T.}~\bibnamefont
  {Mikosch}},\ }\href
  {https://doi.org/https://doi.org/10.1007/978-3-642-33483-2} {\emph {\bibinfo
  {title} {Modelling Extremal Events: for Insurance and Finance}}},\ Stochastic
  Modelling and Applied Probability\ (\bibinfo  {publisher} {Springer Berlin
  Heidelberg},\ \bibinfo {year} {2013})\BibitemShut {NoStop}%
\bibitem [{\citenamefont {Gasull}\ \emph
  {et~al.}(2015{\natexlab{a}})\citenamefont {Gasull}, \citenamefont {Jolis},\
  and\ \citenamefont {Utzet}}]{GASULL2015376}%
  \BibitemOpen
  \bibfield  {author} {\bibinfo {author} {\bibfnamefont {A.}~\bibnamefont
  {Gasull}}, \bibinfo {author} {\bibfnamefont {M.}~\bibnamefont {Jolis}},\ and\
  \bibinfo {author} {\bibfnamefont {F.}~\bibnamefont {Utzet}},\ }\bibfield
  {title} {\bibinfo {title} {On the norming constants for normal maxima},\
  }\href {https://doi.org/https://doi.org/10.1016/j.jmaa.2014.08.025}
  {\bibfield  {journal} {\bibinfo  {journal} {J. Math. Anal. Appl.}\ }\textbf
  {\bibinfo {volume} {422}},\ \bibinfo {pages} {376} (\bibinfo {year}
  {2015}{\natexlab{a}})}\BibitemShut {NoStop}%
\bibitem [{\citenamefont {Gasull}\ \emph
  {et~al.}(2015{\natexlab{b}})\citenamefont {Gasull}, \citenamefont
  {L\'opez-Salcedo},\ and\ \citenamefont
  {Utzet}}]{RePEc:spr:testjl:v:24:y:2015:i:4:p:714-733}%
  \BibitemOpen
  \bibfield  {author} {\bibinfo {author} {\bibfnamefont {A.}~\bibnamefont
  {Gasull}}, \bibinfo {author} {\bibfnamefont {J.}~\bibnamefont
  {L\'opez-Salcedo}},\ and\ \bibinfo {author} {\bibfnamefont {F.}~\bibnamefont
  {Utzet}},\ }\bibfield  {title} {\bibinfo {title} {{Maxima of Gamma random
  variables and other Weibull-like distributions and the Lambert W function}},\
  }\href {https://doi.org/10.1007/s11749-015-0431-9} {\bibfield  {journal}
  {\bibinfo  {journal} {{TEST}}\ }\textbf {\bibinfo {volume} {24}},\ \bibinfo
  {pages} {714} (\bibinfo {year} {2015}{\natexlab{b}})},\ \Eprint
  {https://arxiv.org/abs/1308.5534} {arXiv:1308.5534 [math.ST]} \BibitemShut
  {NoStop}%
\bibitem [{\citenamefont {Abbott}\ \emph
  {et~al.}(2017{\natexlab{b}})\citenamefont {Abbott} \emph
  {et~al.}}]{LIGOScientific:2017ytx}%
  \BibitemOpen
  \bibfield  {author} {\bibinfo {author} {\bibfnamefont {B.~P.}\ \bibnamefont
  {Abbott}} \emph {et~al.} (\bibinfo {collaboration} {LIGO Scientific,
  Virgo}),\ }\bibfield  {title} {\bibinfo {title} {{First narrow-band search
  for continuous gravitational waves from known pulsars in advanced detector
  data}},\ }\href {https://doi.org/10.1103/PhysRevD.96.122006} {\bibfield
  {journal} {\bibinfo  {journal} {Phys. Rev. D}\ }\textbf {\bibinfo {volume}
  {96}},\ \bibinfo {pages} {122006} (\bibinfo {year} {2017}{\natexlab{b}})},\
  \bibinfo {note} {[Erratum: Phys.Rev.D 97, 129903 (2018)]},\ \Eprint
  {https://arxiv.org/abs/1710.02327} {arXiv:1710.02327 [gr-qc]} \BibitemShut
  {NoStop}%
\bibitem [{\citenamefont {Abbott}\ \emph {et~al.}(2019)\citenamefont {Abbott}
  \emph {et~al.}}]{LIGOScientific:2019mhs}%
  \BibitemOpen
  \bibfield  {author} {\bibinfo {author} {\bibfnamefont {B.~P.}\ \bibnamefont
  {Abbott}} \emph {et~al.} (\bibinfo {collaboration} {LIGO Scientific,
  Virgo}),\ }\bibfield  {title} {\bibinfo {title} {{Narrow-band search for
  gravitational waves from known pulsars using the second LIGO observing
  run}},\ }\href {https://doi.org/10.1103/PhysRevD.99.122002} {\bibfield
  {journal} {\bibinfo  {journal} {Phys. Rev. D}\ }\textbf {\bibinfo {volume}
  {99}},\ \bibinfo {pages} {122002} (\bibinfo {year} {2019})},\ \Eprint
  {https://arxiv.org/abs/1902.08442} {arXiv:1902.08442 [gr-qc]} \BibitemShut
  {NoStop}%
\bibitem [{\citenamefont {Middleton}\ \emph {et~al.}(2020)\citenamefont
  {Middleton}, \citenamefont {Clearwater}, \citenamefont {Melatos},\ and\
  \citenamefont {Dunn}}]{Middleton:2020skz}%
  \BibitemOpen
  \bibfield  {author} {\bibinfo {author} {\bibfnamefont {H.}~\bibnamefont
  {Middleton}}, \bibinfo {author} {\bibfnamefont {P.}~\bibnamefont
  {Clearwater}}, \bibinfo {author} {\bibfnamefont {A.}~\bibnamefont
  {Melatos}},\ and\ \bibinfo {author} {\bibfnamefont {L.}~\bibnamefont
  {Dunn}},\ }\bibfield  {title} {\bibinfo {title} {{Search for gravitational
  waves from five low mass X-ray binaries in the second Advanced LIGO observing
  run with an improved hidden Markov model}},\ }\href
  {https://doi.org/10.1103/PhysRevD.102.023006} {\bibfield  {journal} {\bibinfo
   {journal} {Phys. Rev. D}\ }\textbf {\bibinfo {volume} {102}},\ \bibinfo
  {pages} {023006} (\bibinfo {year} {2020})},\ \Eprint
  {https://arxiv.org/abs/2006.06907} {arXiv:2006.06907 [astro-ph.HE]}
  \BibitemShut {NoStop}%
\bibitem [{\citenamefont {Jones}\ and\ \citenamefont
  {Sun}(2021)}]{Jones:2020htx}%
  \BibitemOpen
  \bibfield  {author} {\bibinfo {author} {\bibfnamefont {D.}~\bibnamefont
  {Jones}}\ and\ \bibinfo {author} {\bibfnamefont {L.}~\bibnamefont {Sun}},\
  }\bibfield  {title} {\bibinfo {title} {{Search for continuous gravitational
  waves from Fomalhaut b in the second Advanced LIGO observing run with a
  hidden Markov model}},\ }\href {https://doi.org/10.1103/PhysRevD.103.023020}
  {\bibfield  {journal} {\bibinfo  {journal} {Phys. Rev. D}\ }\textbf {\bibinfo
  {volume} {103}},\ \bibinfo {pages} {023020} (\bibinfo {year} {2021})},\
  \Eprint {https://arxiv.org/abs/2007.08732} {arXiv:2007.08732 [gr-qc]}
  \BibitemShut {NoStop}%
\bibitem [{\citenamefont {Beniwal}\ \emph {et~al.}(2021)\citenamefont
  {Beniwal}, \citenamefont {Clearwater}, \citenamefont {Dunn}, \citenamefont
  {Melatos},\ and\ \citenamefont {Ottaway}}]{Beniwal:2021hvc}%
  \BibitemOpen
  \bibfield  {author} {\bibinfo {author} {\bibfnamefont {D.}~\bibnamefont
  {Beniwal}}, \bibinfo {author} {\bibfnamefont {P.}~\bibnamefont {Clearwater}},
  \bibinfo {author} {\bibfnamefont {L.}~\bibnamefont {Dunn}}, \bibinfo {author}
  {\bibfnamefont {A.}~\bibnamefont {Melatos}},\ and\ \bibinfo {author}
  {\bibfnamefont {D.}~\bibnamefont {Ottaway}},\ }\bibfield  {title} {\bibinfo
  {title} {{Search for continuous gravitational waves from ten H.E.S.S. sources
  using a hidden Markov model}},\ }\href
  {https://doi.org/10.1103/PhysRevD.103.083009} {\bibfield  {journal} {\bibinfo
   {journal} {Phys. Rev. D}\ }\textbf {\bibinfo {volume} {103}},\ \bibinfo
  {pages} {083009} (\bibinfo {year} {2021})},\ \Eprint
  {https://arxiv.org/abs/2102.06334} {arXiv:2102.06334 [astro-ph.HE]}
  \BibitemShut {NoStop}%
\bibitem [{\citenamefont {Isi}\ \emph {et~al.}(2020)\citenamefont {Isi},
  \citenamefont {Mastrogiovanni}, \citenamefont {Pitkin},\ and\ \citenamefont
  {Piccinni}}]{Isi:2020uxj}%
  \BibitemOpen
  \bibfield  {author} {\bibinfo {author} {\bibfnamefont {M.}~\bibnamefont
  {Isi}}, \bibinfo {author} {\bibfnamefont {S.}~\bibnamefont {Mastrogiovanni}},
  \bibinfo {author} {\bibfnamefont {M.}~\bibnamefont {Pitkin}},\ and\ \bibinfo
  {author} {\bibfnamefont {O.~J.}\ \bibnamefont {Piccinni}},\ }\bibfield
  {title} {\bibinfo {title} {{Establishing the significance of continuous
  gravitational-wave detections from known pulsars}},\ }\href
  {https://doi.org/10.1103/PhysRevD.102.123027} {\bibfield  {journal} {\bibinfo
   {journal} {Phys. Rev. D}\ }\textbf {\bibinfo {volume} {102}},\ \bibinfo
  {pages} {123027} (\bibinfo {year} {2020})},\ \Eprint
  {https://arxiv.org/abs/2010.12612} {arXiv:2010.12612 [gr-qc]} \BibitemShut
  {NoStop}%
\bibitem [{\citenamefont {Ashton}\ and\ \citenamefont
  {Prix}(2018)}]{Ashton:2018ure}%
  \BibitemOpen
  \bibfield  {author} {\bibinfo {author} {\bibfnamefont {G.}~\bibnamefont
  {Ashton}}\ and\ \bibinfo {author} {\bibfnamefont {R.}~\bibnamefont {Prix}},\
  }\bibfield  {title} {\bibinfo {title} {{Hierarchical multistage MCMC
  follow-up of continuous gravitational wave candidates}},\ }\href
  {https://doi.org/10.1103/PhysRevD.97.103020} {\bibfield  {journal} {\bibinfo
  {journal} {Phys. Rev. D}\ }\textbf {\bibinfo {volume} {97}},\ \bibinfo
  {pages} {103020} (\bibinfo {year} {2018})},\ \Eprint
  {https://arxiv.org/abs/1802.05450} {arXiv:1802.05450 [astro-ph.IM]}
  \BibitemShut {NoStop}%
\bibitem [{\citenamefont {Keitel}\ \emph {et~al.}(2021)\citenamefont {Keitel},
  \citenamefont {Tenorio}, \citenamefont {Ashton},\ and\ \citenamefont
  {Prix}}]{Keitel2021}%
  \BibitemOpen
  \bibfield  {author} {\bibinfo {author} {\bibfnamefont {D.}~\bibnamefont
  {Keitel}}, \bibinfo {author} {\bibfnamefont {R.}~\bibnamefont {Tenorio}},
  \bibinfo {author} {\bibfnamefont {G.}~\bibnamefont {Ashton}},\ and\ \bibinfo
  {author} {\bibfnamefont {R.}~\bibnamefont {Prix}},\ }\bibfield  {title}
  {\bibinfo {title} {Pyfstat: a python package for continuous
  gravitational-wave data analysis},\ }\href
  {https://doi.org/10.21105/joss.03000} {\bibfield  {journal} {\bibinfo
  {journal} {J. Open Source Softw.}\ }\textbf {\bibinfo {volume} {6}},\
  \bibinfo {pages} {3000} (\bibinfo {year} {2021})}\BibitemShut {NoStop}%
\bibitem [{\citenamefont {Virtanen}\ \emph {et~al.}(2020)\citenamefont
  {Virtanen}, \citenamefont {Gommers}, \citenamefont {Oliphant}, \citenamefont
  {Haberland}, \citenamefont {Reddy} \emph {et~al.}}]{2020SciPy-NMeth}%
  \BibitemOpen
  \bibfield  {author} {\bibinfo {author} {\bibfnamefont {P.}~\bibnamefont
  {Virtanen}}, \bibinfo {author} {\bibfnamefont {R.}~\bibnamefont {Gommers}},
  \bibinfo {author} {\bibfnamefont {T.~E.}\ \bibnamefont {Oliphant}}, \bibinfo
  {author} {\bibfnamefont {M.}~\bibnamefont {Haberland}}, \bibinfo {author}
  {\bibfnamefont {T.}~\bibnamefont {Reddy}}, \emph {et~al.},\ }\bibfield
  {title} {\bibinfo {title} {{{SciPy} 1.0: Fundamental Algorithms for
  Scientific Computing in Python}},\ }\href
  {https://doi.org/10.1038/s41592-019-0686-2} {\bibfield  {journal} {\bibinfo
  {journal} {Nat. Methods}\ }\textbf {\bibinfo {volume} {17}},\ \bibinfo
  {pages} {261} (\bibinfo {year} {2020})}\BibitemShut {NoStop}%
\bibitem [{\citenamefont {Krishnan}\ \emph {et~al.}(2004)\citenamefont
  {Krishnan}, \citenamefont {Sintes}, \citenamefont {Papa}, \citenamefont
  {Schutz}, \citenamefont {Frasca},\ and\ \citenamefont
  {Palomba}}]{PhysRevD.70.082001}%
  \BibitemOpen
  \bibfield  {author} {\bibinfo {author} {\bibfnamefont {B.}~\bibnamefont
  {Krishnan}}, \bibinfo {author} {\bibfnamefont {A.~M.}\ \bibnamefont
  {Sintes}}, \bibinfo {author} {\bibfnamefont {M.~A.}\ \bibnamefont {Papa}},
  \bibinfo {author} {\bibfnamefont {B.~F.}\ \bibnamefont {Schutz}}, \bibinfo
  {author} {\bibfnamefont {S.}~\bibnamefont {Frasca}},\ and\ \bibinfo {author}
  {\bibfnamefont {C.}~\bibnamefont {Palomba}},\ }\bibfield  {title} {\bibinfo
  {title} {Hough transform search for continuous gravitational waves},\ }\href
  {https://doi.org/10.1103/PhysRevD.70.082001} {\bibfield  {journal} {\bibinfo
  {journal} {Phys. Rev. D}\ }\textbf {\bibinfo {volume} {70}},\ \bibinfo
  {pages} {082001} (\bibinfo {year} {2004})},\ \Eprint
  {https://arxiv.org/abs/gr-qc/0407001} {arXiv:gr-qc/0407001} \BibitemShut
  {NoStop}%
\bibitem [{\citenamefont {{Astone}}\ \emph {et~al.}(2014)\citenamefont
  {{Astone}}, \citenamefont {{Colla}}, \citenamefont {{D'Antonio}},
  \citenamefont {{Frasca}},\ and\ \citenamefont
  {{Palomba}}}]{2014PhRvD..90d2002A}%
  \BibitemOpen
  \bibfield  {author} {\bibinfo {author} {\bibfnamefont {P.}~\bibnamefont
  {{Astone}}}, \bibinfo {author} {\bibfnamefont {A.}~\bibnamefont {{Colla}}},
  \bibinfo {author} {\bibfnamefont {S.}~\bibnamefont {{D'Antonio}}}, \bibinfo
  {author} {\bibfnamefont {S.}~\bibnamefont {{Frasca}}},\ and\ \bibinfo
  {author} {\bibfnamefont {C.}~\bibnamefont {{Palomba}}},\ }\bibfield  {title}
  {\bibinfo {title} {{Method for all-sky searches of continuous gravitational
  wave signals using the frequency-Hough transform}},\ }\href
  {https://doi.org/10.1103/PhysRevD.90.042002} {\bibfield  {journal} {\bibinfo
  {journal} {Phys. Rev. D}\ }\textbf {\bibinfo {volume} {90}},\ \bibinfo {eid}
  {042002} (\bibinfo {year} {2014})},\ \Eprint
  {https://arxiv.org/abs/1407.8333} {arXiv:1407.8333 [astro-ph.IM]}
  \BibitemShut {NoStop}%
\bibitem [{\citenamefont {{Miller}}\ \emph {et~al.}(2018)\citenamefont
  {{Miller}}, \citenamefont {{Astone}}, \citenamefont {{D'Antonio}},
  \citenamefont {{Frasca}}, \citenamefont {{Intini}}, \citenamefont {{La
  Rosa}}, \citenamefont {{Leaci}}, \citenamefont {{Mastrogiovanni}},
  \citenamefont {{Muciaccia}}, \citenamefont {{Palomba}}, \citenamefont
  {{Piccinni}}, \citenamefont {{Singhal}},\ and\ \citenamefont
  {{Whiting}}}]{Miller:2018rbg}%
  \BibitemOpen
  \bibfield  {author} {\bibinfo {author} {\bibfnamefont {A.}~\bibnamefont
  {{Miller}}}, \bibinfo {author} {\bibfnamefont {P.}~\bibnamefont {{Astone}}},
  \bibinfo {author} {\bibfnamefont {S.}~\bibnamefont {{D'Antonio}}}, \bibinfo
  {author} {\bibfnamefont {S.}~\bibnamefont {{Frasca}}}, \bibinfo {author}
  {\bibfnamefont {G.}~\bibnamefont {{Intini}}}, \bibinfo {author}
  {\bibfnamefont {I.}~\bibnamefont {{La Rosa}}}, \bibinfo {author}
  {\bibfnamefont {P.}~\bibnamefont {{Leaci}}}, \bibinfo {author} {\bibfnamefont
  {S.}~\bibnamefont {{Mastrogiovanni}}}, \bibinfo {author} {\bibfnamefont
  {F.}~\bibnamefont {{Muciaccia}}}, \bibinfo {author} {\bibfnamefont
  {C.}~\bibnamefont {{Palomba}}}, \bibinfo {author} {\bibfnamefont {O.~J.}\
  \bibnamefont {{Piccinni}}}, \bibinfo {author} {\bibfnamefont
  {A.}~\bibnamefont {{Singhal}}},\ and\ \bibinfo {author} {\bibfnamefont
  {B.~F.}\ \bibnamefont {{Whiting}}},\ }\bibfield  {title} {\bibinfo {title}
  {{Method to search for long duration gravitational wave transients from
  isolated neutron stars using the generalized frequency-Hough transform}},\
  }\href {https://doi.org/10.1103/PhysRevD.98.102004} {\bibfield  {journal}
  {\bibinfo  {journal} {Phys. Rev. D}\ }\textbf {\bibinfo {volume} {98}},\
  \bibinfo {pages} {102004} (\bibinfo {year} {2018})},\ \Eprint
  {https://arxiv.org/abs/1810.09784} {arXiv:1810.09784 [astro-ph.IM]}
  \BibitemShut {NoStop}%
\bibitem [{\citenamefont {Oliver}\ \emph {et~al.}(2019)\citenamefont {Oliver},
  \citenamefont {Keitel},\ and\ \citenamefont {Sintes}}]{Oliver:2019ksl}%
  \BibitemOpen
  \bibfield  {author} {\bibinfo {author} {\bibfnamefont {M.}~\bibnamefont
  {Oliver}}, \bibinfo {author} {\bibfnamefont {D.}~\bibnamefont {Keitel}},\
  and\ \bibinfo {author} {\bibfnamefont {A.~M.}\ \bibnamefont {Sintes}},\
  }\bibfield  {title} {\bibinfo {title} {{Adaptive transient Hough method for
  long-duration gravitational wave transients}},\ }\href
  {https://doi.org/10.1103/PhysRevD.99.104067} {\bibfield  {journal} {\bibinfo
  {journal} {Phys. Rev. D}\ }\textbf {\bibinfo {volume} {99}},\ \bibinfo
  {pages} {104067} (\bibinfo {year} {2019})},\ \Eprint
  {https://arxiv.org/abs/1901.01820} {arXiv:1901.01820 [gr-qc]} \BibitemShut
  {NoStop}%
\bibitem [{\citenamefont {Covas}\ and\ \citenamefont
  {Sintes}(2019)}]{Covas:2019jqa}%
  \BibitemOpen
  \bibfield  {author} {\bibinfo {author} {\bibfnamefont {P.~B.}\ \bibnamefont
  {Covas}}\ and\ \bibinfo {author} {\bibfnamefont {A.~M.}\ \bibnamefont
  {Sintes}},\ }\bibfield  {title} {\bibinfo {title} {{New method to search for
  continuous gravitational waves from unknown neutron stars in binary
  systems}},\ }\href {https://doi.org/10.1103/PhysRevD.99.124019} {\bibfield
  {journal} {\bibinfo  {journal} {Phys. Rev. D}\ }\textbf {\bibinfo {volume}
  {99}},\ \bibinfo {pages} {124019} (\bibinfo {year} {2019})},\ \Eprint
  {https://arxiv.org/abs/1904.04873} {arXiv:1904.04873 [astro-ph.IM]}
  \BibitemShut {NoStop}%
\bibitem [{\citenamefont {Whelan}\ \emph {et~al.}(2015)\citenamefont {Whelan},
  \citenamefont {Sundaresan}, \citenamefont {Zhang},\ and\ \citenamefont
  {Peiris}}]{Whelan:2015dha}%
  \BibitemOpen
  \bibfield  {author} {\bibinfo {author} {\bibfnamefont {J.~T.}\ \bibnamefont
  {Whelan}}, \bibinfo {author} {\bibfnamefont {S.}~\bibnamefont {Sundaresan}},
  \bibinfo {author} {\bibfnamefont {Y.}~\bibnamefont {Zhang}},\ and\ \bibinfo
  {author} {\bibfnamefont {P.}~\bibnamefont {Peiris}},\ }\bibfield  {title}
  {\bibinfo {title} {{Model-Based Cross-Correlation Search for Gravitational
  Waves from Scorpius X-1}},\ }\href
  {https://doi.org/10.1103/PhysRevD.91.102005} {\bibfield  {journal} {\bibinfo
  {journal} {Phys. Rev. D}\ }\textbf {\bibinfo {volume} {91}},\ \bibinfo
  {pages} {102005} (\bibinfo {year} {2015})},\ \Eprint
  {https://arxiv.org/abs/1504.05890} {arXiv:1504.05890 [gr-qc]} \BibitemShut
  {NoStop}%
\bibitem [{\citenamefont {Dergachev}(2010)}]{Dergachev:2010tm}%
  \BibitemOpen
  \bibfield  {author} {\bibinfo {author} {\bibfnamefont {V.}~\bibnamefont
  {Dergachev}},\ }\bibfield  {title} {\bibinfo {title} {{On blind searches for
  noise dominated signals: a loosely coherent approach}},\ }\href
  {https://doi.org/10.1088/0264-9381/27/20/205017} {\bibfield  {journal}
  {\bibinfo  {journal} {Class. Quant. Grav.}\ }\textbf {\bibinfo {volume}
  {27}},\ \bibinfo {pages} {205017} (\bibinfo {year} {2010})},\ \Eprint
  {https://arxiv.org/abs/1003.2178} {arXiv:1003.2178 [gr-qc]} \BibitemShut
  {NoStop}%
\bibitem [{\citenamefont {Goetz}\ and\ \citenamefont
  {Riles}(2011)}]{Goetz:2011bd}%
  \BibitemOpen
  \bibfield  {author} {\bibinfo {author} {\bibfnamefont {E.}~\bibnamefont
  {Goetz}}\ and\ \bibinfo {author} {\bibfnamefont {K.}~\bibnamefont {Riles}},\
  }\bibfield  {title} {\bibinfo {title} {{An all-sky search algorithm for
  continuous gravitational waves from spinning neutron stars in binary
  systems}},\ }\href {https://doi.org/10.1088/0264-9381/28/21/215006}
  {\bibfield  {journal} {\bibinfo  {journal} {Class. Quant. Grav.}\ }\textbf
  {\bibinfo {volume} {28}},\ \bibinfo {pages} {215006} (\bibinfo {year}
  {2011})},\ \Eprint {https://arxiv.org/abs/1103.1301} {arXiv:1103.1301
  [gr-qc]} \BibitemShut {NoStop}%
\bibitem [{\citenamefont {Dergachev}\ and\ \citenamefont
  {Papa}(2019)}]{Dergachev:2019wqa}%
  \BibitemOpen
  \bibfield  {author} {\bibinfo {author} {\bibfnamefont {V.}~\bibnamefont
  {Dergachev}}\ and\ \bibinfo {author} {\bibfnamefont {M.~A.}\ \bibnamefont
  {Papa}},\ }\bibfield  {title} {\bibinfo {title} {{Sensitivity improvements in
  the search for periodic gravitational waves using O1 LIGO data}},\ }\href
  {https://doi.org/10.1103/PhysRevLett.123.101101} {\bibfield  {journal}
  {\bibinfo  {journal} {Phys. Rev. Lett.}\ }\textbf {\bibinfo {volume} {123}},\
  \bibinfo {pages} {101101} (\bibinfo {year} {2019})},\ \Eprint
  {https://arxiv.org/abs/1902.05530} {arXiv:1902.05530 [gr-qc]} \BibitemShut
  {NoStop}%
\bibitem [{\citenamefont {Abbott}\ \emph {et~al.}(2022)\citenamefont {Abbott}
  \emph {et~al.}}]{LIGOScientific:2021ozr}%
  \BibitemOpen
  \bibfield  {author} {\bibinfo {author} {\bibfnamefont {R.}~\bibnamefont
  {Abbott}} \emph {et~al.} (\bibinfo {collaboration} {LIGO Scientific, VIRGO,
  KAGRA}),\ }\bibfield  {title} {\bibinfo {title} {{Search for continuous
  gravitational waves from 20 accreting millisecond X-ray pulsars in O3 LIGO
  data}},\ }\href {https://doi.org/10.1103/PhysRevD.105.022002} {\bibfield
  {journal} {\bibinfo  {journal} {Phys. Rev. D}\ }\textbf {\bibinfo {volume}
  {105}},\ \bibinfo {pages} {022002} (\bibinfo {year} {2022})},\ \Eprint
  {https://arxiv.org/abs/2109.09255} {arXiv:2109.09255 [astro-ph.HE]}
  \BibitemShut {NoStop}%
\bibitem [{\citenamefont {Pletsch}\ and\ \citenamefont
  {Allen}(2009)}]{Pletsch:2009uu}%
  \BibitemOpen
  \bibfield  {author} {\bibinfo {author} {\bibfnamefont {H.~J.}\ \bibnamefont
  {Pletsch}}\ and\ \bibinfo {author} {\bibfnamefont {B.}~\bibnamefont
  {Allen}},\ }\bibfield  {title} {\bibinfo {title} {{Exploiting global
  correlations to detect continuous gravitational waves}},\ }\href
  {https://doi.org/10.1103/PhysRevLett.103.181102} {\bibfield  {journal}
  {\bibinfo  {journal} {Phys. Rev. Lett.}\ }\textbf {\bibinfo {volume} {103}},\
  \bibinfo {pages} {181102} (\bibinfo {year} {2009})},\ \Eprint
  {https://arxiv.org/abs/0906.0023} {arXiv:0906.0023 [gr-qc]} \BibitemShut
  {NoStop}%
\bibitem [{\citenamefont {Wette}\ \emph {et~al.}(2018)\citenamefont {Wette},
  \citenamefont {Walsh}, \citenamefont {Prix},\ and\ \citenamefont
  {Papa}}]{Wette:2018bhc}%
  \BibitemOpen
  \bibfield  {author} {\bibinfo {author} {\bibfnamefont {K.}~\bibnamefont
  {Wette}}, \bibinfo {author} {\bibfnamefont {S.}~\bibnamefont {Walsh}},
  \bibinfo {author} {\bibfnamefont {R.}~\bibnamefont {Prix}},\ and\ \bibinfo
  {author} {\bibfnamefont {M.~A.}\ \bibnamefont {Papa}},\ }\bibfield  {title}
  {\bibinfo {title} {{Implementing a semicoherent search for continuous
  gravitational waves using optimally-constructed template banks}},\ }\href
  {https://doi.org/10.1103/PhysRevD.97.123016} {\bibfield  {journal} {\bibinfo
  {journal} {Phys. Rev. D}\ }\textbf {\bibinfo {volume} {97}},\ \bibinfo
  {pages} {123016} (\bibinfo {year} {2018})},\ \Eprint
  {https://arxiv.org/abs/1804.03392} {arXiv:1804.03392 [astro-ph.IM]}
  \BibitemShut {NoStop}%
\bibitem [{\citenamefont {Abbott}\ \emph
  {et~al.}(2021{\natexlab{a}})\citenamefont {Abbott} \emph
  {et~al.}}]{LIGOScientific:2021quq}%
  \BibitemOpen
  \bibfield  {author} {\bibinfo {author} {\bibfnamefont {R.}~\bibnamefont
  {Abbott}} \emph {et~al.} (\bibinfo {collaboration} {LIGO Scientific, VIRGO,
  KAGRA}),\ }\bibfield  {title} {\bibinfo {title} {{Narrowband searches for
  continuous and long-duration transient gravitational waves from known pulsars
  in the LIGO-Virgo third observing run}},\ }\href@noop {} {\bibfield
  {journal} {\bibinfo  {journal} {arXiv e-print}\ } (\bibinfo {year}
  {2021}{\natexlab{a}})},\ \Eprint {https://arxiv.org/abs/2112.10990}
  {arXiv:2112.10990 [gr-qc]} \BibitemShut {NoStop}%
\bibitem [{\citenamefont {Modafferi}\ \emph {et~al.}(2022)\citenamefont
  {Modafferi}, \citenamefont {Moragues},\ and\ \citenamefont
  {Keitel}}]{narrow_band_proc}%
  \BibitemOpen
  \bibfield  {author} {\bibinfo {author} {\bibfnamefont {L.~M.}\ \bibnamefont
  {Modafferi}}, \bibinfo {author} {\bibfnamefont {J.}~\bibnamefont
  {Moragues}},\ and\ \bibinfo {author} {\bibfnamefont {D.}~\bibnamefont
  {Keitel}} (\bibinfo {collaboration} {LIGO Scientific, VIRGO, KAGRA}),\
  }\bibfield  {title} {\bibinfo {title} {{Search setup for long-duration
  transient gravitational waves from glitching pulsars during LIGO-Virgo third
  observing run}}\ }(\bibinfo {year} {2022})\ \Eprint
  {https://arxiv.org/abs/2201.08785} {arXiv:2201.08785 [gr-qc]} \BibitemShut
  {NoStop}%
\bibitem [{\citenamefont {Abbott}\ \emph
  {et~al.}(2021{\natexlab{b}})\citenamefont {Abbott} \emph {et~al.}}]{O2Data}%
  \BibitemOpen
  \bibfield  {author} {\bibinfo {author} {\bibfnamefont {R.}~\bibnamefont
  {Abbott}} \emph {et~al.} (\bibinfo {collaboration} {{LIGO Scientific
  Collaboration and Virgo Collaboration}}),\ }\bibfield  {title} {\bibinfo
  {title} {Open data from the first and second observing runs of advanced ligo
  and advanced virgo},\ }\href {https://doi.org/10.1016/j.softx.2021.100658}
  {\bibfield  {journal} {\bibinfo  {journal} {SoftwareX}\ }\textbf {\bibinfo
  {volume} {13}},\ \bibinfo {pages} {100658} (\bibinfo {year}
  {2021}{\natexlab{b}})},\ \Eprint {https://arxiv.org/abs/1912.11716}
  {arXiv:1912.11716 [gr-qc]} \BibitemShut {NoStop}%
\bibitem [{\citenamefont {{Prix, Reinhard}}(2006)}]{CFSv2bias}%
  \BibitemOpen
  \bibfield  {author} {\bibinfo {author} {\bibnamefont {{Prix, Reinhard}}},\
  }\href@noop {} {\bibinfo {title} {{F-statistic bias due to
  noise-estimator}}},\ \bibinfo {howpublished}
  {\url{https://dcc.ligo.org/LIGO-T1100551/public}} (\bibinfo {year}
  {2006})\BibitemShut {NoStop}%
\bibitem [{\citenamefont {Steltner}\ \emph {et~al.}(2021)\citenamefont
  {Steltner}, \citenamefont {Papa}, \citenamefont {Eggenstein}, \citenamefont
  {Allen}, \citenamefont {Dergachev}, \citenamefont {Prix}, \citenamefont
  {Machenschalk}, \citenamefont {Walsh}, \citenamefont {Zhu},\ and\
  \citenamefont {Kwang}}]{Steltner:2020hfd}%
  \BibitemOpen
  \bibfield  {author} {\bibinfo {author} {\bibfnamefont {B.}~\bibnamefont
  {Steltner}}, \bibinfo {author} {\bibfnamefont {M.~A.}\ \bibnamefont {Papa}},
  \bibinfo {author} {\bibfnamefont {H.~B.}\ \bibnamefont {Eggenstein}},
  \bibinfo {author} {\bibfnamefont {B.}~\bibnamefont {Allen}}, \bibinfo
  {author} {\bibfnamefont {V.}~\bibnamefont {Dergachev}}, \bibinfo {author}
  {\bibfnamefont {R.}~\bibnamefont {Prix}}, \bibinfo {author} {\bibfnamefont
  {B.}~\bibnamefont {Machenschalk}}, \bibinfo {author} {\bibfnamefont
  {S.}~\bibnamefont {Walsh}}, \bibinfo {author} {\bibfnamefont {S.~J.}\
  \bibnamefont {Zhu}},\ and\ \bibinfo {author} {\bibfnamefont {S.}~\bibnamefont
  {Kwang}},\ }\bibfield  {title} {\bibinfo {title} {{Einstein@Home All-sky
  Search for Continuous Gravitational Waves in LIGO O2 Public Data}},\ }\href
  {https://doi.org/10.3847/1538-4357/abc7c9} {\bibfield  {journal} {\bibinfo
  {journal} {Astrophys. J.}\ }\textbf {\bibinfo {volume} {909}},\ \bibinfo
  {pages} {79} (\bibinfo {year} {2021})},\ \Eprint
  {https://arxiv.org/abs/2009.12260} {arXiv:2009.12260 [astro-ph.HE]}
  \BibitemShut {NoStop}%
\bibitem [{\citenamefont {{Ming}}\ \emph {et~al.}(2019)\citenamefont {{Ming}},
  \citenamefont {{Papa}}, \citenamefont {{Singh}}, \citenamefont
  {{Eggenstein}}, \citenamefont {{Zhu}}, \citenamefont {{Dergachev}},
  \citenamefont {{Hu}}, \citenamefont {{Prix}}, \citenamefont {{Machenschalk}},
  \citenamefont {{Beer}}, \citenamefont {{Behnke}},\ and\ \citenamefont
  {{Allen}}}]{Ming:2019xse}%
  \BibitemOpen
  \bibfield  {author} {\bibinfo {author} {\bibfnamefont {J.}~\bibnamefont
  {{Ming}}}, \bibinfo {author} {\bibfnamefont {M.~A.}\ \bibnamefont {{Papa}}},
  \bibinfo {author} {\bibfnamefont {A.}~\bibnamefont {{Singh}}}, \bibinfo
  {author} {\bibfnamefont {H.~B.}\ \bibnamefont {{Eggenstein}}}, \bibinfo
  {author} {\bibfnamefont {S.~J.}\ \bibnamefont {{Zhu}}}, \bibinfo {author}
  {\bibfnamefont {V.}~\bibnamefont {{Dergachev}}}, \bibinfo {author}
  {\bibfnamefont {Y.}~\bibnamefont {{Hu}}}, \bibinfo {author} {\bibfnamefont
  {R.}~\bibnamefont {{Prix}}}, \bibinfo {author} {\bibfnamefont
  {B.}~\bibnamefont {{Machenschalk}}}, \bibinfo {author} {\bibfnamefont
  {C.}~\bibnamefont {{Beer}}}, \bibinfo {author} {\bibfnamefont
  {O.}~\bibnamefont {{Behnke}}},\ and\ \bibinfo {author} {\bibfnamefont
  {B.}~\bibnamefont {{Allen}}},\ }\bibfield  {title} {\bibinfo {title}
  {{Results from an Einstein@Home search for continuous gravitational waves
  from Cassiopeia A, Vela Jr. and G347.3}},\ }\href
  {https://doi.org/10.1103/PhysRevD.100.024063} {\bibfield  {journal} {\bibinfo
   {journal} {Phys. Rev. D}\ }\textbf {\bibinfo {volume} {100}},\ \bibinfo
  {pages} {024063} (\bibinfo {year} {2019})},\ \Eprint
  {https://arxiv.org/abs/1903.09119} {arXiv:1903.09119 [gr-qc]} \BibitemShut
  {NoStop}%
\bibitem [{\citenamefont {Papa}\ \emph {et~al.}(2016)\citenamefont {Papa},
  \citenamefont {Eggenstein}, \citenamefont {Walsh}, \citenamefont {DiPalma},
  \citenamefont {Allen}, \citenamefont {Astone}, \citenamefont {Bock},
  \citenamefont {Creighton}, \citenamefont {Keitel}, \citenamefont
  {Machenschalk}, \citenamefont {Prix}, \citenamefont {Siemens}, \citenamefont
  {Singh}, \citenamefont {Zhu},\ and\ \citenamefont {Schutz}}]{Papa:2016cwb}%
  \BibitemOpen
  \bibfield  {author} {\bibinfo {author} {\bibfnamefont {M.~A.}\ \bibnamefont
  {Papa}}, \bibinfo {author} {\bibfnamefont {H.~B.}\ \bibnamefont
  {Eggenstein}}, \bibinfo {author} {\bibfnamefont {S.}~\bibnamefont {Walsh}},
  \bibinfo {author} {\bibfnamefont {I.}~\bibnamefont {DiPalma}}, \bibinfo
  {author} {\bibfnamefont {B.}~\bibnamefont {Allen}}, \bibinfo {author}
  {\bibfnamefont {P.}~\bibnamefont {Astone}}, \bibinfo {author} {\bibfnamefont
  {O.}~\bibnamefont {Bock}}, \bibinfo {author} {\bibfnamefont {T.~D.}\
  \bibnamefont {Creighton}}, \bibinfo {author} {\bibfnamefont {D.}~\bibnamefont
  {Keitel}}, \bibinfo {author} {\bibfnamefont {B.}~\bibnamefont
  {Machenschalk}}, \bibinfo {author} {\bibfnamefont {R.}~\bibnamefont {Prix}},
  \bibinfo {author} {\bibfnamefont {X.}~\bibnamefont {Siemens}}, \bibinfo
  {author} {\bibfnamefont {A.}~\bibnamefont {Singh}}, \bibinfo {author}
  {\bibfnamefont {S.~J.}\ \bibnamefont {Zhu}},\ and\ \bibinfo {author}
  {\bibfnamefont {B.~F.}\ \bibnamefont {Schutz}},\ }\bibfield  {title}
  {\bibinfo {title} {{Hierarchical follow-up of subthreshold candidates of an
  all-sky Einstein@Home search for continuous gravitational waves on LIGO sixth
  science run data}},\ }\href {https://doi.org/10.1103/PhysRevD.94.122006}
  {\bibfield  {journal} {\bibinfo  {journal} {Phys. Rev. D}\ }\textbf {\bibinfo
  {volume} {94}},\ \bibinfo {pages} {122006} (\bibinfo {year} {2016})},\
  \Eprint {https://arxiv.org/abs/1608.08928} {arXiv:1608.08928 [astro-ph.IM]}
  \BibitemShut {NoStop}%
\bibitem [{\citenamefont {Ming}\ \emph {et~al.}(2022)\citenamefont {Ming},
  \citenamefont {Papa}, \citenamefont {Eggenstein}, \citenamefont
  {Machenschalk}, \citenamefont {Steltner}, \citenamefont {Prix} \emph
  {et~al.}}]{Ming:2021xtz}%
  \BibitemOpen
  \bibfield  {author} {\bibinfo {author} {\bibfnamefont {J.}~\bibnamefont
  {Ming}}, \bibinfo {author} {\bibfnamefont {M.~A.}\ \bibnamefont {Papa}},
  \bibinfo {author} {\bibfnamefont {H.-B.}\ \bibnamefont {Eggenstein}},
  \bibinfo {author} {\bibfnamefont {B.}~\bibnamefont {Machenschalk}}, \bibinfo
  {author} {\bibfnamefont {B.}~\bibnamefont {Steltner}}, \bibinfo {author}
  {\bibfnamefont {R.}~\bibnamefont {Prix}}, \emph {et~al.},\ }\bibfield
  {title} {\bibinfo {title} {Results from an einstein@home search for
  continuous gravitational waves from g347.3 at low frequencies in {LIGO} o2
  data},\ }\href {https://doi.org/10.3847/1538-4357/ac35cb} {\bibfield
  {journal} {\bibinfo  {journal} {Astrophys. J.}\ }\textbf {\bibinfo {volume}
  {925}},\ \bibinfo {pages} {8} (\bibinfo {year} {2022})},\ \Eprint
  {https://arxiv.org/abs/2108.02808} {arXiv:2108.02808 [gr-qc]} \BibitemShut
  {NoStop}%
\bibitem [{\citenamefont {Hall}(1979)}]{10.2307/3212912}%
  \BibitemOpen
  \bibfield  {author} {\bibinfo {author} {\bibfnamefont {P.}~\bibnamefont
  {Hall}},\ }\bibfield  {title} {\bibinfo {title} {On the rate of convergence
  of normal extremes},\ }\href {http://www.jstor.org/stable/3212912} {\bibfield
   {journal} {\bibinfo  {journal} {Journal of Applied Probability}\ }\textbf
  {\bibinfo {volume} {16}},\ \bibinfo {pages} {433} (\bibinfo {year}
  {1979})}\BibitemShut {NoStop}%
\bibitem [{\citenamefont {Bennett}\ \emph {et~al.}(2013)\citenamefont
  {Bennett}, \citenamefont {Melatos}, \citenamefont {Delaigle},\ and\
  \citenamefont {Hall}}]{Bennett:2013nea}%
  \BibitemOpen
  \bibfield  {author} {\bibinfo {author} {\bibfnamefont {M.~F.}\ \bibnamefont
  {Bennett}}, \bibinfo {author} {\bibfnamefont {A.}~\bibnamefont {Melatos}},
  \bibinfo {author} {\bibfnamefont {A.}~\bibnamefont {Delaigle}},\ and\
  \bibinfo {author} {\bibfnamefont {P.}~\bibnamefont {Hall}},\ }\bibfield
  {title} {\bibinfo {title} {{Reanalysis of $\mathcal{F}$-statistic
  gravitational-wave searches with the higher criticism statistic}},\ }\href
  {https://doi.org/10.1088/0004-637X/766/2/99} {\bibfield  {journal} {\bibinfo
  {journal} {Astrophys. J.}\ }\textbf {\bibinfo {volume} {766}},\ \bibinfo
  {pages} {99} (\bibinfo {year} {2013})},\ \Eprint
  {https://arxiv.org/abs/1302.2635} {arXiv:1302.2635 [astro-ph.SR]}
  \BibitemShut {NoStop}%
\bibitem [{\citenamefont {Sezgin}\ and\ \citenamefont
  {Sankur}(2004)}]{sezgin2004survey}%
  \BibitemOpen
  \bibfield  {author} {\bibinfo {author} {\bibfnamefont {M.}~\bibnamefont
  {Sezgin}}\ and\ \bibinfo {author} {\bibfnamefont {B.}~\bibnamefont
  {Sankur}},\ }\bibfield  {title} {\bibinfo {title} {Survey over image
  thresholding techniques and quantitative performance evaluation},\ }\href
  {https://doi.org/10.1117/1.1631315} {\bibfield  {journal} {\bibinfo
  {journal} {J. Electron. Imaging}\ }\textbf {\bibinfo {volume} {13}},\
  \bibinfo {pages} {146} (\bibinfo {year} {2004})}\BibitemShut {NoStop}%
\bibitem [{\citenamefont {Otsu}(1979)}]{4310076}%
  \BibitemOpen
  \bibfield  {author} {\bibinfo {author} {\bibfnamefont {N.}~\bibnamefont
  {Otsu}},\ }\bibfield  {title} {\bibinfo {title} {A threshold selection method
  from gray-level histograms},\ }\href
  {https://doi.org/10.1109/TSMC.1979.4310076} {\bibfield  {journal} {\bibinfo
  {journal} {IEEE Trans. Syst. Man Cybern.}\ }\textbf {\bibinfo {volume} {9}},\
  \bibinfo {pages} {62} (\bibinfo {year} {1979})}\BibitemShut {NoStop}%
\bibitem [{\citenamefont {Li}\ and\ \citenamefont {Lee}(1993)}]{LI1993617}%
  \BibitemOpen
  \bibfield  {author} {\bibinfo {author} {\bibfnamefont {C.}~\bibnamefont
  {Li}}\ and\ \bibinfo {author} {\bibfnamefont {C.}~\bibnamefont {Lee}},\
  }\bibfield  {title} {\bibinfo {title} {Minimum cross entropy thresholding},\
  }\href {https://doi.org/https://doi.org/10.1016/0031-3203(93)90115-D}
  {\bibfield  {journal} {\bibinfo  {journal} {Pattern Recognit.}\ }\textbf
  {\bibinfo {volume} {26}},\ \bibinfo {pages} {617} (\bibinfo {year}
  {1993})}\BibitemShut {NoStop}%
\bibitem [{\citenamefont {Li}\ and\ \citenamefont {Tam}(1998)}]{LI1998771}%
  \BibitemOpen
  \bibfield  {author} {\bibinfo {author} {\bibfnamefont {C.}~\bibnamefont
  {Li}}\ and\ \bibinfo {author} {\bibfnamefont {P.}~\bibnamefont {Tam}},\
  }\bibfield  {title} {\bibinfo {title} {An iterative algorithm for minimum
  cross entropy thresholding},\ }\href
  {https://doi.org/https://doi.org/10.1016/S0167-8655(98)00057-9} {\bibfield
  {journal} {\bibinfo  {journal} {Pattern Recognit. Lett.}\ }\textbf {\bibinfo
  {volume} {19}},\ \bibinfo {pages} {771} (\bibinfo {year} {1998})}\BibitemShut
  {NoStop}%
\bibitem [{\citenamefont {Prewitt}\ and\ \citenamefont
  {Mendelsohn}(1966)}]{min_threshold}%
  \BibitemOpen
  \bibfield  {author} {\bibinfo {author} {\bibfnamefont {J.~M.~S.}\
  \bibnamefont {Prewitt}}\ and\ \bibinfo {author} {\bibfnamefont {M.~L.}\
  \bibnamefont {Mendelsohn}},\ }\bibfield  {title} {\bibinfo {title} {The
  analysis of cell images},\ }\href
  {https://doi.org/https://doi.org/10.1111/j.1749-6632.1965.tb11715.x}
  {\bibfield  {journal} {\bibinfo  {journal} {Ann. N. Y. Acad. Sci.}\ }\textbf
  {\bibinfo {volume} {128}},\ \bibinfo {pages} {1035} (\bibinfo {year}
  {1966})}\BibitemShut {NoStop}%
\bibitem [{\citenamefont {van~der Walt}\ \emph {et~al.}(2014)\citenamefont
  {van~der Walt}, \citenamefont {{S}ch\"onberger}, \citenamefont
  {{Nunez-Iglesias}}, \citenamefont {{B}oulogne}, \citenamefont {{W}arner},
  \citenamefont {{Y}ager}, \citenamefont {{G}ouillart}, \citenamefont {{Y}u},\
  and\ \citenamefont {the scikit-image contributors}}]{scikit-image}%
  \BibitemOpen
  \bibfield  {author} {\bibinfo {author} {\bibfnamefont {S.}~\bibnamefont
  {van~der Walt}}, \bibinfo {author} {\bibfnamefont {J.~L.}\ \bibnamefont
  {{S}ch\"onberger}}, \bibinfo {author} {\bibfnamefont {J.}~\bibnamefont
  {{Nunez-Iglesias}}}, \bibinfo {author} {\bibfnamefont {F.}~\bibnamefont
  {{B}oulogne}}, \bibinfo {author} {\bibfnamefont {J.~D.}\ \bibnamefont
  {{W}arner}}, \bibinfo {author} {\bibfnamefont {N.}~\bibnamefont {{Y}ager}},
  \bibinfo {author} {\bibfnamefont {E.}~\bibnamefont {{G}ouillart}}, \bibinfo
  {author} {\bibfnamefont {T.}~\bibnamefont {{Y}u}},\ and\ \bibinfo {author}
  {\bibnamefont {the scikit-image contributors}},\ }\bibfield  {title}
  {\bibinfo {title} {scikit-image: image processing in {P}ython},\ }\href
  {https://doi.org/10.7717/peerj.453} {\bibfield  {journal} {\bibinfo
  {journal} {PeerJ}\ }\textbf {\bibinfo {volume} {2}},\ \bibinfo {pages} {e453}
  (\bibinfo {year} {2014})}\BibitemShut {NoStop}%
\end{thebibliography}%

\end{document}